\documentclass[11pt,a4paper]{article}
\usepackage{epsfig}
\usepackage[T1]{fontenc}    
\usepackage{graphics}
\usepackage{graphicx}
\usepackage{pstricks,pst-coil,pst-fill,pst-plot}
\usepackage[fleqn]{amsmath}    
\usepackage{amssymb}    
\usepackage{amsfonts}   
\usepackage{verbatim}   
\usepackage{mathrsfs}   
\usepackage{dsfont}
\usepackage{euscript}
\usepackage{yfonts}
\usepackage{enumerate}     
\usepackage{amsthm}         
\usepackage{txfonts}
\usepackage{marvosym}
\usepackage{stmaryrd}
\usepackage{vmargin}        
\usepackage{wasysym}		
\usepackage{upgreek}

\setmarginsrb{1.8cm}{2cm}{1.8cm}{2cm}{1cm}{1cm}{1cm}{1.6cm}
 \makeatletter
 \@addtoreset{equation}{section}
 \makeatother

\providecommand{\MR}{\relax\ifhmode\unskip\space\fi MR }

\providecommand{\href}[2]{#2}

\usepackage{comment} 

\let\ua=\uparrow
\let\da=\downarrow
\let\tend=\rightarrow

\long\def\symbolfootnote[#1]#2{\begingroup%
\def\thefootnote{\fnsymbol{footnote}}\footnote[#1]{#2}\endgroup}

\newtheorem{theorem}{Theorem}[section]
\newtheorem{prop}[theorem]{Proposition}
\newtheorem*{theorem*}{Theorem}

\newtheorem{defin}[theorem]{Definition}

\newtheorem{lemme}[theorem]{Lemma}

\def\Proof{\medskip\noindent {\it Proof --- \ }}

\def\qed{\hfill\rule{2mm}{2mm}}

\newcommand\beq{\begin{equation}}
\newcommand\enq{\end{equation}}
\newcommand\bem{\begin{multline}}
\newcommand\enm{\end{multline}}

\def\beqa{\begin{eqnarray}}
\def\eeqa{\end{eqnarray}}
\def\ba{\begin{array}}
\def\ea{\end{array}}
\def\det{\operatorname{det}}

\newcommand{\f}[2]{{\ensuremath{%
    \mathchoice%
    {\dfrac{#1}{#2}}
    {\dfrac{#1}{#2}}
    {\frac{#1}{#2}}
    {\frac{#1}{#2}}
}}}
\newcommand{\tf}[2]{\ensuremath{#1/#2}}

\def\a{\alpha}

\def\be{\beta}
\def\ga{\gamma}
\def\Ga{\Gamma}

\def\de{\delta}

\def\De{\Delta}
\def\eps{\epsilon}
\def\veps{\varepsilon}
\def\la{\lambda}

\def\sg{\sigma}
\def\vsg{\varsigma}
\def\Sg{\Sigma}

\def\Ups{\Upsilon}
\def\ups{\upsilon}
\def\th{\theta}

\def\Om{\Omega}
\def\om{\omega}
\def\vp{\varphi}

\newcommand{\mc}[1]{\ensuremath{\mathcal{#1}}}
\newcommand{\mf}[1]{\ensuremath{\mathfrak{#1}}}
\newcommand{\msc}[1]{\ensuremath{\mathscr{#1}}}

\newcommand{\bs}[1]{\ensuremath{\boldsymbol{#1}}}

\DeclareFontFamily{OT1}{pzc}{}
\DeclareFontShape{OT1}{pzc}{m}{it}{<-> s * [1.10] pzcmi7t}{}
\DeclareMathAlphabet{\mathpzc}{OT1}{pzc}{m}{it}

\def \i{ \mathrm i}

\newcommand{\ov}[1]{\ensuremath{\overline{#1}}}
\newcommand{\wt}[1]{\ensuremath{\widetilde{#1}}}
\newcommand{\wh}[1]{\ensuremath{\widehat{#1}}}

\newcommand{\Int}[2]{\ensuremath{\int\limits_{#1}^{#2}}}

\newcommand{\Fint}[2]{\ensuremath{\fint\limits_{#1}^{#2}}}

\newcommand{\sul}[2]{\ensuremath{\sum\limits_{#1}^{#2}}}
\newcommand{\pl}[2]{\ensuremath{\prod\limits_{#1}^{#2}}}

\newcommand{\R}{\ensuremath{\mathbb{R}}}
\newcommand{\Cx}{\ensuremath{\mathbb{C}}}

\newcommand{\Dp}[1]{\ensuremath{\partial_{#1}}}

\newcommand{\ex}[1]{\ensuremath{\e{e}^{#1}}}

\newcommand{\op}[1]{ \boldsymbol{ \texttt{#1} } }

\newcommand{\norm}[1]{\ensuremath{  || #1 || }}

\newcommand{\dd}{\mathrm{d}}
\newcommand{\e}[1]{\ensuremath{\mathrm{#1}}}

\newcommand{\intff}[2]{\ensuremath{ [  #1 \,; #2 ] }}
\newcommand{\intfo}[2]{\ensuremath{ [  #1 \,; #2 [ }}
\newcommand{\intof}[2]{\ensuremath{ ]  #1 \,; #2 ] }}
\newcommand{\intoo}[2]{\ensuremath{ ]  #1 \,; #2 [ }}

\newcommand{\widesim}[2][1.5]{
  \mathrel{\underset{#2}{\scalebox{#1}[1]{$\sim$}}}
}

\begin{document}

\begin{center}
\begin{LARGE}

{\bf Large deviations of energy transfers in nonequilibrium CFT
and asymptotics of non-local Riemann-Hilbert problems}
\end{LARGE}

\vspace{4mm}
{\large Karol K. Koz{\l}owski \footnote{e-mail: karol.kozlowski@ens-lyon.fr}}, 
{\large Krzysztof Gaw\c{e}dzki\footnote{chercheur \'em\'erite, email: kgawedzk@ens-lyon.fr} }
\\[1ex]
Univ Lyon, ENS de Lyon, Univ Claude Bernard Lyon 1, CNRS, Laboratoire de Physique, F-69342 Lyon, France \\[2.5ex]

\end{center}

\centerline{\bf Abstract} \vspace{1cm}
\parbox{12cm}{\small}

A wide class of $1+1$ dimensional unitary conformal field theories
allows for an explicit construction of nonequilibrium "profile
states" interpolating smoothly between different equilibria on the left
and on the right. It has been recently established that the 
generating function for the full counting statistics of energy transfers 
in such states may be expressed in terms 
of the solution to a non-local Riemann-Hilbert problem. 
Following earlier works on 
the statistics of energy transfers, 
in particular the ones of Bernard-Doyon on the "partitioning protocol" 
in conformal field theory, the full counting statistics of energy 
transfers in the profile states was conjectured to satisfy a large 
deviation principle in the limit of long transfer-times. The present 
paper establishes rigorously this conjecture by carrying out 
the long-time asymptotic analysis of the underlying non-local 
Riemann-Hilbert problem.

\vspace{40pt}

\tableofcontents

\section{Introduction}

The aim of the paper is to establish  
a large auxiliary-parameter behaviour of certain biholomorphisms 
that provide the holomorphic structure  
on conformally welded cylinders. Such cylinders are
obtained by %
identifying the boundaries of an infinite strip in the complex plane 
after the composition with a line-diffeomorphism. The original motivation 
for this investigation came from  
specific questions, described below, 
related to a rigorous characterisation of
certain correlation functions 
pertaining to non-equilibrium situations
in a large class of unitary 1+1 dimensional conformal field theories (CFTs). 
The CFT problem was first addressed, on heuristic grounds, in 
\cite{BernardDoyonEnergyFlowLDPInNonEqCFT} in a closely related but more 
singular 
framework. More recently, it was reformulated in a rigorous 
non-singular setup in 
\cite{KozGawedzkiFullCountingStatisticsIThermoLimAndConjecture} 
where a closed formula was proven for the correlator of interest. The 
non-trivial building block of the obtained expression was a biholomorphism 
realising the rectification of a cylinder conformally-welded from a strip. 
What is important for the applications is the dependence of this 
biholomorphism on certain auxiliary parameters describing the 
line-diffeomorphism used to weld together the boundaries of the strip. 
A precise control on that dependence constitutes the main result of 
this work.

\subsection{Asymptotics of conformal maps on the welded 
cylinders}

\begin{figure}[ht]
\begin{center}

\includegraphics[width=.6\textwidth]{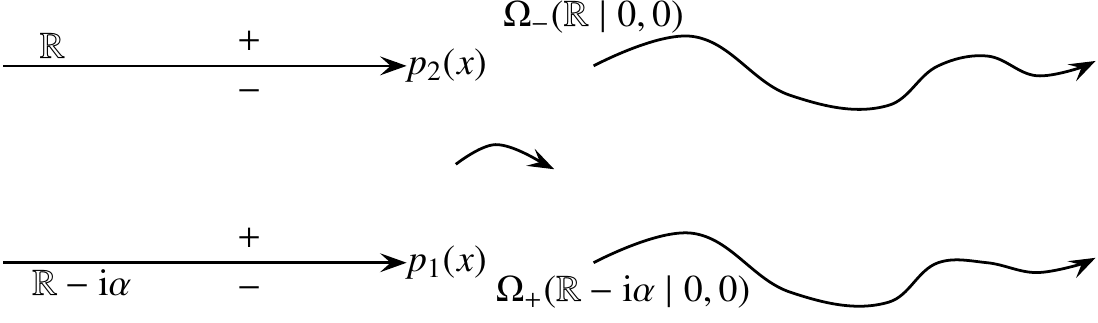}
\caption{ The strip $\mc{S}_{\a}$, parametrisation of its boundary along with its orientation and image thereof through the biholomorphism $\Om$  in the case when $\varkappa^{\pm}=0$. 
\label{Figure Strip S alpha} }
\end{center}
\end{figure}

In order to state the results, we first need to introduce a few notations 
so as to make the setting explicit. 
Let $\a>0$ and let 
\beq
\mc{S}_{\a} \, = \, \Big\{ z \in \Cx  \, : \, -\a \, < \, \Im(z) \, < \, 0\Big\} \;.  
\label{definition strip S alpha}
\enq
refer to the strip of width $\a$ in $\Cx$ located below $\R$. 
Endow the upper and the lower boundaries of the closure 
$\overline{\mc{S}_{\a}}$ of $\mc{S}_{\a}$ with 
the orientation of increasing real parts as depicted in 
Figure \ref{Figure Strip S alpha}. 
These boundaries are parameterised by $p_1(x)$ and $p_2(x)$, where 
$p_1: \R \tend \R-\i \a$ and  $p_2: \R \tend \R$ are smooth diffeomorphisms. 
The welded cylinder is then defined as the manifold obtained from 
$\ov{\mc{S}_{\a}}$ by identifying the points of $\Dp{}\mc{S}_{\a}$ 
parameterised by $p_1(x)$ and $p_2(x)$. 
It comes with the complex structure such that local holomorphic 
functions on 
it
are the smooth ones that are holomorphic 
when restricted to $\mc{S}_{\a}$ \cite{GoluzinGeometricTheory}.
Clearly, 
it is enough to parameterise the boundaries of the strip by taking 
\beq
p_1(x)=g(x)-\i\a \qquad \e{and} \qquad p_2(x)=x
\label{definition sewing diffeomorphisms}
\enq
with $g$ a smooth diffeomorphism of $\R$. The corresponding welded 
cylinder will be denoted as $\mc{S}_{\a,g}$.  
For example, when $g(x)=x$ then the welded cylinder is tautologically 
equivalent to the standard cylinder of circumference $\a$. 
More generally, it is of interest to consider the case where 
$g$ is smooth and such that 
$g-\e{id}$ is constant on the two connected components of 
the complement of some large enough segment 
$\intff{-M}{m}$ of $\R$, 
\textit{viz}. 
\beq
g(x) \; = \; \left\{ \ba{cc}  x+ \varkappa^-  &   x \leq - M \vspace{1mm}\\ 
				  g(x)        &   -M \leq x \leq m \vspace{1mm} \\
			    x+ \varkappa^+  &   x \geq  m  \ea  \right. \;,  
\label{ecriture hypothese generale sur g}
\enq
for some constants $\varkappa^{\pm}$. The welded cylinder 
$\mc{S}_{\a,g}$ for any such $g$ is biholomorphically 
equivalent to the standard cylinder.  The biholomorphism  
realising this equivalence may be constructed by means of solving 
a scalar, non-local Riemann-Hilbert problem with a jump.

\begin{prop}
 \label{proposition ouverture}
Assume that $g$ satisfies \eqref{ecriture hypothese generale sur g}, 
and consider the scalar non-local Riemann-Hilbert problem with a shift 
consisting in finding a holomorphic function 
$z \mapsto \Om(z\mid \varkappa^{+}, \varkappa^{+})$ on $\mc{S}_{\a}$ 
having smooth $-$, resp. $+$, boundary values on $\R$, resp. $\R-\i\a$, 
such that 
\begin{itemize}
 \item $\Om_-$, resp. $\Om_+$, 
is a bijection from $\R$, resp. $\R-\i\a$, 
onto  $\Om_-( \R \mid \varkappa^{+}, \varkappa^{-} )$, 
resp. $\Om_+( \R - \i\a \mid \varkappa^{+}, \varkappa^{-}  )$; 
 \item $\Om_+(g(x)-\i\a \mid \varkappa^{+}, \varkappa^{-} )  \, = \,\Om_-( x \mid \varkappa^{+}, \varkappa^{-}  )  -  \i \a\,$  for $x \in \R$;
 \item $\Om( z \mid \varkappa^{+}, \varkappa^{-} ) \, = \, \wt{\ga}_{\pm} z  
 \, + \, C_{\Om} \de_{\pm, -}\; + \; \e{O}\Big( \ex{\mp \frac{2\pi}{\a} \wt{\ga}_{\pm} z  } \Big)\,$ as $\Re(z) \tend \pm \infty$, 
\end{itemize}
for $\wt{\ga}_{\pm} = \f{-\i\a }{ \varkappa^{\pm}-\i\a }$ and some constant 
$C_{\Om} \in \Cx$. 

Then, the above problem admits a unique solution. Moreover, the latter is 
a biholomorphism from  $\mc{S}_{\a}$ onto its image 
$ \Om( \mc{S}_{\a } \mid \varkappa^{+}, \varkappa^{-} )$.

\end{prop}

When $\varkappa^{+}=\varkappa^{-}=0$, \textit{viz}. when the welding 
diffeomorphism is such that $g-\e{id}$ has compact support, then 
the above proposition may be seen, after composing with obvious 
biholomorphisms, as a direct consequence of the material discussed 
in \cite{GakhovBoundaryValueProblems}. However, for general 
$\varkappa^{\pm}$, the techniques of \cite{GakhovBoundaryValueProblems} 
are not sufficient to establish this result and one has to rely  
on the setting developed in the core of this paper.  The proof of 
the above proposition is given in Appendix \ref{Appendix Conformal Map}.

$ \Om(\ \cdot\mid \varkappa^{+}, \varkappa^{-} )$ 
induces a biholomorphism from $\mc{S}_{\a,g}$ onto the standard  
cylinder $\mc{S}_{\a,\e{id}}$ upon identifying the endpoints 
$ \Om_-( \R \mid \varkappa^{+}, \varkappa^{-} ) \ni z 
\equiv  z-\i\a \in \Om_+( \R - \i\a \mid \varkappa^{+}, \varkappa^{-} )$.

\begin{figure}[ht]
\begin{center}

\includegraphics[width=.6\textwidth]{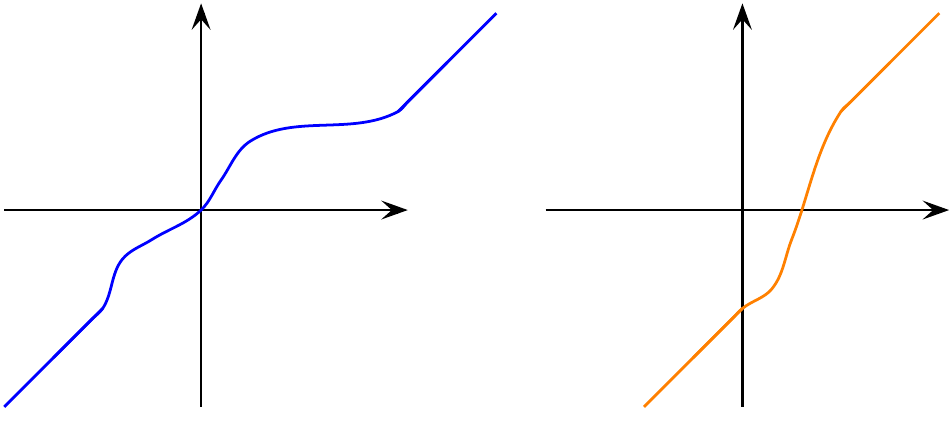}

\caption{ Diffeomorphisms $g_L$ (on the left, blue curve) and $g_R$ (on the right, orange curve).  
\label{Figure diffeomorphism gL et gR et leur raccolement} }
\end{center}
\end{figure}

\begin{figure}[ht]
\begin{center}

\includegraphics[width=.6\textwidth]{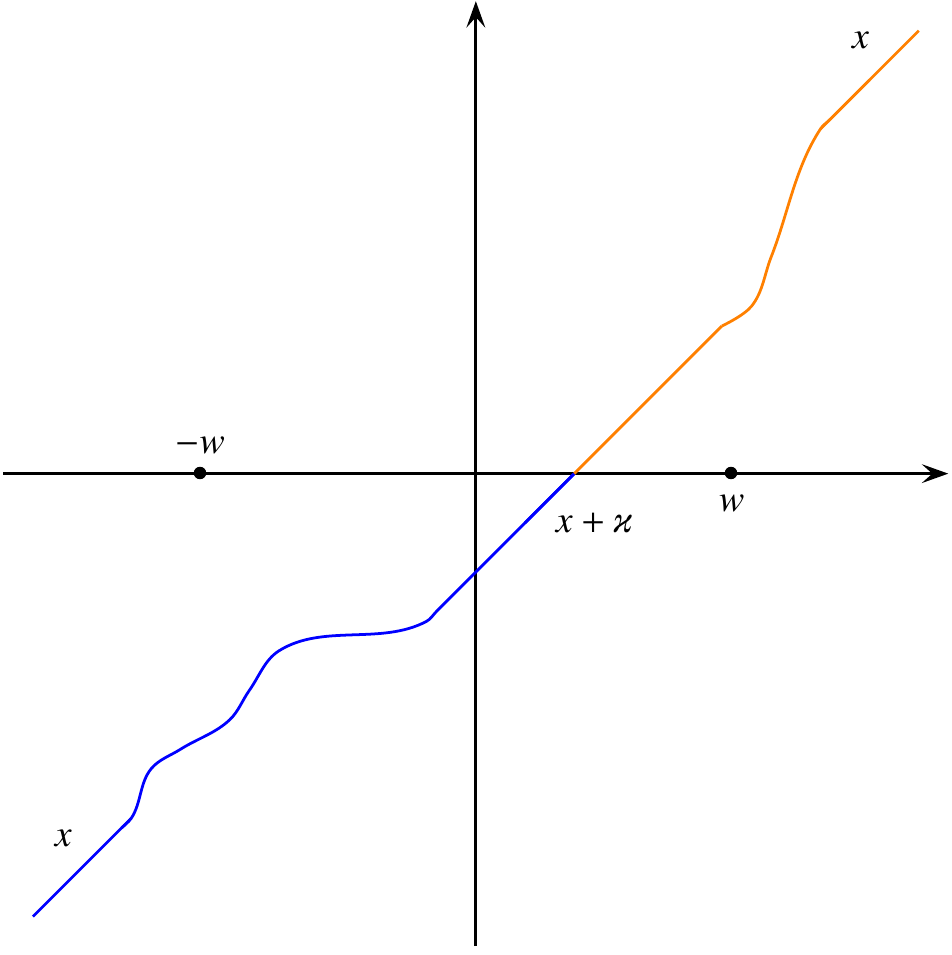}

\caption{ Diffeomorphism $g$ constructed from the gluing of $\color{blue}{g_L}$ and $\color{orange}{g_R}$. 
\label{Figure diffeomorphism g} }
\end{center}
\end{figure}

The main interest of the present work lies in accessing 
the behaviour of the biholomorphism $\Om$ in the case when the 
diffeomorphism of the line $g$  is such that $g-\e{id}$ has compact 
support, \textit{viz}. $\varkappa^{\pm}=0$, and is constructed from 
two diffeomorphisms $g_L$ and $g_R$ of the line in such a way that 
$g$ has a non-trivial behaviour only in the neighbourhood of the points 
$-w$ and $+w$. In order to insist on the vanishing of the constants 
$\varkappa{^\pm}$, we shall henceforth denote this biholomorphism 
by $\chi$, \textit{viz}. $\chi(z)=\Om(z\mid 0, 0)$. 
Such a situation is depicted in Figures \ref{Figure diffeomorphism gL et gR et leur raccolement}-\ref{Figure diffeomorphism g}. To be more precise
about the structure of $g$, pick $M_R, M_L>0$ and let $\varkappa \in \R$. Then, let $g_{L/R}$ be smooth diffeomorphisms of the real line taking the 
piecewise form 
\beq
g_L(x) \;  = \; \left\{ \ba{ccc}  x && x  \, \leq \,  -M_L  \vspace{2mm} \\ 
				  g_L(x) && -M_L \, < \,  x  \, < \,  M_L  \vspace{2mm}  \\ 
				 x +\varkappa &\; & M_L  \, \leq  \,  x 	  \ea \right. 
\quad \e{and} \quad 
g_R(x) \;  = \; \left\{ \ba{ccc}  x +\varkappa &\; & x \,  \leq \,  -M_R  \vspace{2mm}  \\ 
				  g_R(x)& & -M_R \, < \,  x \,  < \,  M_R  \vspace{2mm}  \\ 
				 x  &&  M_R \, \leq  \,  x 	  \ea \right.  \; .
\enq
Then, the diffeomorphism $g$ of interest is defined as 

\beq
g(x)  \, = \,    \left\{ \ba{cc}  x    &    x < -M_L - w  \vspace{2mm} \\ 
				g_L(x+w)-w  \qquad & - M_L - w \leq x \leq  M_L  - w  \vspace{2mm} \\ 
				x+\varkappa  & M_L - w \leq x \leq  w  - M_R     \vspace{2mm} \\  
				  g_R(x-w) +w  \qquad & w - M_R < x < w + M_R   \vspace{2mm}  \\ 
				 x  & M_R+w < x	  \ea \right.
\enq

The non-local Riemann-Hilbert problem of Proposition 
\ref{proposition ouverture} takes for such a $g$ the slightly simpler 
form below. 

\begin{defin}
\label{Definition RHP non loc pour chi}

Given a smooth diffeomorphism $g$ of $\R$ such that $g-\e{id}$ has 
compact support, the non-local Riemann-Hilbert problem for $\chi$
consists in finding a holomorphic function $z\mapsto\chi(z)$ 
on $\mc{S}_{\a}$ such that   

\begin{itemize}
 
 \item it has smooth $-$, resp. $+$, boundary values on $\R$, 
resp. $\R-\i\a$; 

 \item $\chi_+(g(x)-\i\a )  \, = \,\chi_-( x )  -  \i \a\,$  for $x \in \R$;
 
 \item $\chi( z ) \, = \,   z \, + \, C_{\chi} \de_{\pm, -}\; + \; \e{O}\Big( \ex{\mp \frac{2\pi}{\a} z  } \Big)\,$ as $\Re(z) \tend \pm \infty$, 
 
\end{itemize}
for some constant $C_{\chi} \in \Cx$.

\end{defin}

As will be discussed in the following, the interest in this specific form 
of the diffeomorphism $g$ and in the $w\tend +\infty$ behaviour it begets 
to $\chi$ stems from certain questions occurring in 1+1 dimensional unitary 
conformal field theories. In order to state the main technical achievement 
of this work, Theorem \ref{Theoreme principal} below, we  
need to introduce  two auxiliary non-local Riemann-Hilbert problems with 
shift that are associated with the welding diffeomorphisms $g_L$ and $g_R$. 
First, however, define
\beq
\ga \, = \, \f{ -\varkappa  }{ \varkappa - \i \a }  \qquad \e{and} \qquad \wt{\ga} \, = \, \ga + 1 \, = \, \f{ -\i \a  }{ \varkappa - \i \a } \;. 
\enq
and denote by $\mc{O}\big( \mc{S}_{\a} \big)$ the space of 
holomorphic functions on $\mc{S}_{\a}$. The left Riemann-Hilbert 
problem consists in finding 
$\chi^{(L)} \in \mc{O}\big( \mc{S}_{\a} \big)$ that admits smooth $-$, resp. $+$, boundary values on $\R$, resp. $\R-\i\a$, and such that 
\begin{itemize}
 \item for some constant $C_{\chi^{(L)}}$, $\,\chi^{(L)}(z) \, 
= \, C_{\chi^{(L)}} \, + \, \e{O}\Big( \ex{ \f{2\pi}{\a } z } \Big)$  
\,when $\Re(z) \tend -\infty$ and up to the boundary;
 \item $\chi^{(L)}(z) \, = \, \ga \cdot 
z  \, + \, \e{O}\Big(  \ex{ - \f{2\pi \wt{\ga} }{\a } z } \Big)$  
when $\Re(z) \tend +\infty$ and up to the boundary;
 \item $\chi^{(L)}_+(g_L(x)-\i \a )\, = \, \chi^{(L)}_-(x)\, 
+ \, x \, - \, g_L(x) $.
\end{itemize}

The right Riemann-Hilbert problem consists in finding  $\chi^{(R)} 
\in \mc{O}\big( \mc{S}_{\a} \big)$ that admits smooth $-$, resp. 
$+$, boundary values on $\R$, resp. $\R-\i\a$, and such that 
\begin{itemize}
 \item $\chi^{(R)}(z) \, = \, \e{O}\Big( \ex{ - \f{2\pi}{\a } z } \Big)$  
\,when $\Re(z) \tend +\infty$ and up to the boundary;
 \item for some constant $C_{\chi^{(R)}}$, 
$\,\chi^{(R)}(z) \, = \, \ga  z \, + \, C_{\chi^{(R)}}  \, 
+ \, \e{O}\Big(  \ex{   \f{2\pi \wt{\ga} }{\a } z } \Big)$  
\,when $\Re(z) \tend -\infty$ and up to the boundary;
 \item $\chi^{(R)}_+(g_R(x)-\i \a )\, = \, \chi^{(R)}_-(x)\, + \, x \, - \, g_R(x) $.
\end{itemize}
One may readily convince oneself that $\chi^{(L)}(z)
=\Om(z\mid \varkappa,0)-z$ while $\chi^{(R)}(z)=\Om(z\mid 0, \varkappa)-z$, where in the first case, 
$\Om$ is constructed out of the diffeomorphism $g_L$ and in the second case out of $g_R$.

\begin{theorem}
\label{Theoreme principal}
Let $g$ be as given above in terms of $g_L$ and $g_R$. Then, the 
left/right non-local Riemann-problems for  $\chi^{(L/R)}$ are uniquely 
solvable and the unique solution to the non-local Riemann-Hilbert problem 
with a shift for $\chi$  
described in Definition \ref{Definition RHP non loc pour chi}
admits the large $w$ asymptotic expansion which takes the patch-wise 
form given in Fig. \ref{Figure for grd w pour solution chi}.
There, $\mf{c}=2 \ga w-C_{\chi^{(R)}} $ and $\de\chi^{(R/L)}$ are 
holomorphic in the domains where they appear and enjoy there the uniform 
estimates, in $z\in \mc{S}_{\a}$ and in $w$:
\beq
\de\chi^{(L)}(z) \; = \; \de \mf{c} \, + \, \e{O}\Big( \ex{-\eta w 
+ \eta^{\prime}z }    \Big) \qquad and \qquad 
\de\chi^{(R)}(z) \; = \;\e{O}\Big( \ex{-\eta w-\eta^{\prime}z }   \Big) 
\quad with \quad  \de \mf{c} \, = \, \e{O}\Big( \ex{-\eta w} \Big) \;. 
\label{ecriture control sur rest a droite ou gauche}
\enq
Above, $\eta, \eta^{\prime}>0$ are some constants just as $\de \mf{c}$. 
The remainder functions $\de\chi^{(R/L)}$ are such that $\chi$ is indeed 
smooth across the separating segment $\Ga_0=\intff{ 0 }
{ \varkappa - \i\a }$. Finally, all estimates appearing above are 
differentiable uniformly on $\mc{S}_{\a}$ and up to its boundary.

\begin{figure}[ht]
\begin{center}

\includegraphics[width=.7\textwidth]{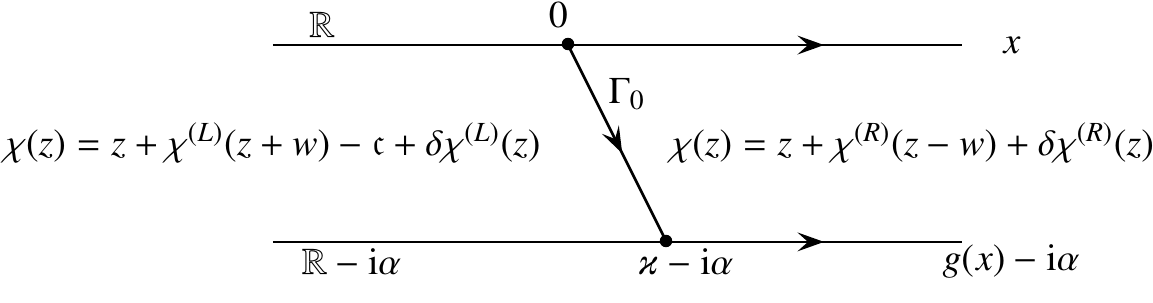}
\caption{ Piecewise expression for the solution $\chi$.}   
\label{Figure for grd w pour solution chi} 
\end{center}
\end{figure}
\end{theorem}

The presence of the two biholomorphisms $\chi^{(L/R)}+\e{id}$ appearing 
in the leading behaviour of $\chi$ to the left/right of $\Ga_{0}$
is certainly natural in that with $w$ growing the two non-trivial pieces 
of $g$ should cease to interact so that, locally,
the overall biholomorphism should only "feel" the effect of $g_L$ or $g_R$. 
The hardest part of the proof Theorem \ref{Theoreme principal}
consists in establishing the differentiable control on the remainder given in 
\eqref{ecriture control sur rest a droite ou gauche}. It is the proof 
of this property, crucial for the application of 
Theorem \ref{Theoreme principal} to conformal field theories, that 
occupies most of this work.

\subsection{Large deviation principle for the full counting 
statistics of energy transfers in 1+1 dimensional conformal 
field theories}
 
Let $x\mapsto \be(x)$ be an inverse temperature profile, \textit{viz}. 
a smooth function bounded from below by a strictly positive constant 
such that $\be^{\prime}(x)$ has compact support and constant sign. 
Furthermore, consider a 1+1 dimensional unitary conformal field theory in 
a finite interval $\intff{-L/4}{L/4}$ with boundary conditions that assure no
energy flux through the endpoints of the interval. Such a theory
is described by a representation of 
the Virasoro algebra on some Hilbert space $\mf{h}$
generated by the energy-momentum tensor of the form 
\cite{CardyEncyclopediaofMathematicaPhysics,KozGawedzkiFullCountingStatisticsIThermoLimAndConjecture} 
\beq
\op{T}(x) \, = \, \f{2\pi }{ L^2 } \sul{ n \in \mathbb{Z} }{}  
\ex{ \tfrac{2\i\pi n}{L}(x+\tfrac{L}{4}) } 
\cdot \Big(\op{L}_n- \f{c}{24} \de_{n,0} \Big) \;, 
\enq
 in which $\op{L}_n$ are generators of the Virasoro algebra. 
Out of these quantities one constructs the operator 
\beq
\op{G}_L(t) \; = \; v \Int{ -\frac{L}{4} }{ \frac{L}{4} } 
\hspace{-1mm} \dd x \,  \be(x){\cal E}(t,x)\qquad\text{for}\qquad
{\cal E}(t,x)=\op{T}(x-vt)
+ \op{T}(-x-vt-\tfrac{L}{2}) 
\enq
(${\cal E}(t,x)$ is the energy density).
 It was proven in the work 
\cite{KozGawedzkiFullCountingStatisticsIThermoLimAndConjecture} that 
the Fourier transform of the probability measure which 
describes the 
 energy transfers in time $t$ between two baths at inverse temperatures 
$\be(-\infty)$ and $ \be(+\infty)$ connected by the interpolating inverse 
temperature profile $\be(x)$ takes the form 
\beq
\Psi_{t,L}(\la) \; = \; \f{  \e{tr}_{\mf{h}} \Big[  \ex{ \f{ \i \la }{\De \be} \op{G}_L(t)   }   \ex{-(\f{ \i \la }{\De \be} +1)\op{G}_L(0)   } \Big]   }
{  \e{tr}_{\mf{h}} \Big[  \ex{- \op{G}_L(0)   } \Big]  }\qquad \e{with} \qquad \De \be \, = \, \Int{\R}{}\be^{\prime}(x) \dd x \;. 
\enq
More details on the well-definiteness of the above expression 
can be found in the above mentioned work. Ref.
\cite{KozGawedzkiFullCountingStatisticsIThermoLimAndConjecture}
studied the thermodynamic limit of $\Psi_{t,L}(\la)$ 
and it was rigorously proven there 
that 
\beq
\lim_{L\tend +\infty} \Psi_{t,L}(\la) \; = \;  \pl{\veps= \pm}{} \Psi_{t}^{(\veps)}(\la) \; , 
\enq
uniformly in $\la$ belonging to compact subsets of $\R$, where 
\beq
\Psi_{t}^{(\veps)}(\la)  \, = \, \exp\Bigg\{ \phi^{(\veps)}(t)\,  -\,  \f{ \i c }{ 24\pi} \Int{0}{ \tfrac{\la}{\De \be}} \dd s  \Int{\R}{} \dd x  
\, \xi_{t}^{(\veps)}(x) \cdot \bigg\{  \mc{S}\big[\,  
\chi_{s,t;-}^{(\veps)} \big](x) - \f{2\pi^2}{\a^2}   
\Big(\Dp{x}\chi_{s,t;-}^{(\veps)}(x) \Big)^2   \bigg\}     \Bigg\}\;. 
\label{ecriture expression integrale pour Psit}
\enq
The last formula contains several ingredients. First of all, 
\beq
\mc{S}[f]= \f{f'''(x) }{f^{\prime}(x) } \, 
- \, \f{ 3 }{ 2 } \bigg(  \f{ f^{\prime\prime}(x) }{ f^{\prime}(x) } 
\bigg)^2    
\enq
is the Schwartzian derivative, $\a$ is a constant built up from 
the $\pm \infty$ limits of the inverse-temperature profile  
and $\phi^{(\veps)}(t)$ is an explicit, smooth and bounded function 
of $t$. Furthermore, $\xi_{t}^{(\veps)}$ is a smooth compactly 
supported function depending on the auxiliary time parameter $t$. 
One is also given two smooth diffeomorphism of the line $g_{s,t}^{(\veps)}$ 
such that  $g_{s,t}^{(\veps)}-\e{id}$ has compact support. 
These depend smoothly on two auxiliary parameters: the time $t$ and a 
real variable 
$s$. To each diffeomorphism $g_{s,t}^{(\veps)}$ one then associates the 
corresponding solution $\chi_{s,t}^{(\veps)}$ to the non-local 
Riemann-Hilbert problem of the strip $\mc{S}_{\a}$, as introduced in 
Definition \ref{Definition RHP non loc pour chi}. Then, 
$\chi_{s,t;-}^{(\veps)}$ stand for its $-$ boundary values on $\R$. 
This concludes the description of the building blocks  of the 
thermodynamic limit $\Psi_{t}^{(\veps)}(\la)$.

The explicit construction of the functions $\xi_{t}^{(\veps)}$ and  
$g_{s,t}^{(\veps)}$ can be found in  
\cite{KozGawedzkiFullCountingStatisticsIThermoLimAndConjecture}. Here, 
we only remind the properties which are crucial for establishing  
the results given in Theorem 
\ref{Theorem limite grand tps pour TF de PDF energy transfers} below.  
Namely, there exist real parameters $\varkappa^{\veps}$ and segments 
$\mc{I}^{(\veps)}_L$, $\mc{I}^{(\veps)}_{\e{bk}}$ and 
$\mc{I}^{(\veps)}_R$ having disjoint interiors such that: 
\begin{itemize}
 \item $\e{supp}\Big[ \xi_{t}^{(\veps)} \Big] \subset   
\mc{I}^{(\veps)}_L \cup \mc{I}^{(\veps)}_{\e{bk}} \cup 
\mc{I}^{(\veps)}_R$; 
 \item $\e{supp}\Big[ g_{s,t}^{(\veps)}-\e{id} \Big] \subset   
\mc{I}^{(\veps)}_L \cup \mc{I}^{(\veps)}_{\e{bk}} 
\cup \mc{I}^{(\veps)}_R$; 
\item $  \e{diam}\big( \mc{I}^{(\veps)}_L \big)$ and 
$ \e{diam}\big(  \mc{I}^{(\veps)}_R \big)$ are $t$ independent;
%
%
\item $  \e{diam}\big( \mc{I}^{(\veps)}_{\e{bk}} \big) 
= \ell^{\veps} t-C$  for some $\ell^{\veps},C >0$ ;
\item ${ \xi_{t}^{(\veps)}}_{\mid \mc{I}^{(\veps)}_{\e{bk}} } 
= -\varkappa^{\veps} \; \; $  and   $\; \; 
{(g_{s,t}^{(\veps)}-\e{id}) }_{\mid \mc{I}^{(\veps)}_{\e{bk}} } 
= \,\varkappa^{\veps} s $;
\item ${\xi_{t}^{(\veps)}}_{\mid \mc{I}^{(\veps)}_{L,R} }$
and $\ ({g_{s,t}^{(\veps)}-\e{id}})_{\mid \mc{I}^{(\veps)}_{L,R}}$
have $t$-independent shape on those intervals.
\end{itemize}

As it is apparent from \eqref{ecriture expression integrale pour Psit}, 
the thermodynamic limit of $\Psi_{t,L}(\la)$ depends on time $t$. 
One is interested in obtaining a large deviation principle, when 
$t\tend+\infty$, for the thermodynamic limit of the associated 
probability measure. The rate function governing this large deviation 
principle may be deduced from the Legendre transform in $\i\la$
of the limiting functions $\underset{t\tend + \infty}{\lim} 
\big\{ t^{-1} \ln \Psi_{t}^{(\veps)}(\la) \big\}$, $\veps=\pm$. 
 In order to control this limit and compute it, one needs all the 
information that have been established in 
Theorem \ref{Theoreme principal} given above. 
In fact, a direct application of this theorem shows that the Schwarzian 
derivative term  
contributes as $\e{O}(1)$ when $t\tend +\infty$
while the only linear in $t$ behaviour of the integral giving rise 
to $\ln \Psi_{t}^{(\veps)}(\la)$ is generated from the constant 
term in the behaviour of $\Dp{x}\chi_{s,t;-}^{(\veps)}(x)$ for $x\in
\mc{I}^{(\veps)}_{\e{bk}}$ sufficiently far away from the endpoints
of that segment. After straightforward calculations, one gets  
\begin{theorem}
\label{Theorem limite grand tps pour TF de PDF energy transfers}
%
%
%
%
\beq
\underset{t\tend + \infty}{\lim} \big\{ t^{-1} \ln \Psi_{t}^{(\veps)}(\la) \big\} \; = \; -  \f{ c \pi }{12 \a } \cdot \f{ \varkappa^{\veps} \ell^{\veps} \la }{ \varkappa^{\veps}   \la - \i \a \De \be } \;. 
\label{LDratefunction}
\enq
\end{theorem}
This theorem concludes the proof of the large deviation principle 
stated in \cite{KozGawedzkiFullCountingStatisticsIThermoLimAndConjecture}. 
The above form of large deviations for the energy transfers coincides
with the one 
anticipated in \cite{BernardDoyonEnergyFlowLDPInNonEqCFT},
see also \cite{BernardDoyonNon-equilibrium steady-states}.
\vskip 0.3cm

\noindent{\bf Remark.} An examination of the arguments leading to
Theorem \ref{Theoreme principal} shows that the convergence
in \eqref{LDratefunction} is uniformly differentiable in  $\lambda$ belonging to compact subsets of $\R$.

 \subsection{Outline of the paper}
 
 The paper is organised as follows. Section \ref{Section RHP non local 
sur Cylindre cousu} establishes the unique solvability of a class 
of non-local Riemann-Hilbert problems on welded cylinders. 
Subsection \ref{SousSection defnitions generale pour RHP non local} 
provides the definition of  
a 
class of Riemann-Hilbert 
problems 
that will be considered there. Then, various technical results relative 
to the original setting of the Riemann-Hilbert problems are established,
in particular an improvement of the decay at $\infty$, the correspondence 
with solutions to linear integral equations and the existence of smooth 
boundary values for the solution. The non-local Riemann-Hilbert problem 
in the optimal setting is then outlined in Subsection 
\ref{SousSection RHP non local lisse}. Finally, Subsection 
\ref{Sous section solvabilite unique du RHP global}
establishes the unique solvability of the non-local Riemann-Hilbert 
problems of interest. This is done by proving the invertibility 
of the operator $\e{id}-\op{K}$ which drives the linear integral 
equations %
that 
are satisfied by the boundary values of the solution. 
The preliminary notations for this result are established in Subsubsection \ref{SousSousSection Notation preliminaires}. 
The reduction of the operator $\e{id}-\op{K}$ to  
$\e{id}-\op{M} $ with compact $\op{M}$
is carried out in Subsubsection \ref{SousSousSection reduction 1-K a 1-M}
and, finally, the sought
invertibility
is established in  Subsubsection \ref{SousSousSection invertibilite 
de id - K}.

Section \ref{Section RHP non locaux modeles} studies three auxiliary 
special 
non-local Riemann-Hilbert problems which play a role in the large-$w$ 
asymptotic analysis of the solution to the Riemann-Hilbert problem stated 
in Proposition \ref{proposition ouverture} in the presence of the 
$w$-dependent welding diffeomorphism $g$ as described above. 
Subsection \ref{SousSection Transfo Cauchy sur Cylindre cousu} discusses 
properties of the Cauchy transform on a 
welded cylinder, 
Subsection \ref{SousSection RHP modele gauche} establishes the existence 
of the solution to the Riemann-Hilbert problem for the function $\chi^{(L)}$ 
described 
above while Subsection \ref{SousSection RHP modele droit} does it for 
the one associated with $\chi^{(R)}$. 

Section \ref{Section RHP non local glocal} establishes Theorem 
\ref{Theoreme principal}. 
The proof given there heavily relies on  
technical results,  
relative to the uniform in large $w$ 
invertibility of the integral operator $\e{id}-\op{K}_{\e{tot}}$ which 
drives the integral equations satisfied by the boundary values of 
the solution $\chi$. Those are established throughout Section \ref{Section Invertibility of the operator id moins K}. 
Subsection \ref{SousSection decomposition de K tot} provides a convenient decomposition of the integral kernel of $\op{K}_{\e{tot}}$. 
Various technical properties issuing from this decomposition are then established throughout Subsections \ref{SousSection preliminary estimates on B tot}, \ref{SousSection finer direct sum partition of hilbert space}, 
 \ref{SousSection Decomposition in the - sector}, \ref{SousSection Decomposition in the + sector} and \ref{SousSection Decomposition in the 0 sector}. 
Finally, the uniform in $w$ invertibility of $\e{id}-\op{K}_{\e{tot}}$ is established in Subsection \ref{SousSection final form of integral equation}.

Several auxiliary results  
are postponed to the appendices. Appendix \ref{Appendix Conformal Map}
briefly outlines the proof of Proposition \ref{proposition ouverture}. 
Appendix \ref{Appendix section inversion operateurs id - L ups ups} 
provides details on the inversion of certain Wiener-Hopf equations on 
the half-line while Appendix \ref{Appendix Secion inversion operateur L0} 
discusses the inversion of a truncated Wiener-Hopf operator arising in 
the analysis of Section \ref{Section RHP non local glocal}. This last 
result is achieved by solving a local $2\times 2$ matrix Riemann-Hilbert 
problem. Finally, Appendix \ref{Appendix Auxiliary Lemma} establishes 
a technical Lemma useful for certain estimates obtained in Section 
\ref{Section RHP non local glocal}.

 \subsection{Notations}
 \label{SubSection notations}

\begin{itemize}

%
%
%
%
%
%
%
%
%

\item Given an open subset $U\subset \Cx$, $\,\mc{O}(U)$ stands for 
the ring of holomorphic functions on $U$. 
  
\item Given an open subset $U\subset \Cx$, and a function $f$ defined 
on $ U\setminus \ga$, with $\ga$ an oriented curve in $U$, we denote 
by $f_{\pm}(s)$ the boundary values - if these exists in an appropriate 
sense - of $f(z)$ on $\ga$ when the argument $z$ approaches the point 
$s \in \Sg$ non-tangentially and from the left ($+$) or the right ($-$) 
side of the curve. Furthermore, if one deals with vector or matrix-valued 
function, then this notation is to be understood entry-wise.  

\item $\mathbb{H}^{\pm} = \{z \in \mathbb{C}\,\,:\,\,\mathrm{Im}\,
(\pm z) > 0\}$ is the upper/lower half-plane, and $\mathbb{R}^{\pm} 
= \{x \in \mathbb{R}\,\,:\,\,\pm x \geq 0\}$ is the
positive/negative real axis.

\item Given a set $A$, $\,\ov{A}$ stands for its closure and 
$\bs{1}_{A}$ stands for the indicator function of $A$.

\item Given a ring $\mc{R}$, $\,\mc{M}_{n}(\mc{R})$ stands for the 
space of $n \times n$  matrices over this ring.

\item Given two functions $f,g$ defined in an open neighbourhood $U$ 
of a point $\bs{y}=(y_1,\dots, y_n)$, the relation 
$f(\bs{x})=\e{O}\Big( g(\bs{x}) \Big)$ 
means that there exists $M>0$ such that $|f(\bs{x})|\, 
\leq \, M |g(\bs{x})|$ holds in a neighbourhood 
of $\bs{y}$. The $\e{O}$-relation is said to be differentiable if, for all $k_a\geq 0$, 
\beq
\Dp{x_1}^{k_1}\dots\Dp{x_n}^{k_n}f(\bs{x})
=\e{O}\Big( \e{max}\big\{ |\Dp{x_1}^{m_1}\dots\Dp{x_n}^{m_n}g(\bs{x})| 
\; : \; 0\leq m_a \leq k_a \big\} \Big) \;
\enq
holds in a neighbourhood of $\bs{y}$.

 \item For matrix valued functions, a relation $M(\bs{x})=\e{O}\big( N(\bs{x}) \big)$ 
is to be understood entrywise, \textit{viz}. $M_{ab}(\bs{x})=\e{O}\big( N_{ab}(\bs{x}) \big)$.

\item Let $\Ga$ be Euler's $\Ga$-function. We  
use the shorthand convention
\beq
\Ga\left( \ba{cc}  a_1,\dots, a_n \\ b_1,\dots b_m \ea \right) \; = \; \f{ \pl{k=1}{n}\Ga(a_k)  }{  \pl{k=1}{m}\Ga(b_k) } \;. 
\label{convention notation produit fonctions gamma}
\enq

\end{itemize}

\section{Non-local Riemann--Hilbert problems on 
welded cylinders}
\label{Section RHP non local sur Cylindre cousu}

\subsection{General definitions and considerations}
\label{SousSection defnitions generale pour RHP non local}

Throughout this section, we shall focus on the "affine at $\infty$" 
setting, \textit{viz}. a situation where there exist reals 
$\varkappa^{\pm}$ such that the diffeomorphism $g$ appearing in the 
welding of the strip $\mc{S}_{\a}$ \eqref{definition strip S alpha}, 
\textit{c.f.} Fig.~\ref{Figure Strip S alpha}, behaves as 
$g(x)  =x+ \varkappa^{\pm}$ when $\pm x>M$, for $M$ large enough, 
namely 
\beq
g(x) \; = \; \left\{ \ba{cc}  x+ \varkappa^-  &   x \leq - M \vspace{1mm}\\ 
				  g(x)        &   |x| \leq M \vspace{1mm} \\
			    x+ \varkappa^+  &   x \geq  M  \ea  \right. \;. 
\enq
The main purpose of this section is to establish the unique solvability, 
along with certain other properties, of a class of non-local 
Riemann-Hilbert problems that arise in later subsections. Thus, we start 
by introducing the class of problems of interest.  

The diffeomorphisms $p_1, p_2$ realising the welding of the strip $\mc{S}_{\a}$ as in 
Fig.~\ref{Figure Strip S alpha} are denoted as in 
\eqref{definition sewing diffeomorphisms}. 
Next, we assume being given a  smooth function $G_{\Xi}$ on $\R$ which 
has the structure:
\beq
G_{\Xi}(x) \, = \, G^{(\e{c})}_{\Xi}(x) \, + \, \mc{G}_{\Xi}(x) \, 
-  \, \mc{G}_{\Xi}( g(x)-\i\a) \;.  
\label{GXix}
\enq
There  $G^{(\e{c})}_{\Xi}$ is smooth with compact support and 
$\e{supp}[ G^{(\e{c})}_{\Xi} ] \subset \intff{-M}{M}$. Furthermore, 
\beq
\mc{G}_{\Xi} \quad  \e{is} 
\,\e{smooth}\,\e{on}\quad\ov{\mc{S}}_{\a}\quad\e{and}\ 
 \e{analytic}\, \e{on} \quad  
\mc{S}_{\a} \cap \Big\{ z \in \Cx \; : \; |\Re(z)| > \tf{M}{2} \Big\}
\label{ecriture propriete fonction shift holomorphe}
\enq
and vanishes exponentially fast 
at $\Re(z) \tend \pm \infty$, \textit{viz}. there exists $\varrho>0$ 
such that $\mc{G}_{\Xi}(z)  \, = \, \e{O}\Big( \ex{\mp \varrho z} \Big) $  
uniformly on $ \ov{\mc{S}}_{\a} \cap \big\{ z \in \Cx \; : \; |\Re(z)| 
> \tf{M}{2} \big\}$.

One may associate with this setting the  
following non-local Riemann-Hilbert problem on the strip $\mc{S}_{\a}$. 
Find  $\Xi \in \mc{O}(\mc{S}_{\a})$ such that 
\begin{itemize}
 \item  $\Xi_+\circ p_1 \in L^{2}_{\e{loc}}(\R)\;$, $\;\;\; \Xi_-\circ p_2   \in L^{2}_{\e{loc}}(\R)$ ;
 \item $\Xi_{+}\big( p_1(x)\big) \, = \, \Xi_{-}\big( p_2(x)\big)\, 
+ \, G_{\Xi}(x)$ 
\ \,for $x \in \R$; 
\item there exist constants $C_{\Xi}, C_{-1}$  such that 
\beq
\Xi(z) =  C_{\Xi}\, \cdot
\de_{\pm; -} \, + \,  \f{C_{-1}}{z} \, + \, \e{O}( z^{-2}  ) 
\qquad \e{when} \qquad \Re(z) \tend \pm \infty \; 
\label{ecritue DA pm infty de Xi}
\enq
uniformly up to the boundary of $\mc{S}_{\a}$.
\end{itemize}

\subsubsection*{$\bullet$ Improved asymptotic decay at infinity}
\label{SousSousSection amelioration de la decroissance a infini}

\begin{lemme}
\label{Lemme decroissance reguliere en infinity du RHP} 
 
 Any solution to the above Riemann-Hilbert problem   
decays exponentially fast at infinity as 
\beq
\Xi(z) \, = \,  C_{\Xi}\cdot 
  \de_{\pm; -}  \, - \, \mc{G}_{\Xi}(z) \; + \;  
\e{O}\Big(  \ex{ \mp \f{2\pi}{\a} \wt{\ga}_{\pm} z } \Big)  
\quad when \quad \Re(z)  \, \tend \pm \infty \;, \qquad with 
\qquad \wt{\ga}_{\pm}  \; = \; \f{ - \i \a  }{ -\i \a + \varkappa^{\pm} }   
\label{definition decroissance Xi a infini}
\enq
this uniformly up to the boundary.
\end{lemme}

\Proof 
To improve the bounds on the asymptotic
behaviour, we first introduce the curve built out of oriented segments
\beq
\msc{C}_{r,\a} \; = \; \intff{ -r + \varkappa^- 
- \i \a }{  r + \varkappa^+ - \i \a } \cup  \intff{ r 
+ \varkappa^+ - \i \a }{ r  } \cup \intff{r}{-r} 
\cup \intff{ -r }{ -r + \varkappa^- - \i \a } \;. 
\enq
For any $z \in \mc{S}_{\a}$, it holds 
\beq
 \Xi(z) \,  = \, \lim_{r \tend + \infty} 
\Int{  \msc{C}_{r,\a}    }{} \f{  \wt{\ga}_+ \, \dd s }{ \i \a }  
\f{  \Xi(s) }{1- \ex{ \f{2\pi \wt{\ga}_+ }{ \a } (z-s) }   }  \;. 
\enq
Writing explicitly the various integrations and using that 
$g(\intff{-r}{r})= \intff{ -r + \varkappa^-  }{ r + \varkappa^+  }$ 
provided that $r$ is large enough, leads to 
\bem
\Xi(z) \,  = \, \lim_{r \tend + \infty} \Bigg[ 
\Int{  -r  }{r} \f{  \wt{\ga}_+ \, \dd s }{ \i \a }  
\bigg[ \f{  \Xi_+( g(s)-\i\a)  g^{\prime}(s) }{ 1-
\ex{ \f{2\pi \wt{\ga}_+ }{ \a } (z-g(s)+\i\a) }   }   
 \, - \, \f{  \Xi_-(s) }{ 1- \ex{ \f{2\pi \wt{\ga}_+ }{ \a } 
(z-s) }     } \bigg]  \\
 \; + \; 
 \bigg\{    \Int{-r}{-r+\varkappa^--\i\a } 
+  \Int{r+\varkappa^+ - \i\a }{r} \bigg\}  
\f{  \wt{\ga}_+ \, \dd s }{ \i \a }  
\f{  \Xi(s) }{ 1-\ex{ \f{2\pi \wt{\ga}_+ }{ \a } (z-s) }     }    
\Bigg] \;. 
\end{multline}
One may now take the $r\tend + \infty$ limit. The contribution 
from $ \intff{ -r }{ -r + \varkappa^- - \i \a } $ goes to zero because 
the numerator is bounded while the denominator blows up exponentially 
fast in $r$. The  contribution from $ \intff{ r + \varkappa^+ 
- \i \a }{ r } $ goes to zero because the denominator approaches 1 
while the numerator goes to zero uniformly on this bounded segment.  
Finally, the integral over $\intff{-r}{r}$ converges in the limit since, 
for $s \tend \pm\infty$, the integrand is a $\e{O}\big( s^{-2} \big)$ 
owing to the form of the uniform up to the boundary asymptotic expansion 
of $\Xi$, \textit{c.f.} \eqref{ecritue DA pm infty de Xi}.
This thus yields
\beq
 \Xi(z) \,  = \,   \Int{\R}{}   \f{  \wt{\ga}_+ \, \dd s }{ \i \a }  
\Bigg\{ \f{  \Xi_+( g(s)-\i\a)  g^{\prime}(s) }{1- 
\ex{ \f{2\pi \wt{\ga}_+ }{ \a } (z-g(s) + \i\a) }   }   
 \, - \, \f{  \Xi_-(s) }{1- \ex{ \f{2\pi \wt{\ga}_+ }{ \a } (z-s) }   } 
\Bigg\} \;. 
\enq

Then, by using the jump condition, the fast decay of $G_{\Xi}$ at 
infinity and the fact that, for $x>M$
\beq
g(x) \, = \, x \, + \, \varkappa^+  \quad \e{while} \quad 
\wt{\ga}_+(-\i\a +\varkappa^+) \, = \, -\i\a \; ,
\enq
one gets the representation 
\beq
\Xi(z) \, = \,  \Int{-\infty}{M}   \f{  \wt{\ga}_+ \, 
\dd s }{ \i \a }  \Xi_+( g(s)-\i\a) 
\Bigg\{ \f{    g^{\prime}(s) }{1- \ex{ \f{2\pi \wt{\ga}_+ }{ \a } 
(z-g(s) + \i\a) }   }   \, - \,
  \f{ 1  }{1- \ex{ \f{2\pi \wt{\ga}_+ }{ \a } (z-s) }   }    \Bigg\} 
\; + \;  \Int{\R }{}   \f{  \wt{\ga}_+ \, \dd s }{ \i \a }   
\f{ G_{\Xi}(s)    }{1- \ex{ \f{2\pi \wt{\ga}_+ }{ \a } (z-s) }   }   \;. 
\enq
The last integral may be recast in a form which allows one to readily 
extract the asymptotic behaviour at $\Re(z)\tend + \infty$. For that 
purpose, one observes that 
\bem
\Int{\R }{}   \f{  \wt{\ga}_+ \, \dd s }{ \i \a }   
\f{ \mc{G}_{\Xi}(s)-\mc{G}_{\Xi}(g(s)-\i\a)    }{1- \ex{ \f{2\pi 
\wt{\ga}_+ }{ \a } (z-s) }   }  \; = \; 
\Int{-\infty }{M}   \f{  \wt{\ga}_+ \, \dd s }{ \i \a }   
\f{ \mc{G}_{\Xi}(s)-\mc{G}_{\Xi}(g(s)-\i\a)    }{1- \ex{ \f{2\pi 
\wt{\ga}_+ }{ \a } (z-s) }   }   \\ 
\; - \;  \Int{ M + \varkappa^+ - \i\a }{ M }   \f{  \wt{\ga}_+ 
\, \dd s }{ \i \a }   \f{ \mc{G}_{\Xi}(s)    }{1- \ex{ \f{2\pi 
\wt{\ga}_+ }{ \a } (z-s) }   }
\; +  \;
\Int{ \msc{C}^{\prime}_M }{ }  \f{  \wt{\ga}_+ \, \dd s }{ \i \a }   
\f{ \mc{G}_{\Xi}(s)    }{1- \ex{ \f{2\pi \wt{\ga}_+ }{ \a } (z-s) }   }
\end{multline}
where 
\beq
\msc{C}^{\prime}_M\, = \, \intof{+\infty - \i \a }{ M + \varkappa^+ 
-\i\a } \cup \intff{M+\varkappa^+-\i\a}{M} \cup \intfo{M}{+\infty} \;. 
\enq
The last integral can be taken by residues, hence leading to 
\bem
\Xi(z) \, = \,    -\mc{G}_{\Xi}(z) \, + \, \Int{-\infty}{M}   
\f{  \wt{\ga}_+ \, \dd s }{ \i \a }  \Xi_+( g(s)-\i\a) 
\Bigg\{ \f{    g^{\prime}(s) }{1- \ex{ \f{2\pi \wt{\ga}_+ }{ \a } 
(z-g(s) + \i\a) }   }   \, - \,
  \f{ 1  }{1- \ex{ \f{2\pi \wt{\ga}_+ }{ \a } (z-s) }   }    \Bigg\} \; 
+ \;  \Int{\R }{}   \f{  \wt{\ga}_+ \, \dd s }{ \i \a }   
\f{ G_{\Xi}^{(\e{c})}(s)    }{1- \ex{ \f{2\pi \wt{\ga}_+ }{ \a } (z-s) 
}   }   \\
\, + \, \Int{-\infty }{M}   \f{  \wt{\ga}_+ \, \dd s }{ \i \a }   
\f{ \mc{G}_{\Xi}(s)-\mc{G}_{\Xi}(g(s)-\i\a)    }{1- \ex{ \f{2\pi 
\wt{\ga}_+ }{ \a } (z-s) }   } 
\; - \;  \Int{ M + \varkappa^+ - \i\a }{ M }   \f{  \wt{\ga}_+ 
\, \dd s }{ \i \a }   \f{ \mc{G}_{\Xi}(s)    }{1- \ex{ \f{2\pi 
\wt{\ga}_+ }{ \a } (z-s) }   } \;. 
\label{ecriture representation Xi reguliere pour asymptotiques a droite}
\end{multline}
The form of the asymptotic expansion at $\Re(z)\tend + \infty$ is readily 
deduced from this representation.

Quite similarly to the previous case, one infers the integral 
representation 
\bem
\Xi(z) \,  = \, \lim_{r \tend + \infty} \Int{  \msc{C}_{r,\a}    }{} 
\f{  \wt{\ga}_- \, \dd s }{ \i \a }  \f{  \Xi(s) }{\ex{ \f{2\pi 
\wt{\ga}_- }{ \a } (s-z) } -1  }  \\
 \,  = \, \lim_{r \tend + \infty} \Bigg[ \Int{  -r  }{r} 
\f{  \wt{\ga}_- \, \dd s }{ \i \a }  \bigg[ \f{  \Xi_+( g(s)-\i\a)  
g^{\prime}(s) }{ \ex{ \f{2\pi \wt{\ga}_- }{ \a } (g(s)-z - \i\a) } -1  }   
 \, - \, \f{  \Xi_-(s) }{ \ex{ \f{2\pi \wt{\ga}_- }{ \a } (s-z) } 
- 1    } \bigg] \; + \; 
 \bigg\{    \Int{-r}{-r+\varkappa^--\i\a } +  \Int{r+\varkappa^+ 
- \i\a }{r} \bigg\}   \f{  \wt{\ga}_- \, \dd s }{ \i \a }  
\f{  \Xi(s) }{ \ex{ \f{2\pi \wt{\ga}_- }{ \a } (s-z) } - 1    }    \Bigg] \;. 
\label{ecriture rep convenable pour cptment moins infty de Xi}
\end{multline}
The integral over the segment $\intff{-r}{-r+\varkappa_--\i\a }$ 
produces $C_{\Xi}$ plus terms vanishing when $r\tend +\infty$ due 
to the form of the $\Re(z)\tend -\infty$ asymptotics of $\Xi$,
\textit{c.f.} \eqref{ecritue DA pm infty de Xi}. 
The integral over $\intff{r+\varkappa_+ - \i\a }{r}$ vanishes since  
$\Xi$ is bounded in that direction and the denominator blows up 
exponentially fast. 
Finally, the integrand of the first integral appearing in 
\eqref{ecriture rep convenable pour cptment moins infty de Xi} is 
in $L^{1}(\R)$ due to the asymptotic behaviour of $\Xi$ at infinity.

Then, proceeding analogously as in the $\Re(z)\tend + \infty$ case  
one  gets  
\beq
 \Xi(z) \,  = \, C_{\Xi}\, + \,   \Int{  -M   }{  + \infty } 
\f{  \wt{\ga}_- \, \dd s }{ \i \a }  \Xi_+( g(s)-\i\a) \cdot  
\bigg\{ \f{   g^{\prime}(s) }{ \ex{ \f{2\pi \wt{\ga}_- }{ \a } 
(g(s)-z - \i\a) } -1  }   
 \, - \, \f{  1 }{ \ex{ \f{2\pi \wt{\ga}_- }{ \a } (s-z) } - 1    } 
\bigg\} \; + \; 
\Int{  \R    }{   } \f{  \wt{\ga}_- \, \dd s }{ \i \a } 
\f{ G_{\Xi}(s) }{ \ex{ \f{2\pi \wt{\ga}_- }{ \a } (s-z) } - 1    }  \;. 
\enq
Furthermore, one has 
\bem
\Int{\R }{}   \f{  \wt{\ga}_- \, \dd s }{ \i \a }   
\f{ \mc{G}_{\Xi}(s)-\mc{G}_{\Xi}(g(s)-\i\a)    }{ \ex{ \f{2\pi 
\wt{\ga}_- }{ \a } (s-z) }-1   }  \; = \; 
\Int{-M }{+\infty}   \f{  \wt{\ga}_- \, \dd s }{ \i \a }   
\f{ \mc{G}_{\Xi}(s) - \mc{G}_{\Xi}(g(s)-\i\a)    }{\ex{ \f{2\pi 
\wt{\ga}_- }{ \a } (s-z) } -1  }   \\ 
\; - \;  \Int{-M}{-M+\varkappa^--\i\a}   \f{  \wt{\ga}_- \, 
\dd s }{ \i \a }   \f{ \mc{G}_{\Xi}(s)    }{ \ex{ \f{2\pi 
\wt{\ga}_- }{ \a } (s-z) } -1  }
\; +  \;
\Int{ \msc{C}^{\prime\prime}_M }{ }  \f{  \wt{\ga}_- \, 
\dd s }{ \i \a }   \f{ \mc{G}_{\Xi}(s)    }{  \ex{ \f{2\pi 
\wt{\ga}_- }{ \a } (s-z) } -1  }
\end{multline}
where 
\beq
\msc{C}^{\prime\prime}_M\, = \, \intfo{-\infty}{-M} \cup 
\intof{-M}{ -M + \varkappa^- -\i\a } \cup \intff{-M 
+ \varkappa^ - - \i\a }{ -\infty- \i \a }    \;. 
\enq
This yields 
\bem
 \Xi(z) \,  = \, C_{\Xi} \, - \,  \mc{G}_{\Xi}(z) \, + \,   
\Int{  -M   }{  + \infty } \f{  \wt{\ga}_- \, \dd s }{ \i \a }  
\Xi_+( g(s)-\i\a) \cdot 
\bigg\{ \f{   g^{\prime}(s) }{ \ex{ \f{2\pi 
\wt{\ga}_- }{ \a } (g(s)-z - \i\a) } -1  }   
 \, - \, \f{  1 }{ \ex{ \f{2\pi \wt{\ga}_- }{ \a } (s-z) } - 1    } 
\bigg\} \; + \; 
\Int{  \R    }{   } \f{  \wt{\ga}_- \, \dd s }{ \i \a } 
\f{ G_{\Xi}^{(\e{c})}(s) }{ \ex{ \f{2\pi \wt{\ga}_- }{ \a } 
(s-z) } - 1    } \\
\, + \, \Int{-M }{+\infty}   \f{  \wt{\ga}_- \, \dd s }{ \i \a }   
\f{  \mc{G}_{\Xi}(s)- \mc{G}_{\Xi}(g(s)-\i\a)    }{\ex{ \f{2\pi 
\wt{\ga}_- }{ \a } (s-z) } -1  } 
\; - \;  \Int{-M}{-M+\varkappa_--\i\a}   \f{  \wt{\ga}_- \, 
\dd s }{ \i \a }   \f{  \mc{G}_{\Xi}(s)    }{ \ex{ \f{2\pi 
\wt{\ga}_- }{ \a } (s-z) } -1  } \;. 
\label{ecriture representation Xi reguliere pour asymptotiques a gauche}
\end{multline}
Thus, the asymptotic behaviour at $\Re(z)\tend -\infty$ follows, 
along with its uniformness up to the boundary. 
 \qed

One should note that the representation 
\eqref{ecriture representation Xi reguliere pour asymptotiques a droite} 
clearly indicates that the $\pm$ boundary values $\Xi_{\pm}(x)$ for 
$x \geq M$ \textit{only} depend on the boundary values  
$\Xi_{+}( g(y)-\i\a)$, with $y<M$. 
The fact that the boundary values $\Xi_{+}(x-\i\a)$ and $\Xi_-(x)$ 
are smooth when $x\geq M$ is also clear from this representation. 
A similar property can be inferred from 
\eqref{ecriture representation Xi reguliere pour asymptotiques a gauche} 
relatively to the properties of the boundary values when $x<-M$.

\subsubsection*{$\bullet$ Correspondence with integral equations}
\label{SousSousSection Correspondance avec les LIE}

\vspace{3mm}

We now establish a one-to-one correspondence between solutions to 
the non-local Riemann Hilbert problem for $\Xi$ and solutions to 
certain linear integral equations on the space 
\beq
\mc{E}(\R) \; = \; \bigg\{ f \in L^{2}_{\e{loc}}(\R) \, ; \, 
\exists\;\; C_f\; \e{and} \; \eta>0  \quad     f(x)\, = \, C_{f}\, 
\de_{\pm;-} \, + \, \e{O}\Big( \ex{\mp \eta x} \Big)     
\qquad \e{for} \qquad x\tend \pm \infty  \bigg\} \;.
\label{definition espace E de R}
\enq

\begin{lemme}
\label{Lemme eqn integrale type sinh pour theta} 
 
Let $\Xi$ be a solution to the non-local Riemann-Hilbert problem 
for $\Xi$. Then $\Xi$ has smooth boundary values on $\Dp{}\mc{S}_{\a}$. 
Moreover, given any $\tau>\a$, the function $\th(x) \, = \, 
\Xi_+(g(x)-\i\a)$ belongs to $\mc{E}(\R)$
and solves the linear integral equation on $\mc{E}(\R)$
\beq
\big(\e{id}\, - \, \op{K}\big)[\th](x) \, = \, \f{ 1 }{ 2 } 
\Big\{  G_{\Xi}(x)  \, + \,  \mc{H}_{\R}\big[G_{\Xi}\big](x) \Big\} \, 
- \, \op{K}_{12}[G_{\Xi}](x) \;, 
\label{ecriture de eqn lin noyau sinh pour caracteriser les solutions}
\enq
where 
\beq
 \mc{H}_{\R}\big[f\big](x)\, = \,  \Fint{\R }{} \f{   \dd y }{  \i \tau }  
  \f{     f(y) }{ \sinh\Big[ \f{\pi}{\tau}(y-x) \Big]  } \; ,  
\enq
with the principal value prescription for the integral, 
is the sinh-Hilbert transform on $L^2(\R)$ and  
\beq
\op{K} \; = \; \op{K}_{12}\,+\, \op{K}_{21} \, + \, \op{K}_{11}  
\label{definition operateur K sinh}
\enq
is built up in terms of the three integral operators
\beq
\op{K}_{12}[h](x) \; = \; - \Int{ \R }{}  \f{ \dd y }{2 \i \tau }
\f{ h(y)  }{ \sinh\Big[ \tfrac{\pi}{\tau} (y-g(x)+\i\a ) \Big]  }  
\quad ,  \quad 
\op{K}_{21}[h](x) \; = \;  \Int{ \R }{}  \f{ \dd y }{ 2 \i \tau } 
\f{  h(y) g^{\prime}(y)  }{ \sinh\Big[ \tfrac{\pi}{\tau} ( g(y)-x 
- \i \a ) \Big]  }
\label{definition operateurs K12 et K21 sinh}
\enq
and 
\beq
\op{K}_{11}[h](x) \; = \;  \Int{ \R }{}  \f{ \dd y }{ 2  \i \tau } 
h(y) \bigg\{ \f{  g^{\prime}(y)  }{ \sinh\Big[  \tfrac{\pi}{\tau} 
(g(y)-g(x)) \Big]  } \, - \,  \f{ 1  }{ \sinh\Big[  \tfrac{\pi}{\tau} 
(y-x) \Big] }     \bigg\} \;. 
\label{definition operateur K11 sinh}
\enq

Reciprocally, any solution $\th \in \mc{E}(\R)$ to the linear integral 
equation 
\eqref{ecriture de eqn lin noyau sinh pour caracteriser les solutions} 
gives rise to a solution to the non-local Riemann--Hilbert problem for 
$\Xi$.

\end{lemme}

\Proof 

The asymptotic behaviour of $\Xi$ at infinity ensures that, for any 
$z \in \mc{S}_{\a}$,  
one has:
\beq
\Xi(z) \, = \, \Int{ \Dp{}\mc{S}_{\a} }{} \f{ \dd y }{ 2 \i\tau } 
\f{ \Xi(y)  }{ \sinh\Big[  \tfrac{\pi}{\tau}(y-z) \Big] } \;. 
\enq
Then, by setting  $\th_1(x) \; = \; \Xi_{+}\big( p_1(x)\big)$ and  
$\th_2(x) \; = \; \Xi_{-}\big( p_2(x)\big)$, one gets 
\beq
\Xi(z) \, = \, -\Int{\R}{} \f{ \dd y }{ 2 \i \tau } 
\f{ \th_2(y)   }{  \sinh\Big[  \tfrac{\pi}{\tau}(y-z) \Big]  } \; + \; 
\Int{\R}{} \f{ \dd y }{ 2 \i \tau } \f{ \th_1(y) g^{\prime}(y)  }
{  \sinh\Big[  \tfrac{\pi}{\tau}(g(y)-z- \i \a )\Big] } \;. 
\label{ecriture rep int solution RHP pour noyau Cauchy Sinh}
\enq
Furthermore, the above integral representation leads to the  
following relations for the $-$, resp. $+$, boundary values on $\R$, 
resp. $\R-\i\a$:
\beq
\left\{ \ba{ccc}  \tfrac{1}{2} \th_2(x) & = &  \op{K}_{21}
[\th_1](x) \;- \; {\displaystyle \Fint{ \R }{} } \f{ \dd y }{  2 \i \tau  } 
\f{ \th_2(y)    }{ \sinh\Big[  \tfrac{\pi}{\tau} (y-x) \Big]  }  
\vspace{2mm}  \\
\tfrac{ 1 }{ 2 } \th_1(x) & = &  \op{K}_{12}[\th_2](x)  \, +\,  
{\displaystyle \Fint{ \R }{} } \f{ \dd y }{  2 \i \tau } \f{ \th_1(y) \, 
g^{\prime}(y)  }{ \sinh\Big[  \tfrac{\pi}{\tau} ( g(y) - g(x) ) \Big]  }  
\ea \right.  \;. 
\enq
Then, adding up the two above equations and using explicitly the form 
of the jump conditions $\th_1(x)=\th_2(x)+G_{\Xi}(x)$, 
one gets the linear integral equation 
\eqref{ecriture de eqn lin noyau sinh pour caracteriser les solutions}. 
The latter allows one to represent $\th_1$ as  
\beq
\th_{1}(x) \, = \, \op{K}[\th_1](x) \, + \,  \f{ 1 }{ 2 } 
\Big\{  G_{\Xi}(x)  \, + \,  \mc{H}_{\R}\big[G_{\Xi}-G_{\Xi}(x)\big](x) 
\Big\} \, - \, \op{K}_{12}[G_{\Xi}](x)
\enq
from which the smoothness of $\Xi_{+}(g(x)-\i\a)$ on $\R$ is manifest. 
The latter ensures that $\Xi_{+}(x-\i\a)$ is smooth as well. 
The smoothness of $\Xi_{-}$ follows from $\Xi_{-}(x) \, = \, 
\th_1(x) \, - \,   G_{\Xi}(x)$. 
Finally, one has that $\th_1\in \mc{E}(\R)$ as can be inferred from 
the asymptotic behaviour \eqref{definition decroissance Xi a infini} 
established in Lemma \ref{Lemme decroissance reguliere en infinity du RHP}
and the hypotheses on $\mc{G}_{\Xi}$ that are outlined  
after \eqref{ecriture propriete fonction shift holomorphe}. 
\vspace{2mm}

Reciprocally, let $\th$ be a solution to the linear integral equation 
\eqref{ecriture de eqn lin noyau sinh pour caracteriser les solutions} 
on $\mc{E}(\R)$. Then, since the integral kernels $K_{ab}$ and 
$G_{\Xi}$ are all smooth, so is $\th$. Next, one defines a holomorphic 
function $\wt{\Xi}$ on $\mc{S}_{\tau-\a}$ as
\beq
\wt{\Xi}(z) \, = \, -\Int{\R}{} \f{ \dd y }{ 2 \i \tau } 
\f{\th(y) g^{\prime}(y)    }{  \sinh\Big[  \tfrac{\pi}{\tau}(g(y)-z) 
\Big]  } \; + \; 
\Int{\R}{} \f{ \dd y }{ 2 \i \tau } \f{  \th(y) \,
- \, G_{\Xi}(y)  }{  \sinh\Big[  \tfrac{\pi}{\tau}(y-z 
+\i\a )\Big] } \;. 
\enq
It is direct to check that $\wt{\Xi}$ admits smooth $-$, reps. $+$, 
boundary values on $\R$, resp. $\R-\i(\tau-\a)$. 
Furthermore, the asymptotic behaviour of $\th$ at infinity ensures that 
\beq
\wt{\Xi}(z) \, = \, C_{\wt{\Xi}} \de_{\pm;-} \;  
+ \; \e{O}\Big( \ex{\mp \eta z}\Big) \qquad \e{as} \quad \Re(z) 
\tend \pm \infty
\enq
for some $\eta>0$ and some constant $ C_{\wt{\Xi}}$. Moreover, direct 
calculations using the linear integral equation satisfied by $\th$ 
ensure that $\wt{\Xi}_-(g(x)) =\wt{\Xi}_+(x-\i(\tau-\a))$. 
Hence, $\wt{\Xi}$ satisfies the non-local Riemann-Hilbert problem 
outlined in the beginning of Subsection 
\ref{SousSection defnitions generale pour RHP non local}
under the replacement $g \hookrightarrow g^{-1}$ and corresponding 
to a vanishing shift function $G_{\wt{\Xi}}=0$. As established 
below, such homogeneous non-local Riemann-Hilbert problems admit 
only zero solutions. In particular, this entails that 
$\wt{\Xi}_-(g(x)) =\wt{\Xi}_+(x-\i(\tau-\a))=0$. These two equations provide 
one with an additional set of two singular linear integral equations 
satisfied by $\th$, namely 
\beqa
\f{1}{2}\th(x)  & =  &  \Fint{}{} \f{ \dd y }{ 2 \i \tau } 
\f{\th(y) g^{\prime}(y)    }{  \sinh\Big[  \tfrac{\pi}{\tau}(g(y)-g(x)) 
\Big]  } \; + \; 
\op{K}_{12}[\th - G_{\Xi}](x) \label{eqn int sing pour th type 1}\\
 \f{1}{2}\th(x) \,- \, \f{1}{2}G_{\Xi}(x) & =  & -  \Fint{}{} 
\f{ \dd y }{ 2 \i \tau } \f{\th(y) \,- \, 
G_{\Xi}(y)  }{  \sinh\Big[  \tfrac{\pi}{\tau}(g(y)-g(x)) \Big]  }\; + \; 
\op{K}_{21}[\th](x) \;. 
\label{eqn int sing pour th type 2}
\eeqa
This being settled, one introduces the holomorphic function $\Xi$ 
on $\mc{S}_{\a}$ such that
\beq
\Xi(z) \, = \, -\Int{\R}{} \f{ \dd y }{ 2 \i \tau } \f{ \th(y) \,
- \, G_{\Xi}(y)   }{  \sinh\Big[  \tfrac{\pi}{\tau}(y-z) \Big]  } \; + \; 
\Int{\R}{} \f{ \dd y }{ 2 \i \tau } \f{ \th(y) 
g^{\prime}(y)  }{  \sinh\Big[  \tfrac{\pi}{\tau}(g(y)-z-\i\a )\Big] } \;. 
\enq
As before, $\Xi$ admits smooth $\pm$ boundary values on $\mc{S}_{\a}$ 
and enjoys the asymptotic behaviour at infinity 
\beq
\Xi(z) \, = \, C_{\Xi} \de_{\pm;-} \;  + \; \e{O}\Big( 
\ex{\mp \eta z}\Big) \qquad \e{as} \quad \Re(z) \tend \pm \infty
\enq
for some $\eta>0$ and some constant $C_{\Xi}$. The two equations 
\eqref{eqn int sing pour th type 1}-\eqref{eqn int sing pour th type 2}
ensure that $\th(x)= \Xi_+(g(x)-\i\a)$ and $\Xi_-(x) =\th(x) \,
- \, G_{\Xi}(x)$.  
All in all, this ensures that $\Xi$ is indeed a solution 
of the non-local Riemann-Hilbert problem for $\Xi$. \qed

\subsection{Non-local Riemann-Hilbert problem in 
a smooth setting}
\label{SousSection RHP non local lisse}

The results obtained in Lemmata \ref{Lemme decroissance reguliere en 
infinity du RHP} and \ref{Lemme eqn integrale type sinh pour theta} ensure 
that the original Riemann-Hilbert problem for $\Xi$ may be equivalently 
formulated in a setting which involves much better behaved function. 

The Riemann-Hilbert problem for $\Xi$ of interest consists in finding 
\begin{itemize}
 \item $\Xi \in \mc{O}(\mc{S}_{\a})$ having smooth $-$, resp. $+$, 
boundary values on $\R$, resp. $\R-\i\a$, in particular,  
$\Xi_+(p_{1}(x))$ and $\Xi_{-}\big( p_2(x)\big)$ are both smooth on $\R$. 
 \item $\Xi_{+}\big( p_1(x)\big) \, = \, \Xi_{-}\big( p_2(x)\big)\, 
+ \, G_{\Xi}(x)$, with $x \in \R$; 
\item there exists a constants $C_{\Xi}$ and $\eta>0$  such that 
\beq
\Xi(z) =  C_{\Xi} \de_{\pm; -} \, + \,  \e{O}\Big( \ex{-\eta |\Re(z)| } 
\Big) \qquad \e{when} \qquad \Re(z) \tend \pm \infty \; 
\enq
with an asymptotic expansion that is valid uniformly up to the boundary. 
\end{itemize}

\subsection{Unique solvability of the homogeneous non-local 
Riemann-Hilbert problem and invertibility of $\e{id}-\op{K}$}
\label{Sous section solvabilite unique du RHP global}

Lemma  \ref{Lemme eqn integrale type sinh pour theta} established that 
any solution to the non-local Riemann-Hilbert problem for $\Xi$ gives 
rise to a solution to the linear integral equation driven by 
$\e{id}-\op{K}$ on $\mc{E}(\R)$. In fact, given two solutions 
$\Xi_1, \Xi_2$ to the Riemann-Hilbert problem for $\Xi$, their 
difference $\de\Xi=\Xi_1-\Xi_2$ satisfies the Riemann-Hilbert problem 
for $\Xi$ associated with a vanishing shift function and thus 
$\de\th(x)=\de\Xi_-(x)=\de\Xi_+(g(x)-\i\a)$ gives rise to a solution 
to the homogeneous integral equation 
\beq
\Big( \e{id}-\op{K}\Big)[\de \th] \, = \, 0 \;. 
\label{ecriture eqn int Id moins K homogene}
\enq
Thus, the unique solvability of the non-local Riemann-Hilbert problem 
for $\Xi$, or, equivalently, the fact that only $0$ solves the  non-local 
Riemann-Hilbert problem associated with the zero shift function, is ensured 
once that invertibility\symbolfootnote[2]{In fact, it is enough that 
\eqref{ecriture eqn int Id moins K homogene} does not admit solutions 
in the class of functions corresponding to boundary values of 
holomorphic functions solving the zero shift non-local Riemann-Hilbert 
problem for $\Xi$. However, as shown below, this is in fact equivalent 
to the invertibility.} of  $\e{id}-\op{K}$ on $\mc{E}(\R)$ is satisfied. 
The arguments developed in the literature which allow 
to establish this property, \textit{see} \cite{GakhovBoundaryValueProblems}
for the details, build strongly on the compactness of the operator 
$\op{K}$, which, however, 
does not hold in the present setting.  Consequently, 
one has to recourse to a more sophisticated reasoning 
in order to establish the invertibility of $\e{id}-\op{K}$.  

The main idea in the present case consists in splitting the operator 
$\op{K}$ introduced in \eqref{definition operateur K sinh} 
into three pieces
\beq
\op{K}\, = \, \op{L}^{++} \, + \, \op{L}^{--}  \, + \, \op{B} \;. 
\label{decomposition operateur K en L pm pls compact}
\enq
This splitting is such that the operators $\op{L}^{\pm\pm}$ have a purely 
continuous spectrum while $\op{B}$ is compact and hence has a 
pointwise spectrum. Upon establishing the invertibility of the 
operators $\e{id}-\op{L}^{\pm\pm}$, the decomposition 
\eqref{decomposition operateur K en L pm pls compact}  
reduces the question of  
invertibility of $\e{id}-\op{K}$ to the invertibility of 
an auxiliary operator $\e{id}-\op{M}$, where $\op{M}$ is compact. Once 
this stage of the analysis is reached, the remainder of the reasoning  
will be carried out within the standard techniques outlined in 
\cite{GakhovBoundaryValueProblems}. 

In the decomposition \eqref{decomposition operateur K en L pm pls 
compact}, the operator $\op{B}$ is defined as 
\beq
\op{B} \, = \, \op{K}_{11} \, + \, \de\op{K}_{12} \,  
+\, \de\op{K}_{21} \, .
\label{definition operateur B}
\enq
The integral kernels of the operators $\de\op{K}_{12}$ and 
$\de\op{K}_{21}$ are given by 
\beq
\de\op{K}_{12}(x,y) \, = \,  -   \f{1 }{2 \i \tau } 
\Bigg\{ \f{ 1  }{ \sinh\Big[\tfrac{\pi}{\tau} (y-g(x)+\i\a ) 
\Big]  } \, - \,
\sul{\ups=\pm }{} \f{  \bs{1}_{\R^{\ups}\times 
\R^{\ups}}(x,y)  }{ \sinh\Big[ \tfrac{\pi}{\tau} 
(y - x - \varkappa^{\ups} + \i\a  ) \Big]  }   \Bigg\}
\label{noyau integral operateur delta K12}
\enq
and 
\beq
\de\op{K}_{21}(x,y) \, = \,     \f{1 }{2 \i \tau } 
\Bigg\{ \f{ g^{\prime}(y)  }{ \sinh\Big[\tfrac{\pi}{\tau} 
(g(y) - x - \i\a ) \Big]  } \, - \,
\sul{\ups=\pm }{} \f{  \bs{1}_{\R^{\ups}\times 
\R^{\ups}}(x,y)  }{ \sinh\Big[ \tfrac{\pi}{\tau} 
(y + \varkappa^{\ups} - x  - \i\a  ) \Big]  }   \Bigg\} \;. 
\label{noyau integral operateur delta K21}
\enq
Finally, the operators $\op{L}^{\pm\pm}$ appearing in 
\eqref{decomposition operateur K en L pm pls compact} are integral 
operators on $L^{2}(\R^{\pm})$ with 
integral kernels  
$L^{\ups\ups}(x,y)=L^{\ups}(x-y)\cdot 
\bs{1}_{\R^{\ups}\times \R^{\ups}}(x,y)\,$ 
for the difference-dependent
\beq
L^{\ups}(x-y) \, = \,     \f{1 }{2 \i \tau } 
\Bigg\{   \f{  1  }{ \sinh\Big[ \tfrac{\pi}{\tau} 
(y + \varkappa^{\ups} - x  - \i\a  ) \Big]  }
\, - \,  \f{  1  }{ \sinh\Big[ \tfrac{\pi}{\tau} 
(y - \varkappa^{\ups} - x  + \i\a  ) \Big]  }   \Bigg\} \;. 
\label{definition noyau integral L++}
\enq

We now establish that $\op{B}$ is a compact, Hilbert-Schmidt  operator. 
\begin{prop}
\label{Proposition Caractere Hilbert Schmidt operateur B}
 $\op{B}$ is a Hilbert-Schmidt operator on $L^{2}(\R)$. Its integral 
kernel $B(x,y)$  is smooth on $\R^{\ups}\times\R^{\ups^{\prime}}$, $\ups, \ups^{\prime} \in \{\pm\}$  and enjoys the bounds
\beq
B(x,y) \; = \; \e{O}\Big(\ex{-\f{\pi}{\tau} \big( |x|+|y| \big) } \Big)  \;. 
\label{ecriture borne noyau integral B decroissance exponentielle}
\enq
The remainder appearing above is differentiable in the sense that 
$\Dp{x}^k \Dp{y}^{\ell}B(x,y) \; 
= \; \e{O}\Big(\ex{-\f{\pi}{\tau} \big( |x|+|y| \big) } \Big) $ 
on $\R^{\ups}\times\R^{\ups^{\prime}}$ 
for integers $k, \ell\in \mathbb{N}$. The control is however not uniform 
with respect to the order of the derivatives. 

\end{prop}
\Proof 

Smoothness of $B(x,y)$ on $\R^{\ups}\times\R^{\ups^{\prime}}$ is 
evident from 
\eqref{definition operateur B}-\eqref{noyau integral operateur delta K21}. 
The Hilbert-Schmidt nature of $\op{B}$ is a direct consequence of 
the bounds 
\eqref{ecriture borne noyau integral B decroissance exponentielle}. 
Finally, in order to establish \eqref{ecriture borne noyau integral 
B decroissance exponentielle} one bounds each of the three building 
blocks of the operator  $\op{B}$ separately. 

First of all consider $K_{11}(x,y)$. For $|y|\leq M$ and $x \tend 
\pm \infty$ one obviously gets $K_{11}(x,y) \, = \,  
\e{O}\Big(\ex{-\f{\pi}{\tau} |x|  } \Big)$. 
A similar bound holds for $|x|\leq M$ and $y \tend \pm \infty$: 
$K_{11}(x,y) \, = \,  \e{O}\Big(\ex{-\f{\pi}{\tau} |y|  } \Big)$. 
Finally, observe that, if  $|x|\geq M$ and  $|y|\geq M$ and $x y >0$ 
then obviously $K_{11}(x,y)=0$. While, for $x y <0$, one gets that 
$K_{11}(x,y) \, = \,  \e{O}\Big(\ex{-\f{\pi}{\tau} |x-y|  } 
\Big)= \e{O}\Big(\ex{-\f{\pi}{\tau} (|x|+|y|)  } \Big) $. 
All in all,
\beq
K_{11}(x,y) \; = \; \e{O}\Big(\ex{-\f{\pi}{\tau} \big( |x|+|y| \big) } 
\Big) \; . 
\enq

Regarding to $\de K_{12}(x,y)$. Assume that $(x,y) 
\in ( \R^{\pm} )^2$ then, by construction $\de K_{12}(x,y)=0$ 
if $|x| \geq M$, while for $|x|\leq M$, 
$\de K_{12}(x,y) \, = \,  \e{O}\Big(\ex{-\f{\pi}{\tau} |y|  } \Big)$. 
It remains to focus on the situation when $xy<0$. Then, taken 
the difference form of the kernel, for large arguments one gets that 
$\de K_{12}(x,y) \, = \,  \e{O}\Big(\ex{-\f{\pi}{\tau} |x-y|  } 
\Big) \, = \, \e{O}\Big(\ex{-\f{\pi}{\tau} (|x|+|y|)  } \Big)$. Thus, 
all in all
\beq
\de K_{12}(x,y) \; = \; \e{O}\Big(\ex{-\f{\pi}{\tau} 
\big( |x|+|y| \big) } \Big) \; . 
\enq

Finally, regarding to $\de K_{21}(x,y)$ the reasoning 
 is quite analogous. Assume that $(x,y) \in ( \R^{\pm} )^2$. 
Then, by construction $\de K_{12}(x,y)=0$ if $|y| \geq M$, while 
for $|y|\leq M$, $\de K_{12}(x,y) \, = \,  \e{O}\Big(\ex{-\f{\pi}{\tau} 
|x|  } \Big)$. When $xy<0$ the difference form of the kernel leads to 
$\de K_{12}(x,y) \, = \, \e{O}\Big(\ex{-\f{\pi}{\tau} (|x|+|y|)  } \Big)$. 
Thus
\beq
\de K_{21}(x,y) \; = \; \e{O}\Big(\ex{-\f{\pi}{\tau} 
\big( |x|+|y| \big) } \Big) \; . 
\enq
Furthermore, it is clear from the reasonings above that the remainders 
are differentiable on $\R^{\ups}\times\R^{\ups^{\prime}}$. 

Taken all together, this yields the claimed bounds on the integral 
kernel $B(x,y)$.   \qed

\subsubsection{Preliminary notations}
\label{SousSousSection Notation preliminaires}

It will appear useful, at various instances, to introduce the 
basic building block function
\beq
\mf{m}_{\zeta}(x) \, = \, \f{ 1 }{ 2 \i \tau   \sinh\Big[ 
\f{\pi}{\tau} (x  - \i\zeta  ) \Big]   }  \qquad \e{so}\; 
\e{that} \qquad 
\mc{F}\big[\mf{m}_{ \pm \zeta}  \big](k) \, 
= \,  \f{ \pm \ex{ \mp k\zeta} }{ 1 + \ex{ \mp k\tau} } \;, 
\label{definition fonction frak d de base}
\enq
provided that $0< \Re\big( \zeta \big)  < \tau$.  
Here and in the following, we define the Fourier transform, whenever 
it makes sense,  
with the convention
\beq
\mc{F}[f](k) \, = \, \Int{\R}{} \dd x f(x) \ex{\i k x}  \;. 
\enq

Since one can express the difference dependent integral kernels 
appearing in $\op{K}$ as 
\beq
L^{\ups}(x) \, = \,  \mf{m}_{\a+\i\varkappa^{\ups} } (x) \, 
- \, \mf{m}_{-\a-\i\varkappa^{\ups} } (x)  \; , 
\enq
with $\ups \in \{\pm \}$, one  
infers that 
\beq
 \mc{F}\big[   L^{\ups}  \big](k)  \; 
= \; \f{ \cosh\big[k(\tfrac{\tau}{2}-\a-\i\varkappa^{\ups} ) 
\big] }{    \cosh\big[\tfrac{k \tau}{2}  \big]  }
\enq
so that 
\beq
1 \, - \, \mc{F}\big[   L^{\ups}  \big](k)  \; = \;
 2 \f{ \sinh\big[\tfrac{k}{2}(\a+\i\varkappa^{\ups} ) \big] \cdot 
 \sinh\big[\tfrac{k}{2}(\tau-\a-\i\varkappa^{\ups} )   \big] }
{    \cosh\big[\tfrac{k \tau}{2}  \big]  } \;. 
\enq
Consider the function $k\mapsto 1 \, - \, \mc{F}\big[ L^{\ups}  \big](k) $ on $\R+\i \ups  v$ with $0<v \ll 1$. 
It is non-vanishing on this line. Furthermore, introduce
\beq
\mc{B}_{\ua}^{(\pm)} \; = \; \Big\{ z \in \Cx \; : \;  
\Im(z) \, > \, \pm v \Big\} \qquad \e{and} 
\qquad \mc{B}_{\da}^{(\pm)} \; = \; 
\Big\{ z \in \Cx \; : \;  \Im(z) \, < \, \pm v \Big\}   \;. 
\label{definition domaines B up down plus moins}
\enq
Then, $1 \, - \, \mc{F}\big[   L^{\ups}  \big]$ admits 
the Wiener-Hopf factorisation 
\beq
1 \, - \, \mc{F}\big[   L^{\ups}  \big](k)  
\; = \; \f{ \a_{\ua}^{(\ups)}(k)  }{ \a_{\da}^{(\ups)}(k)   } 
\label{ecriture eqn de WH pour 1-L eps eps}
\enq
such that 
\begin{itemize}
\item $\a_{\ua}^{(\ups)}\in \mc{O}\Big( \ov{ \mc{B}_{\ua}^{(\ups)}} 
\Big) $ and $\a_{\da}^{(\ups)}\in \mc{O}
\Big( \ov{ \mc{B}_{\da}^{(\ups)}}\setminus \{ 0 \} \Big)$;
\item $\a_{\ua/\da}^{(\ups)}(k) \; \tend  \; 1$ \quad when  \quad $ k\tend  \infty$  \quad with \quad  $k \in \ov{ \mc{B}_{\ua/\da}^{(\ups)} }$  \;. 
\end{itemize}
The Wiener-Hopf factors are given explicitly in terms of Gamma functions as  
\beq
 \a_{\ua}^{(\ups)}(k)  \; = \; - \i k  \f{ \sqrt{2\pi A^{\ups}    
B^{\ups} } \cdot 
\big[ C^{\ups}\big]^{\i k  C^{\ups} }   }
{ \big[ A^{\ups}\big]^{\i k  A^{\ups} } \cdot 
  \big[ B^{\ups}\big]^{\i k  B^{\ups} }     }
\Ga \left( \ba{c}   \tfrac{1}{2} -\i  C^{\ups} k   \\ 
 1  - \i  B^{\ups} k \, ,  1 -\i  A^{\ups} k   \ea   \right)
\label{forme explicite alpha up kappa ups}
\enq
and 
\beq
 \a_{\da}^{(\ups)}(k)  \; = \;   \i k  \sqrt{  \f{ A^{\ups}    
B^{\ups} }{ 2\pi  } } 
 \f{    \big[ C^{\ups}\big]^{\i k  C^{\ups} }   }
{ \big[ A^{\ups}\big]^{\i k  A^{\ups} } \cdot   
\big[ B^{\ups}\big]^{\i k  B^{\ups} }     }
\Ga \left( \ba{c}     \i  A^{\ups} k \, ,   \i  B^{\ups} k   \\ 
   \tfrac{1}{2}  +  \i  C^{\ups} k    \ea   \right) \;. 
\label{forme explicite alpha da kappa ups}
\enq
There, we adopted the conventions introduced in 
\eqref{convention notation produit fonctions gamma} and made 
use of the  
following shorthand notations:
\beq
A^{\ups} \, = \, \f{ \a +\i\varkappa^{\ups} }{2\pi} \quad, 
\qquad B^{\ups} \, = \, \f{ \tau-\a -\i\varkappa^{\ups} }{2\pi} 
\quad \e{and} \quad  
 C^{\ups} \, = \, \f{ \tau }{2\pi}  \;. 
\enq

Note that $ \a_{\ua/\da}^{(\ups)}$ admit meromorphic continuations 
to  $\mc{B}_{\da/\ua}^{(\ups)}$. Furthermore, 
$ \a_{\ua}^{(\ups)}$ has a simple zero at $k=0$ and this is its only 
zero in some open neighbourhood of $\R$. Also, $ \a_{\da}^{(\ups)}$ 
admits a simple pole at $k=0$
and it is its only simple pole in some open neighbourhood of $\R$. 
For further convenience, we parameterise this local behaviour as 
\beq
 \a_{\ua}^{(\ups)}(k)  \; \widesim{k\tend 0}  k \, 
\a_{0}^{(\ups)} \qquad \e{and} \qquad \a_{\da}^{(\ups)}(k)  \; 
\widesim{k\tend 0}  \f{\wt{\a}_{0}^{\, (\ups)} }{ k  }\;.  
\enq
One has
\beq
\a_{0}^{(\ups)}\wt{\a}_{0}^{\, (\ups)}=-1\,.
\label{for b(0)=-1}
\enq

\subsubsection{Preparatory decomposition for $\,\e{id}-\op{K}$}
\label{SousSousSection reduction 1-K a 1-M}

We are now in position to discuss the invertibility of the operator 
$\e{id}-\op{K}$ on the space $\mc{E}(\R)$ as defined in 
\eqref{definition espace E de R}, 
with $\op{K}$ as introduced in \eqref{definition operateur K sinh}  
and rewritten in \eqref{decomposition operateur K en L pm pls compact}.
Assume that one is given a solution $f\in \mc{E}(\R) $ to  
$\big(\e{id}-\op{K}\big)[f]=h$ 
with $h$ having an exponential fall-off at $\pm \infty$, \textit{viz}. 
$h(x) = \e{O}\Big(\ex{- \eta |x| }\Big)$ for some $\eta>0$. 
Then, one may recast the equation in a matrix form relatively to 
the decomposition $\mc{E}(\R)\, = \, \mc{E}(\R^-)\oplus \mc{E}(\R^+)$ 
with 
\beq
\mc{E}(\R^{\sg}) \; = \; \Big\{ f\cdot \bs{1}_{\R^{\sg}} \; : 
\; f \in \mc{E}(\R) \Big\} \;, 
\enq
as
\beq
\left( \ba{cc}  \e{id} \, - \, \op{L}^{--} \, - \,  
\op{B}^{--} &   - \op{B}^{-+}  \\ 
	  -  \op{B}^{+-}  & 				\e{id} \, 
- \, \op{L}^{++} \, - \,  \op{B}^{++}   	
\ea \right) \cdot \left(\ba{c}  f^{-} \\ f^{+}   \ea  \right) 
\; = \;   \left(\ba{c}  h^{-} \\ h^{+}   \ea  \right)  \;. 
\enq
Or, more explicitly 
\beqa
 \Big( \e{id} \, - \, \op{L}^{--} \Big)[f^-]  & 
=  &   \op{B}^{-+}[f^+]\, + \,  \op{B}^{--}[f^-]\, 
+ \, h^{-} \, \equiv \, H^-   \;, \\
 \Big( \e{id} \, - \, \op{L}^{++} \Big)[f^+]  & 
=  &   \op{B}^{++}[f^+]\, + \,  \op{B}^{+-}[f^-]\, 
+ \, h^{+} \, \equiv \, H^+  \;. 
\eeqa
It follows from the above and Proposition 
\ref{Proposition Caractere Hilbert Schmidt operateur B} that the 
functions $f^{\pm}$ and $H^{\pm}$
do belong to the classes considered in Subsections 
\ref{Soussection inversion L ++}-\ref{Soussection inversion L --} 
of the Appendix. Then,  the results from these sections entail that,   
for $v>0$ and small enough
\beqa
\mc{F}\big[ f^{-} \big](k) & = & 
  -    \Int{\R + \i v }{} \f{ \dd s }{2\i\pi}   
\f{  \a_{\da}^{(-)}(k)   \cdot
   \mc{F}\big[ H^- \big] (s) }
{   \a_{\ua}^{(-)}(s) \cdot 
(s-k) }
\quad , \qquad k \in \R - \i v \;, \label{ecriture rep Fourier f moins}\\ 
 \mc{F}\big[ f^{+} \big](k) & = &    
\Int{\R - \i v }{} \f{ \dd s }{2\i\pi}  
\f{  \a_{\da}^{(+)}(s)  \cdot    
\mc{F}\big[ H^+ \big] (s) }{  \a_{\ua}^{(+)}(k) \cdot 
( s-k ) } 
\quad \; \;  , \qquad k \in \R + \i v   \;. 
\label{ecriture rep Fourier f plus}
\eeqa

One may recast the system of equations subordinate to 
\eqref{ecriture rep Fourier f moins}-\eqref{ecriture rep Fourier f plus} 
as a matrix integral equation on \newline $L^2(\R - \i v )
\oplus L^2(\R + \i v)$ on the  unknown vector 
\beq
\bs{u} \, = \, \left( \ba{c} u^- \\ u^+ \ea \right) 
\quad \e{with} \quad  u^{\sg} \, 
= \, \mc{F}[f^{\sg}]\, , \;\; \sg=\pm \; . 
\enq
 For that purpose one observes that 
\beq
\mc{F}\Big[ \op{B}^{\eps \sg} [f^{\sg}] \Big](k) \; 
= \; \Int{ \R+ \i \sg v}{} \hspace{-2mm}\dd k \, 
\wh{B}^{\eps \sg}(k,s) \mc{F}[f^{\sg}](s) 
\enq
in which 
\beq
\wh{B}^{\eps \sg}(k,s) \; = \; \Int{ \R^{\eps}\times \R^{\sg} }{}  
\hspace{-2mm} \f{ \dd x \dd y}{2\pi} \ex{\i k x - \i s y } B(x,y) \;. 
\label{definition Fourier transform de B sur demi axes}
\enq
The fact that the Fourier transforms $\wh{B}^{\eps \sg}(k,s)$ 
are well-defined for $(k,s) \in \big\{ \R - \i \eps v\big\}
\times \big\{ \R+\i \sg v\big\}$ is a direct consequence of 
Proposition \ref{Proposition Caractere Hilbert Schmidt operateur B}. 
In fact, one has that $\wh{B}^{\eps \sg}(k,s)$ is analytic in $k,s$ 
belonging to a tubular neighbourhood of $\R^2$. Moreover, 
the asymptotics established in Proposition 
\ref{Proposition Caractere Hilbert Schmidt operateur B} entail that, 
for some constant $C>0$,  
\beq
\Big| \wh{B}^{\eps \sg}(k,s) \Big| \; \leq  \;  
\f{ C }{ (1+|s|)\cdot
(1+|k|) }
\label{ecriture estimee decroissance asymptotiques hat B}
\enq
uniformly throughout this tubular neighbourhood. 
Taken 
\eqref{ecriture rep Fourier f moins}-\eqref{ecriture rep Fourier f plus},  
it appears convenient to introduce 
\beq
M^{+\sg}(k,t)  \; =  \;  \f{1}{ \a_{\ua}^{(+)}(k) } \cdot   
\Int{ \R - \i v }{}   \f{ \dd s }{ 2\i \pi }  
\f{ \a_{\da}^{(+)}(s) }{ s-k }   \wh{B}^{+\sg}(s,t)    
\label{ecriture noyau integral M + sg}
\enq
and 
\beq
M^{-\sg}(k,t)  \; = \; - \a_{\da}^{(-)}(k) \cdot   
\Int{ \R + \i v }{}  \f{ \dd s }{ 2\i \pi }  
\f{ \big\{  \a_{\ua}^{(-)}(s) \big\}^{-1}  }{ s-k }   
\wh{B}^{-\sg}(s,t)     \;. 
\label{ecriture noyau integral M - sg}
\enq
It is direct to check that $M^{\eps\sg}$ is smooth on 
$\big\{ \R+ \i \eps v \big\}\times \big\{ \R + \i \sg v \big\}$. 
Furthermore, it follows from Lemma 
\ref{Lemme bornes sur croissance via transfo Cauchy}
that there exists a constant $C$ such that 
\beq
\big| M^{ \eps \sg}(k,t) \big| \; \leq  \; \f{ C \cdot 
\ln (1+|k|)  }{ (1+|k|) \cdot
(1+|t|)  }  \qquad \e{for} 
\qquad (k,t)\in \big\{ \R+ \i \eps v \big\}\times 
\big\{ \R + \i \sg v \big\}  \;. 
\label{borne sur croissance de M}
\enq
This entails that the operator $\op{M}$ on $L^2(\R - \i v )
\oplus L^2(\R  + \i v)$ given in matrix form 
\beq
\op{M} \; = \; \left(\ba{cc}  \op{M}^{--} & \op{M}^{-+}  \\ 
				\op{M}^{+-} & \op{M}^{++}    
\ea \right)
\label{ecriture operateur integral matriciel M}
\enq
is Hilbert-Schmidt. Indeed, the Hilbert-Schmidt norm of interest 
takes the form 
\beq
\norm{ \op{M} }_{HS}^2 \; = \; \sul{  \eps, \sg }{} 
\Int{ \R +\i \eps v }{} \dd k  \Int{ \R +\i \sg v }{} 
\hspace{-2mm}  \dd s \, \big| M^{\eps \sg}(k,s) \big|^2 
\enq
and its finiteness follows from the bounds 
\eqref{borne sur croissance de M}.

Then, introducing 
\beqa
\mf{d}^{+}[h](k) & = &       \Int{\R - \i v }{} 
\f{ \dd s }{2\i\pi}  \f{  \a_{\da}^{(+)}(s)  \cdot   
\mc{F}\big[ h^+ \big] (s) }{  \a_{\ua}^{(+)}(k) 
\cdot ( s-k ) }     \;,  \\
\mf{d}^{-}[h](k) & = &   -    \Int{\R + \i v }{} 
\f{ \dd s }{2\i\pi}   \f{  \a_{\da}^{(-)}(k)   \cdot   
\mc{F}\big[ h^- \big] (s) }{   \a_{\ua}^{(-)}(s) \cdot 
(s-k) }  \; , 
\eeqa
one ends up with the linear integral equation 
\beq
\Big(\e{id} - \op{M} \Big)\big[ \bs{u} \big] \; 
= \; \bs{\mf{d}}[h] \qquad \e{with} \qquad \bs{u} \, 
= \, \left( \ba{c} u^- \\ u^+ \ea \right) \qquad \e{and} 
\qquad   \bs{\mf{d}}[h] \, 
= \, \left( \ba{c} \mf{d}^{-}[h] \\ \mf{d}^{+}[h] \ea \right)  \;. 
\enq

Reciprocally, given any solution $\bs{u}$ to the above equation, 
by using the analyticity properties of the functions 
$\mf{d}^{\sg}[h]$ and the integral kernels $M^{\sg \ups}(k,s)$, 
it is direct to infer from the representation 
\beq
u^{\sg}(k)\, =\, \op{M}^{\sg +}[u^{+}](k) \, 
+ \, \op{M}^{\sg -}[u^{-}](k)+\mf{d}^{\sg}[h](k)
\enq
that $u^{\sg}\in \mc{O}(\mathbb{H}^{\sg})$ and that 
$u^{\sg}(k)=C^{\sg}/k+\e{O}(k^{-2})$ on $\mathbb{H}^{\sg}$. 
This entails that the functions    
\beq
\psi^{(\sg)}(x) \, = \, \Int{ \R +\i\sg v}{} 
\f{ \dd k }{ 2\pi } \ex{-\i k x } u^{\sg}(k)
\enq
are supported on $\R^{\sg}$ and enjoy the asymptotic behaviour 
when $x \tend \sg \infty$:
\beq
\psi^{(\sg)}(x) \; = \; C_{ \psi^{(\sg)} } \; 
+ \; \e{O}\Big(  \ex{-\sg v x}  \Big) \;,  
\enq
for some constants $C_{  \psi^{(\sg)} }$. Then, by carrying 
backwards the reasonings described in Subsections  
\ref{Soussection inversion L ++}-\ref{Soussection inversion L --}, 
one infers that 
\beq
 \Big( \e{id} \, - \, \op{L}^{\sg\sg} \Big)[\psi^{(\sg)}](x)   
=     \op{B}^{\sg +}[\psi^{(+)}](x)\, 
+ \,  \op{B}^{\sg -}[ \psi^{(-)} ](x) \, + \, h^{\sg}(x) \, 
\label{ecriture equation sur psi sigma}
\enq
for $\sg=\pm$ and $x \in \R^{\sg}$. Upon setting 
$\psi=\psi^{(+)}+\psi^{(-)}$, one gets that equation 
\eqref{ecriture equation sur psi sigma} can be recast as 
$\big( \e{id} \, - \, \op{K} \big)[\psi]=h$. 
As a consequence, since constants are in the kernel of 
$ \e{id} \, - \, \op{K} $, it follows that the function 
$\th=\psi-C_{\psi^{(+)}} $ solves 
\beq
\big( \e{id} \, - \, \op{K} \big)[\th]=h \qquad \e{with} 
\qquad \th(x) \, = \, C_{\th} \de_{\pm, -}   \; 
+ \; \e{O}\Big(  \ex{- v |x|}  \Big)  \qquad \e{when} \quad x 
\tend \pm \infty \;. 
\label{solution theta de eqn in lin RHP inhomogene non local}
\enq

\subsubsection{Invertibility of $\,\e{id}-\op{K}$ and unique solvability 
of the Riemann-Hilbert problem}
\label{SousSousSection invertibilite de id - K}

\begin{prop}
\label{Proposition solubilite unique RHP non local}
 The operator $\op{M}$ on $L^2(\R - \i v )\oplus L^2(\R  + \i v)$ 
defined through 
\eqref{ecriture noyau integral M + sg},\eqref{ecriture noyau integral M - sg} and \eqref{ecriture operateur integral matriciel M}
 satisfies $\det[\e{id}-\op{M}]\not=0$. Moreover, the non-local 
Riemann-Hilbert problem for $\Xi$ is uniquely solvable and 
the operator $\e{id}-\op{K}$ on $\mc{E}(\R)$ is invertible. 
 
\end{prop}

Once that developments of Sub-Section 
\ref{SousSousSection reduction 1-K a 1-M} have been laid down, 
the proof closely follows the reasoning outlined in 
\cite{GakhovBoundaryValueProblems}.

\Proof

We first establish that the non-local Riemann-Hilbert problem for 
$\Xi$ associated with a zero shift function, \textit{viz}. corresponding 
to $G_{\Xi}=0$, has only the trivial solution $\Xi=0$. Let  
$\Xi$ be a non-vanishing solution to this zero shift problem. 
Thus, $\Xi^{n}$ also solves this problem for any $n\in \mathbb{N}$. 
Setting $\th=\Xi_+\circ p_{2}$, one infers from Lemma 
\ref{Lemme eqn integrale type sinh pour theta} that $\th^{n}$ has 
the asymptotic behaviour  when $x \tend \pm \infty$ : $\th(x) \, 
= \, C_{\th}^{n} \de_{\pm, -}   \; + \; \e{O}\Big(  \ex{- v |x|}  
\Big)$ for some $v>0$ and solves $\big( \e{id} \, - \, \op{K} 
\big)[\th^n]=0$. Since $\Xi$ is  non-identically vanishing, $\th$ is  
non-identically vanishing as well. By building on the earlier  
considerations, one infers that 
\beq
 \bs{u}_{n}=\left( \ba{cc}  \mc{F}\big[ \big(\th^n\big)^-\big] 
\vspace{1mm} \\ \mc{F}\big[ \big(\th^n\big)^+\big]   \ea  \right) 
\in \e{ker}\Big(\e{id} - \op{M} \Big) \;. 
\enq
We now establish that the  $ \bs{u}_{1},\dots,\bs{u}_{k}$ are linearly 
independent for any $k$. Let $c_a$ be such that 
$\sum_{n=1}^{k}c_n \bs{u}_n=0$. Component-wise this yields 
$\sum_{n=1}^{k} c_n \mc{F}\big[ \big(\th^n\big)^{\pm} \big] \, = \, 0$. 
Hence, by taking the inverse Fourier transform of the sum of these 
two relations, one gets that  $\sum_{n=1}^{k} c_n \th^n \, = \, 0$. 
Since the function $\th$ is non-zero, it also cannot be constant owing 
to its asymptotics at $+\infty$. 
Since $\th$ is non-constant, there exists $x_0$ such that 
$\th^{\prime}(x_0)\not=0$. Thus, $\th$ is a diffeomorphism in 
the neighbourhood of $x_0$. This entails that there exist a sequence 
$x_1,x_2,\dots$ of pairwise distinct reals such that 
$\th(x_a)\not=\th(x_b)$ for any $a\not=b$. One then infers from 
the relation $\sum_{n=1}^{k} c_n \th^n \, = \, 0$ the 
system of equations 
\beq
\sul{n=1}{k}c_n\th^n (x_s) \, = \, 0 \; , \quad s=1,\dots, k \;. 
\enq
However, the latter has only trivial solutions owing to the 
invertibility of the associated Vandermonde matrix which stems 
from the condition  $\th(x_a)\not=\th(x_b)$. 

The linear independence of $ \bs{u}_{1},\dots,\bs{u}_{k}$ for any $k$ 
thus entails that $\e{ker}[\e{id} - \op{M}]$ cannot be finite  
dimensional  
contradicting the compactness of $\op{M}$.
Thus the non-local Riemann-Hilbert problem for $\Xi$ associated with 
a zero shift has only the trivial solution $\Xi=0$. 

We now establish that, $\e{ker}[\e{id} - \op{M}]=0$. If not, then 
let  $\bs{u}\in \e{ker}\Big(\e{id} - \op{M} \Big)$, $\bs{u}\not=\bs{0}$. 
Since $\bs{u} \in L^2(\R-\i v) \oplus L^2(\R+\i v)$, the large-$k$ 
asymptotic expansion of $\wh{B}^{\sg\eps}(k,s)$ which follows from 
integration by parts and differentiability of the remainder in 
Proposition \ref{Proposition Caractere Hilbert Schmidt operateur B}
\beq
\wh{B}^{\sg\eps}(k,s) \, \simeq \, \sul{ \ell \geq 0 }{} 
\f{  \vp_{\ell}(s) }{  k^{\ell} } \qquad \e{with} \qquad 
\big|  \vp_{\ell}(s) \big| \, \leq \, \f{C_{\ell} }{ 1+|s| }\;, 
\enq
entails that $u^{\sg}(k)$ admits the asymptotic expansion   
\beq
u^{\sg}(k) \, \simeq \, \sul{ \ell\geq 0  }{} k^{-\ell} 
c_{\ell}^{(\sg)}  \qquad \e{for} \qquad \Re(k)\tend \pm \infty \;. 
\enq
Furthermore, the very structure of the integral kernels of 
the operator $\op{M}$ ensures that $u^{+}\in \mc{O}
\big( \mc{B}^{(-)}_{\ua}\setminus \{0\}\big)$, resp. 
$u^{-}\in \mc{O}\big( \mc{B}^{(+)}_{\da}\setminus \{0\}\big)$. 
Also, $u^{\sg}$ admits a simple pole at $0$. Thus, by taking 
the inverse Fourier transform, one gets a solution $\th$ 
to \eqref{solution theta de eqn in lin RHP inhomogene non local} 
with $h=0$. By following the reasoning outlined in the proof 
of Lemma \ref{Lemme eqn integrale type sinh pour theta}, this solution 
 gives rise to a solution $\wt{\Xi}$ to the non-local Riemann-Hilbert 
problem for $\Xi$ having zero shift  and subordinate to the 
replacement $g \hookrightarrow g^{-1}$ in the 
welding diffeomorphism. 
But then, by the above, this Riemann-Hilbert problem has only 
trivial solutions. This allows one to introduce, following the proof  
of Lemma \ref{Lemme eqn integrale type sinh pour theta},
a holomorphic function $\Xi$ on $\mc{S}_{\a}$ solving the 
non-local Riemann-Hilbert problem for $\Xi$ with a zero shift 
and such that $\th(x)=\Xi_+(g(x)-\i\a)$. Since this problem has 
only trivial solutions, $\th=0$ and thus, by going backwards,  
$\bs{u}=\bs{0}$ as well,  
which is a contradiction.

We have just established that $\e{ker}[\e{id} - \op{M}]=0$. 
Thus, since $\op{M}$ is compact, $\e{id}-\op{M}$ is invertible and 
in particular $\det[\e{id}-\op{M}]\not=0$. 
The latter, by virtue of the construction described earlier on, 
ensures that $\e{id}-\op{K}$ is invertible as well. \qed

\section{Special non-local Riemann-Hilbert problems}
\label{Section RHP non locaux modeles}

In the following, we shall consider two smooth diffeomorphisms 
of $\R$, $g_L$ and $g_R$ which both satisfy $g^{\prime}_{L/R}>0$
and such that 
\beq
g_L(x) \;  = \; \left\{ \ba{ccc}  x && x  \, < \,  -M_L  
\vspace{2mm} \\ 
				  g_L(x) && -M_L \, < \,  x  \, 
< \,  M_L  \vspace{2mm}  \\ 
				 x +\varkappa & & M_L  \, < \,  x 	  
\ea \right. 
\quad \e{and} \quad 
g_R(x) \;  = \; \left\{ \ba{ccc}  x +\varkappa & & x \,  < \,  -M_R  
\vspace{2mm}  \\ 
				  g_R(x)& & -M_R \, < \,  x \,  < \,  
M_R  \vspace{2mm}  \\ 
				 x  &&  M_R \, < \,  x 	  
\ea \right.  \; , 
\label{definition gL et gR}
\enq
for some $M_L, M_R>0$. The purpose of this section is to establish 
the unique solvability of two non-local Riemann-Hilbert problems 
with shifts associated with the diffeomorphisms $g_{L/R}$. Prior 
to that, however, we shall establish some properties of the Cauchy 
transform on a  
welded strip which will play some role in 
later steps of the analysis.

\subsection{Cauchy transform on a welded strip}
\label{SousSection Transfo Cauchy sur Cylindre cousu}

In the following, we shall set 
\beq
\ga \, = \, \f{ -\varkappa  }{ \varkappa - \i \a }  
\qquad \e{and} \qquad \wt{\ga} \, = \, \ga + 1 \, = 
\, \f{ -\i \a  }{ \varkappa - \i \a }
\label{definition gamma et tilde gamma}
\enq
where $\varkappa \in \R$. Observe that the constant $\ga$ 
is such that $f(z) = \ga z$ satisfies to the jump condition 
\beq
f_+(x+\varkappa-\i \a ) \; = \; f_-(x) \, - 
\, \varkappa \quad x \in \R \;. 
\enq

In this subsection, we establish the main properties of a Cauchy 
transform which has the appropriate symmetry to deal properly with 
the welding diffeomorphims $p_{1;\e{sev}}(x) 
= x+\varkappa-\i \a$ from $\R$ onto $\R-\i \a$.

\begin{lemme}
 \label{Lemme Transformee de Cauchy sur Cylinder cousu}
 
 Let $\Ups$ be a holomorphic function on $\{ z \in \Cx \; : 
\; -\a <  \Im(z) < 0 \; \; y- 2 |\varkappa| < \Re(z) < 
y + \varkappa + 2 |\varkappa| \}$ for some $y\in \R$, having 
continuous $-$, resp. $+$, boundary values on the upper, resp. 
lower,  pieces of this domain and satisfying   $\Ups_+(x+\varkappa
-\i \a ) \; = \; \Ups_-(x)$. Then, the Cauchy transform 
\beq
 \op{C}_{ \Ga }[\Ups](z) \,  = \, \Int{ \Ga }{} 
\f{  \wt{\ga} \, \dd s }{ \i \a }  \f{  \Ups(s) }{ \ex{ 
\f{2\pi \wt{\ga} }{ \a } (s-z) } \, - \, 1  } 
 \, ,
\label{definition transformee Cauchy cylindre cousu}
\enq
with  $\Ga=\intff{  y }{  y - \i \a + \varkappa}$ satisfies 
the  
following non-local Riemann-Hilbert problem on $\mc{S}_{\a}$:
\begin{itemize}
 \item $ \op{C}_{ \Ga }[\Ups] \in \mc{O}(\mc{S}_{\a} 
\setminus \Ga)$ and has holomorphic $-$, resp. $+$, boundary values 
on $\Ga$;
\item $\op{C}_{ \Ga ;+}[\Ups](s) 
- \op{C}_{ \Ga ; -}[\Ups](s)=\Ups(s)$ for $s \in \Ga$;
\item $\op{C}_{ \Ga}[\Ups](x+\varkappa - \i\a ) 
= \op{C}_{ \Ga}[\Ups](x)$  for $x \in \R$;
\item up to the boundary $\Dp{}\mc{S}_{\a}$, it holds 
\beq
\op{C}_{ \Ga}[\Ups](z) \, = \, - \de_{\pm, +} \Int{ \Ga }{} 
\f{  \wt{\ga} \, \dd s }{ \i \a }   \Ups(s)    \, 
+ \, \e{O}\Big(   \ex{\mp \f{2\pi \wt{\ga} }{ \a } z  } \Big)   
\quad when \qquad \Re(z) \tend \pm \infty   \;. 
\enq
\end{itemize}

\end{lemme}

\Proof 

Most of the statements are rather evident, the non-local jump condition 
on the boundary following from 
\beq
\op{C}_{ \Ga}[\Ups](x+\varkappa - \i\a ) - \op{C}_{ \Ga}[\Ups](x)  \; 
= \;  \Int{ \Ga }{} \f{  \wt{\ga} \, \dd s }{ \i \a }  \Ups(s) \cdot 
\Bigg\{  \f{ 1 }{ \ex{ \f{2\pi  }{ \a } [ \wt{\ga} (s-x) 
-  \wt{\ga}( \varkappa-\i\a)]  } \, - \, 1  }   \, - \,   
\f{  1 }{ \ex{ \f{2\pi \wt{\ga} }{ \a } (s-x) } \, - \, 1  }  
\Bigg\}  \; = \; 0\;,
\enq
owing to $\wt{\ga}( \varkappa-\i\a)=-\i\a$.

Furthermore, observe that $\op{C}_{ \Ga}[\Ups]$ has cuts on $\Cx$ along 
the curves $\Ga + \i \frac{\a}{ \wt{\ga} } \mathbb{Z}$, \textit{viz}. 
at the points 
\beq
z \, = \, y+  (-\i \a + \varkappa)t + \i n \a -n \varkappa 
\quad \e{with} \quad t \in \intff{0}{1} \quad, \quad n \in \mathbb{Z}\;, 
\enq
which form the line in $\Cx$ passing through $\Ga$.  
For $n\not=0$, none of these points  
 is contained in $\mc{S}_{\a}$, thus $\Ga$
is indeed the sole discontinuity curve for $\op{C}_{ \Ga}[\Ups]$ 
in the strip $\mc{S}_{\a}$. 

The fact that $z \mapsto \op{C}_{ \Ga;\pm}[\Ups](z)$ are holomorphic 
in a neighbourhood of $\Ga$ follows, for $z \in \e{Int}(\Ga)$, from 
a contour deformation in 
\eqref{definition transformee Cauchy cylindre cousu} made possible by  
the fact that $\Ups$ is analytic in the neighbourhood of $\Ga$. 
Holomorphicity in the vicinity of the endpoints of $\Ga$ needs an 
extra care. 

\begin{figure}[ht]
\begin{center}

\includegraphics[width=.5\textwidth]{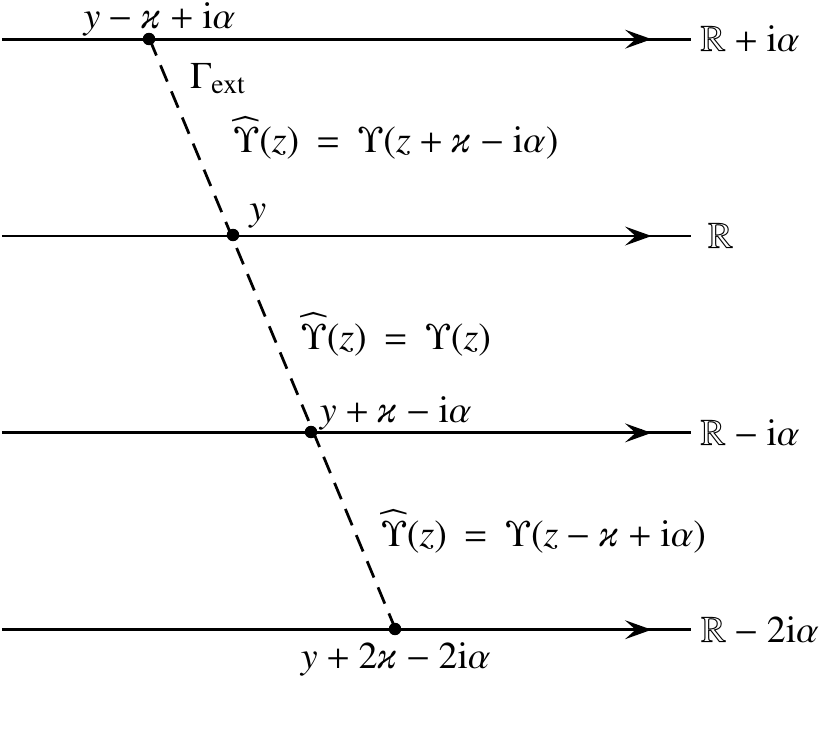}
\caption{ The extended strip with the extended curve  $\Ga_{\e{ext}}$ 
such that $\Ga \, = \, \Ga_{\e{ext}}\cap \mc{S}_{\a}$. 
\label{Figure Domaine prolonge pour Holomorphie transformee Cauchy}}
\end{center}
\end{figure}

In the domain depicted in Figure 
\ref{Figure Domaine prolonge pour Holomorphie transformee Cauchy}, 
one defines, in the neighbourhood of the curve $\Ga_{\e{ext}}$
an analytic function $\wh{\Ups}$. The fact that it is analytic 
follows from the jump conditions on $\Dp{}\mc{S}_{\a}$ satisfied by $\Ups$.

The curve $\Ga$ may be parameterised as 
\beq
y + (-\i\a +\varkappa)t \quad \e{with} \;  t \in \intff{0}{1}\,, 
\qquad \e{namely} \quad  y   -\i \f{ \a t }{\wt{\ga} }\, , 
\quad \e{with} \;  t \in \intff{0}{1} \; . 
\enq
Then, one has that 
\beqa
 \op{C}_{ \Ga }[\Ups](z) &  = & - \Int{ 0}{1} \dd t  
\f{  \Ups \big( y +  (-\i\a +\varkappa) t \big) }{ \ex{ 
\f{2\pi \wt{\ga} }{ \a } (y-z) } \ex{-2\i\pi t}  \, - \, 1  } \\ 
& = & - \Int{ 0 }{ \f{1}{2} } \dd t  \f{  \Ups( y +   
\big(-\i\a +\varkappa) t \big) }{ \ex{ \f{2\pi \wt{\ga} }{ \a } (y-z) } 
\ex{-2\i\pi t}  \, - \, 1  } 
 - \Int{ -\f{1}{2} }{ 0 } \dd t  \f{  \Ups \big( y +   
(-\i\a +\varkappa) t  + (-\i\a +\varkappa) \big) }
{ \ex{ \f{2\pi \wt{\ga} }{ \a } (y-z) } \ex{-2\i\pi t}  \, - \, 1  }   \\
& = &\Int{ \wh{\Ga} }{} \f{  \wt{\ga} \, \dd s }{ \i \a }  
\f{  \wh{\Ups}(s) }{ \ex{ \f{2\pi \wt{\ga} }{ \a } (s-z) } \, - \, 1  } 
\eeqa
in which $\wh{\Ga}=\intff{y \, + \, \tf{(\i\a -\varkappa)}{2} }
{ y \, - \, \tf{(\i\a -\varkappa)}{2} }$. The above representation  
produces manifestly holomorphic $\pm$ boundary values around $z=0$. 
A similar analysis allows one to conclude relatively to the point 
$y+ \varkappa - \i \a$. \qed

\subsection{Left  Riemann-Hilbert problem}
\label{SousSection RHP modele gauche}

The problem consists in finding $\chi^{(L)} \in \mc{O}
\big( \mc{S}_{\a} \big)$ such that $\chi^{(L)}$ admits smooth $-$, 
resp. $+$, boundary values on $\R$, resp. $\R-\i\a$, such that 
\begin{itemize}
 \item $\chi^{(L)}(z) \, = \, C_{\chi^{(L)}} \, 
+ \, \e{O}\Big( \ex{ \f{2\pi}{\a } z } \Big)$  when $\Re(z) 
\tend -\infty$ and up to the boundary;
 \item $\chi^{(L)}(z) \, = \, \ga \cdot z  \, + \, \e{O}
\Big(  \ex{ - \f{2\pi \wt{\ga} }{\a } z } \Big)$  when 
$\Re(z) \tend +\infty$ and up to the boundary;
 \item $\chi^{(L)}_+(g_L(x)-\i \a )\, 
= \, \chi^{(L)}_-(x)\, + \, x \, - \, g_L(x) $.
\end{itemize}
Above, $\ga$ and $\wt{\ga}$ are as introduced in 
\eqref{definition gamma et tilde gamma} while $g_L$ is as given 
by \eqref{definition gL et gR}.

\begin{prop}

The left non-local Riemann-Hilbert problem stated above admits 
a unique solution. 

\end{prop}

\Proof 

First, introduce the holomorphic function on an open neighbourhood 
$\mc{S}_{\a}$
\beq
\om^{(L)}(z) \; = \; \f{ \ga z  }{  \ex{ -\f{2\pi}{\tau}z } + 1 }\;, 
\enq
where $ \tau> 2 \a$. The function $\om^{(L)}(z)$ may be decomposed as 
\beq
\om^{(L)}(z) \; = \; \ga z \; + \; \om^{(L)}_{R}(z) \qquad \e{with} 
\qquad   \om^{(L)}_{R}(z) \, = \,- \f{ \ga z  }{  \ex{ \f{2\pi}{\tau}z } 
+ 1 } \;. 
\label{ecriture decomposition OmL avec gamma z et OmegaLR}
\enq
As a consequence, the  
following estimates hold 
\beq
\om^{(L)}(z) \; = \; \left\{  \ba{cc}   \e{O}\Big( z  
\ex{  \f{2\pi}{\tau}z }  \Big)   & \Re(z) \tend - \infty   
\vspace{3mm}  \\  
  \ga z \, + \, \e{O}\Big( z  \ex{  -\f{2\pi}{\tau}z }  \Big)   
& \Re(z) \tend  +\infty   \ea \right.  \;. 
\enq
 The decomposition \eqref{ecriture decomposition OmL avec gamma 
z et OmegaLR} entails that $\om^{(L)}$ satisfies 
\beq
\om^{(L)}(x+\varkappa -\i\a) \, - \,  \om^{(L)}(x) \; 
= \; - \varkappa  \, + \, \om^{(L)}_R(x+\varkappa -\i\a) \, 
-\, \om^{(L)}_R(x)\qquad \text{for}\quad x \in \R \;. 
\enq

\begin{figure}[ht]
\begin{center}

\includegraphics[width=.5\textwidth]{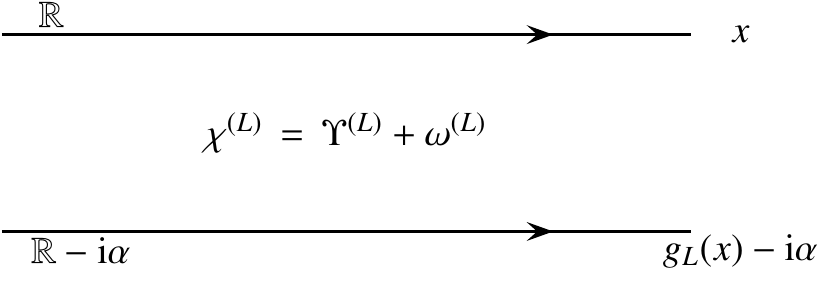}
\caption{ The substitution for the Riemann-Hilbert problem 
for $\chi^{(L)}$. \label{Figure Substitution RHP chiL} }
\end{center}
\end{figure}

Then, one makes the substitution in the Riemann-Hilbert problem 
for $\chi^{(L)}$ as described in Figure 
\ref{Figure Substitution RHP chiL}.

What results is the following  Riemann-Hilbert problem for
$\Ups^{(L)}$: 
\begin{itemize}
 \item $\Ups^{(L)}\in \mc{O}(\mc{S}_{\a})$ ; 
 
  \item $\Ups^{(L)}(z) \, = \,   \, - \,  \om^{(L)}_{R} (z) \, 
+ \, \e{O}\Big(\ex{  -  \f{2\pi \wt{\ga} }{\a } z } \Big)$   
\,when $\Re(z) \tend +\infty$ and up to the boundary;
 \item $\Ups^{(L)}(z) \, = \,C_{\chi^{(L)}}  - \om^{(L)}(z)\, 
+ \,   \e{O}\Big( \ex{   \f{2\pi  }{\a } z } \Big)$  \,when 
$\Re(z) \tend -\infty$ and up to the boundary;

 \item $\Ups^{(L)}$ admits  smooth $-$, resp. $+$, boundary values 
on $\R$, resp. $\R-\i\a$;

 \item $\Ups^{(L)}_+(g_L(x)-\i \a )\, = \, \Ups^{(L)}_-(x)\, 
+ \, G_{\Ups^{(L)}}(x)$, where the jump function takes the form 
\beq
G_{ \Ups^{(L)} }\big( x \big) \, = \, \left\{ \ba{cc}  \om^{(L)}(x) \, 
- \,  \om^{(L)}(x-\i \a)   & x <-M_L \vspace{2mm} \\   
						  x\, - \, g_L(x)   \, 
+ \, \om^{(L)}(x) \, - \,  \om^{(L)}(g_L(x)-\i \a)    & -M_L\,  
\leq \,  x  \, \leq \,  M_L  \vspace{2mm}  \\ 
						  \om^{(L)}_R(x) \, 
- \,  \om^{(L)}_R(x+\varkappa-\i \a)   & M_L < x

						 \ea \right. \;. 
\enq
\end{itemize}
It follows from the stated properties of $\om^{(L)}$ that 
$G_{ \Ups^{(L)}}$ 
has the form \eqref{GXix} for $g=g_L$, $M=M_L$ and
\beq
\mc{G}_{\Ups^{(L)}}(z)=\om^{(L)}_R(z)+\varphi_L(\Re(z))\,\gamma z\,,
\enq
where $\varphi_L(x)$ is a smooth interpolating function equal 
to $1$ for $x<-M_L/2$
and to $0$ for $x>M_L/2$ so that $\mc{G}_{\Ups^{(L)}}$ is analytic on 
$\mc{S}_{\a}$ for $|\Re(z)|>M_L/2$ with exponential falloff  at infinity.
By virtue of Proposition \ref{Proposition solubilite unique RHP 
non local} the non-local Riemann-Hilbert problem for $\Ups^{(L)}$ 
is uniquely solvable, and hence, so is the one for $\chi^{(L)}$.

\vspace{4mm}

We point out that $\chi^{(L)}$ readily allows one to build the solution 
to the non-local Riemann-Hilbert problem associated with a shifted 
function. Namely, for any $w \in \R $, let 
\beq
G_L(x) \, = \, g_L(x+w)-w \, = \,    
\left\{ \ba{cc}  x & x < -M_L-w  \vspace{2mm} \\ 
				  g_L(x+w) -w  \quad & -M_L-w< x 
< M_L-w  \vspace{2mm}  \\ 
				 x +\varkappa & M_L-w < x	  
\ea \right. \; . 
\enq
Then $\Xi^{(L)}(z) \; = \; \chi^{(L)}(z+w)$ solves the 
Riemann-Hilbert problem: 
\begin{itemize}
\item $\Xi^{(L)}  \in \mc{O}\big( \mc{S}_{\a} \big)$ and admits 
smooth $-$, resp. $+$, boundary values on $\R$, resp. $\R-\i\a$;
\item $\Xi^{(L)}(z) \, = \, C_{\Xi^{(L)}}  \, + \, \e{O}
\Big( \ex{ \f{2\pi}{\a } (z+w) } \Big)$  \,when $\Re(z) \tend 
-\infty$ and up to the boundary;
 \item $\Xi^{(L)}(z) \, = \, \ga \cdot (z+w)  \, + \, \e{O}
\Big(  \ex{ - \f{2\pi \wt{\ga} }{\a } (z+w) } \Big)$  \,when 
$\Re(z) \tend +\infty$ and up to the boundary;
 \item $\Xi^{(L)}_+(G_L(x)-\i \a )\, = \, \Xi^{(L)}_-(x)\, 
+ \, x \, - \, G_L(x) $. 
\end{itemize}

We stress that the remainders at $\Re(z)\tend  \pm \infty$ appearing 
above are uniform in $w$.

\subsection{Right Riemann-Hilbert problem}
\label{SousSection RHP modele droit}

The right Riemann-Hilbert problem consists in finding  $\chi^{(R)} 
\in \mc{O}\big( \mc{S}_{\a} \big)$ such that 
\begin{itemize}
\item $\chi^{(R)}$ admits smooth $-$, resp. $+$, boundary values 
on $\R$, resp. $\R-\i\a$;

 \item $\chi^{(R)}(z) \, = \, \e{O}\Big( \ex{ - \f{2\pi}{\a } z } 
\Big)$  \,when $\Re(z) \tend +\infty$ and up to the boundary;
 \item $\chi^{(R)}(z) \, = \, \ga \cdot z \, + \, C_{\chi^{(R)}}  \, 
+ \, \e{O}\Big(  \ex{   \f{2\pi \wt{\ga} }{\a } z } \Big)$  
\,when $\Re(z) \tend -\infty$ and up to the boundary;
 \item $\chi^{(R)}_+(g_R(x)-\i \a )\, = \, \chi^{(R)}_-(x)\, 
+ \, x \, - \, g_R(x) $.
\end{itemize}

We remind that $\ga$ and $\wt{\ga}$ have been introduced in 
\eqref{definition gamma et tilde gamma} while $g_R$ is given by 
\eqref{definition gL et gR}. 
\begin{prop}

The Riemann-Hilbert problem for $\chi^{(R)}$ admits a unique solution. 
 
\end{prop}

\Proof  
Let 
\beq
\om^{(R)}(z) \; = \;  \f{ \ga  z  }{1+\ex{ \f{2\pi}{\tau} z }   }    \;, 
\enq
where $ \tau > 2\a$. The function $\om^{(R)}(z)$ may be decomposed as 
\beq
\om^{(R)}(z) \; = \; \ga z \; + \; \om^{(R)}_{L}(z) \qquad \e{with} 
\qquad   \om^{(R)}_{L}(z) \, = \,- \f{ \ga z  }{  \ex{ -\f{2\pi}{\tau}z } 
+ 1 } \;. 
\label{ecriture decomposition OmR avec gamma z et OmegaRL}
\enq
As a consequence,  
one has
\beq
\om^{(R)}(z) \; = \; \left\{  \ba{cc}   \ga z \, 
+ \,  \e{O}\Big( z  \ex{  \f{2\pi}{\tau}z }  \Big)   
& \Re(z) \tend - \infty   \vspace{3mm}  \\  
  \e{O}\Big( z  \ex{  -\f{2\pi}{\tau}z }  \Big)   
& \Re(z) \tend  +\infty   \ea \right.  \;. 
\enq
 The decomposition 
\eqref{ecriture decomposition OmL avec gamma z et OmegaLR} entails 
that $\om^{(R)}$ satisfies 
\beq
\om^{(R)}(x+\varkappa -\i\a) \, -  \, \om^{(R)}(x) \; = \; 
- \varkappa  \, + \, \om^{(R)}_L(x+\varkappa -\i\a) \, - \,  
\om^{(R)}_L(x) \;. 
\enq
\begin{figure}[ht]
\begin{center}

\includegraphics[width=.5\textwidth]{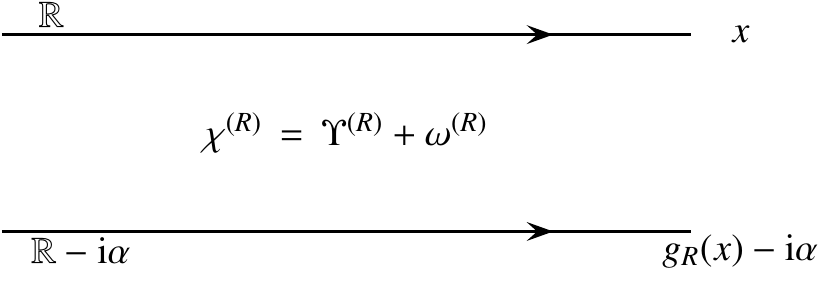}
\caption{ The substitution for the Riemann-Hilbert problem 
for $\chi^{(R)}$. \label{Figure Substitution RHP chiR} }
\end{center}
\end{figure}
Upon implementing the substitution in the Riemann-Hilbert problem 
for $\chi^{(R)}$ as described in Figure 
\ref{Figure Substitution RHP chiR},  one gets that 
$\Ups^{(R)}\in \mc{O}(\mc{S}_{\a})$ solves the Riemann-Hilbert problem 
\begin{itemize}

   \item $\Ups^{(R)}(z) \, = \, - \om^{(R)}(z)  \, 
+ \, \e{O}\Big(\ex{   - \f{   2\pi  }{\a } z } \Big)$   
\,when $\Re(z) \tend +\infty$ and up to the boundary;
 \item $\Ups^{(R)}(z) \, = \, C_{\chi^{(R)}}  \, 
- \om^{(R)}_L(z)\, + \,   \e{O}\Big( \ex{   
\f{2\pi \wt{\ga}  }{\a } z } \Big)$  \,when $\Re(z) 
\tend -\infty$ and up to the boundary;

 \item $\Ups^{(R)}$ admits smooth $-$, resp. $+$, boundary 
values on $\R$, resp. $\R-\i\a$; 

 \item $\Ups^{(R)}_+(g_R(x)-\i \a )\, = \, \Ups^{(R)}_-(x)\, 
+ \, G_{\Ups^{(R)}}(x)$, where the jump function takes the form 
\beq
G_{ \Ups^{(R)} }\big( x \big) \, = \, \left\{ \ba{cc}  
\om^{(R)}_L(x) \, - \,  \om^{(R)}_L(x + \varkappa - \i \a)  
& x <-M_R \vspace{2mm} \\   
						  x\, - \, g_R(x)   
\, + \, \om^{(R)}(x) \, - \,  \om^{(R)}(g_R(x)-\i \a)    & -M_R\,  
\leq \,  x  \, \leq \,  M_R  \vspace{2mm}  \\ 
						 \om^{(R)}(x) \, 
- \,  \om^{(R)}(x-\i \a)   & M_R < x

						 \ea \right.\;. 
\enq
\end{itemize}
It follows from the stated properties of $\om^{(R)}$ that  
$G_{\Ups^{(R)}}$ has the form (2.2) for $g = g_R$, $M=M_R$ and
\beq
\mc{G}_{\Ups^{(R)}}(z)=\om^{(R)}_L(z)+\varphi_R(\Re(z))\,\gamma z\,,
\enq
where $\varphi_R(x)$ is a smooth interpolating function equal 
to $0$ for $x<-M_R/2$
and to $1$ for $x>M_R/2$ so that $\mc{G}_{\Ups^{(R)}}$ is analytic on 
$\mc{S}_{\a}$ for $|\Re(z)|>M_R/2$ with exponential falloff  at infinity.
Again, these properties entail by virtue of Proposition 
\ref{Proposition solubilite unique RHP non local} that the non-local 
Riemann-Hilbert problem for $\Ups^{(R)}$ is uniquely solvable,
and hence, so is the one for $\chi^{(R)}$.  \qed

\vspace{4mm}

The solution $\chi^{(R)}$ gives rise to the solution of the 
non-local Riemann-Hilbert problem associated with a shifted function. 
Namely, for any $w \in \R $, let 
\beq
G_R(x) \, = \, g_R(x-w)+w \, = \,    \left\{ \ba{cc}  x+\varkappa  
& x < -M_R+w  \vspace{2mm} \\ 
				  g_R(x-w) +w  \quad & -M_R+w< x 
< M_R+w  \vspace{2mm}  \\ 
				 x  & M_R+w < x	  \ea \right.
\enq
Then $\Xi^{(R)}(z) \; = \; \chi^{(R)}(z-w)$ solves the Riemann-Hilbert 
problem: 
\begin{itemize}
\item $\Xi^{(R)}  \in \mc{O}\big( \mc{S}_{\a} \big)$ and admits 
$L^{2}(\R)$ $-$, resp. $+$, boundary values on $\R$, resp. $\R-\i\a$,

\item $\Xi^{(R)}(z) \, = \, \e{O}\Big(\ex{   - \f{   2\pi  }{\a } 
(z-w) } \Big)$   \,when $\Re(z) \tend +\infty$ and up to the boundary;
 \item $\Xi^{(R)}(z) \, = \, \ga \cdot (z-w) \, + \, C_{\Xi^{(R)}}  \, 
+ \,\e{O}\Big(\ex{    \f{   2\pi \wt{\ga} }{\a } (z-w) } \Big)$   
\,when $\Re(z) \tend -\infty$ and up to the boundary;
 \item $\Xi^{(R)}_+(G_R(x)-\i \a )\, = \, \Xi^{(R)}_-(x)\, + \, x \, 
- \, G_R(x) $.
\end{itemize}

\section{Asymptotic behaviour of the global non-local 
Riemann-Hilbert problem}
\label{Section RHP non local glocal}

From now on we fix two positive reals $M_L, M_R >0$ and consider 
the function $g$ defined as 
\beq
g(x)  \, = \,    \left\{ \ba{cc}  x    &    x < -M_L - w  
\vspace{2mm} \\ 
				g_L(x+w)-w  \qquad & - M_L - w 
\leq x \leq  M_L  - w  \vspace{2mm} \\ 
				x+\varkappa  & M_L - w \leq x 
\leq  w  - M_R     \vspace{2mm} \\  
				  g_R(x-w) +w  \qquad & w - M_R < x 
< w + M_R   \vspace{2mm}  \\ 
				 x  & M_R+w < x	  \ea \right.
\label{definition fonction g globale}
\enq
where $g_L$ and $g_R$ correspond to the functions introduced in 
\eqref{definition gL et gR}.

Below, we shall establish the main theorem of this work,
Theorem \ref{Theoreme principal}.
\vskip 0.2cm

\noindent{\it Proof of Theorem \ref{Theoreme principal}}.
\vskip 0.1cm

The substitution $\chi(z)=\wt{\chi}(z)+z$  
turns the non-local Riemann-Hilbert problem with shift for $\chi$ into 
the one of finding $\wt{\chi} \in \mc{O}(\mc{S}_{\a})$ such that 
\begin{itemize}
 \item $\wt{\chi}$ has smooth $-$, resp. $+$, boundary values on $\R$, 
resp. $\R-\i\a$;
\item $\wt{\chi}(z)=C_{\chi} \de_{\pm, -} \, + \, \e{O}\Big( 
\ex{ \mp \f{2\pi }{\a } z } \Big)$ for $\Re(z) \tend \pm \infty$, 
this up to the boundary and for some constant $C_{\chi}$;
 \item $\wt{\chi}_{+}\big( g(x)-\i \a \big) \, = \, \wt{\chi}_{-}
\big( x \big)\, + \, x-g(x)$, with $x \in \R$. 
\end{itemize}
The unique solvability of the Riemann-Hilbert problem for $\wt{\chi}$ 
is a direct consequence of Proposition  
\ref{Proposition solubilite unique RHP non local}. 
This thus establishes the unique solvability of the non-local 
Riemann-Hilbert problem with shift for $\chi$. 
However, the approach that was adopted for establishing 
Proposition  \ref{Proposition solubilite unique RHP non local} does not 
allow one for a uniform in $w$ control on the solution. To achieve it, 
first introduce 
\beq
\Psi(z) \, = \, \op{C}_{\Ga_0}\big[  \de \Xi \big] (z)
\enq
in which $\Ga_0=\intff{0}{-\i \a + \varkappa}$ while $\de \Xi 
= \Xi^{(L)}- \Xi^{(R)}$ where $\Xi^{(L/R)}$ are the unique solutions 
to the left/right shifted Riemann--Hilbert problems
that were discussed in Subsections 
\ref{SousSection RHP modele gauche}-\ref{SousSection RHP modele droit}. 
Note that, owing to the large $z$ asymptotics of the solutions 
$\chi^{(L/R)}$, it holds
\beq
\de \Xi(s) \, = \, \ga(s+w) + \e{O}\bigg(  \ex{   -\f{2\pi 
\wt{\ga}  }{\a } (s+w) } \bigg) \, - \, \ga(s-w) \, - \,  
C_{ \Xi^{(R)} } \, - \, \e{O}\bigg(  \ex{    
\f{2\pi \wt{\ga}  }{\a } (s-w) }  \bigg) \; = \; 
2 \ga w \, - \,  C_{\Xi^{(R)}} \, + \, \e{O}\bigg(  
\ex{  - \f{2\pi \wt{\ga}  }{\a } w }  \bigg)
\label{ecriture DA de delta Xi sur ctr Gamma0}
\enq
uniformly in $s \in \Ga_0$ and where $\wt{\ga}$ is as defined in 
\eqref{definition gamma et tilde gamma}. Furthermore, the jump condition 
$\de\Xi_+(x+\varkappa-\i \a)=\de \Xi_-(x)$, valid in some fixed, 
$w$-independent,  neighbourhood of $x=0$, ensures that one can invoke 
Lemma \ref{Lemme Transformee de Cauchy sur Cylinder cousu}
so as to conclude that $\Psi$ satisfies 

\begin{itemize}
 \item $\Psi \in \mc{O}(\mc{S}_{\a} \setminus \Ga)$ and has 
holomorphic $-$, resp. $+$, boundary values on $\Ga_0$; 
 \item $\Psi_+(x)-\Psi_-(x) \, = \, \de \Xi(x) \quad$ for $x \in \Ga_0$; 
\item $\Psi(x+\varkappa - \i\a ) \, = \,  \Psi(x)$  for $x \in \R$;

 \item $\Psi(z) \, = \, \de_{\pm, +}c(w)  \, + \, \e{O}\Big(   
w  \ex{\mp \f{2\pi \wt{\ga} }{ \a } z  } \Big)  $ \,when $  
\Re(z) \tend \pm \infty$ this up to the boundary  $\Dp{}\mc{S}_{\a}$, 
and  uniformly in $w$ with 
\beq
c(w) \, = \, - \Int{ \Ga_0}{} \f{  \wt{\ga} \, \dd s }{ \i \a }    
\de \Xi(s)  \; = \; 2 \ga   w \, - \,  C_{\Xi^{(R)}}  \; + \; 
\e{O}\bigg(  \ex{  - \f{2\pi \wt{\ga}  }{\a } w }  \bigg) \;. 
\enq
\end{itemize}

\begin{figure}[ht]
\begin{center}

\includegraphics[width=.5\textwidth]{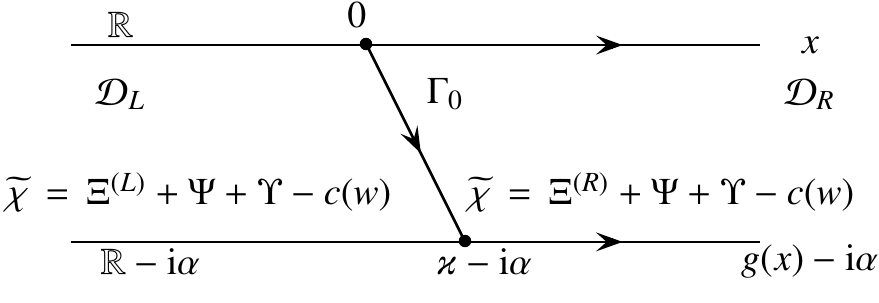}
\caption{ The substitution for the Riemann-Hilbert problem for $\chi$.} 
\label{Figure Substitution RHP chi} 
\end{center}
\end{figure}

Finally, the asymptotic expansion 
\eqref{ecriture DA de delta Xi sur ctr Gamma0} for $\de\Xi$ which holds 
uniformly on $\Ga_0$ allows one to get uniform in $w$ estimates for 
$\Psi$. Indeed, this entails that 
\beq
\Psi(z) \, = \, \Big(2 \ga   w \, - \,  C_{\Xi^{(R)}} \Big)\, 
\bs{1}_{\mc{D}_R}(z) \, +  \,  \op{C}_{\Ga_0}\big[  \de \Xi -2 \ga   w \, 
+ \,  C_{\Xi^{(R)}}  \big] (z)
\enq
where $\mc{D}_R$ is the domain depicted in Fig. 
\ref{Figure Substitution RHP chi} and we used that $ \op{C}_{\Ga_0}
\big[  1  \big] (z)=\bs{1}_{\mc{D}_R}(z)$. 
The second term may be estimated as 
\beq
\op{C}_{\Ga_0}\big[  \de \Xi -2 \ga   w \, + \,  C_{\Xi^{(R)}}  
\big] (z) \; = \; \left\{ \ba{cc}   \e{O}\Big( \ex{  - \f{2\pi 
\wt{\ga}  }{\a } w }   \big(1  +  \ex{  - \f{2\pi \wt{\ga}  }{\a } z } 
\big)  \Big)    &  z \in \mc{D}_R   \vspace{1mm} \\
 \e{O}\Big( \ex{  - \f{2\pi \wt{\ga}  }{\a } (w-z) }   \Big)  & z 
\in \mc{D}_L  \ea  \right. \;. 
\enq
These estimates are uniform up to the boundary of $\mc{S}_{\a}$ and 
up to $\Ga_0$, as follows from the local holomorphicity of $\de \Xi$ 
around $\Ga_0$ and the fact that it satisfies in this neighbourhood 
$\de \Xi_{+}(s+\varkappa- \i\a)\,=\,\de \Xi_{-}(s)$.

Then, one makes the substitution in the Riemann--Hilbert problem 
for $\wt{\chi}$ as described in Figure \ref{Figure Substitution RHP chi}.
One gets that, by construction, $\Ups$ is continuous across $\Ga_0$. 
It thus solves the non-local Riemann-Hilbert problem 
\begin{itemize}
 \item $\Ups \in \mc{O}(\mc{S}_{\a})$ having $L^{2}(\R)$ $-$, resp. 
$+$, boundary values on $\R$, resp. $\R-\i\a$;
\item $\Ups(z)\, =\, C_{\Ups} \de_{\pm;-} \,  +\, \e{O}\Big( 
\ex{ -\eta|\Re(z)| } \Big)$ \,for $\Re(z) \tend \pm \infty$, 
this up to the boundary  for some $C_{\Ups}$ and $\eta>0$; 
 \item $\Ups_{+}\big( g(x)-\i \a \big) \, = \, \Ups_{-}\big( x \big)\, 
+ \, G_{\Ups}(x)$, with $x \in \R$,  
\end{itemize}
where the jump function reads $G_{\Ups}(x)=\Psi(x)-\Psi(g(x)-\i\a)$. 
Observe further that for $x\in \intff{M_L-w}{-M_R+w}$ it holds 
$g(x)=x+\varkappa$ and thus $G_{\Ups}(x)=0$ by virtue of the periodicity 
of $\Psi$. Furthermore, the estimates for $\Psi$ at infinity entail 
that  for $x \in \R \setminus  \intff{M_L-w}{-M_R+w}$
\beq
 G_{\Ups}(x) \, = \, \e{O}\Big(   w  \ex{\mp \f{2\pi \wt{\ga} }{ \a } x  } 
\Big)  \quad  \pm x >0
\enq
which is uniformly exponentially small in $w$ and has an exponential 
fall-off in $x$ at infinity.  
This will allow to control the behaviour of the function $\Ups$ both in $z$ 
and in $w$. The argument goes as follows.

By Proposition  
\ref{Proposition solubilite unique RHP non local} and Lemma 
\ref{Lemme eqn integrale type sinh pour theta}, the function 
$Y_1(x)=\Ups_{+}(g(x)-\i\a)$ corresponds to the unique solution to 
\beq
\Big( \e{id} - \op{K}_{\e{tot}}\Big)[Y_1] \, = \, \f{1}{2} 
\Big\{ G_{\Ups}+\mc{H}[G_{\Ups}] \Big\} \; - \; 
\op{K}_{\e{tot};12}\big[G_{\Ups}\big]
\enq
where $\op{K}_{\e{tot}}$, $\op{K}_{\e{tot};12}$ and $\mc{H}$ are 
the integral operators introduced in Lemma 
\ref{Lemme eqn integrale type sinh pour theta}
whose integral kernels are expressed in terms of the function $g$ 
given in \eqref{definition fonction g globale}.
Due to the properties of $G_{\Ups}$, the results
that will be 
established in Section 
\ref{Section Invertibility of the operator id moins K}, 
in particular Theorem \ref{Theorem bornage solution eqn integrale}, 
ensure that
\beq
Y_1(x) \; = \;  C_{Y_1}\de_{\pm,-} \; + \; \e{O}\Big( 
\ex{-w \eta^{\prime} - \eta |x| }  \Big) \qquad \e{as} 
\qquad x \tend \pm \infty \,
\enq
uniformly in $w$ and for some $\eta, \eta^{\prime}>0$ and with 
$ C_{Y_1} =  \e{O}\Big( \ex{-w \eta^{\prime}  }  \Big)$. Moreover, 
it holds $\norm{ Y_1 }_{L^{\infty}(\R) } \;  =\,   \e{O}
\Big( \ex{-w \eta^{\prime}  }  \Big) $. 
Thus, since 
\beq
\Ups(z) \; = \;  -\Int{\R}{} \f{ \dd y }{ 2 \i \tau } 
\f{ Y_1(y)-G_{\Ups}(y)   }{  \sinh\Big[ \tfrac{\pi}{\tau}(y-z) 
\Big]  } \; + \; 
\Int{\R}{} \f{ \dd y }{ 2 \i \tau } \f{ Y_1(y) g^{\prime}(y)  }
{  \sinh\Big[ \tfrac{\pi}{\tau}(g(y)-z- \i \a )\Big] } \;, 
\enq
one gets that 
\beq
\Ups(z) \; = \; C_{Y_1}\de_{\pm,-} \; + \; \e{O}\Big( 
\ex{-w \eta^{\prime} - \eta |z| }  \Big) \qquad \e{as} 
\qquad \Re(z) \tend \pm \infty  \;, 
\enq
uniformly throughout $\mc{S}_{\a}$ and in $w$.  
The comparison of the decompositions of Figures 
\ref{Figure for grd w pour solution chi}  and 
\ref{Figure Substitution RHP chi} permits then to end the proof
of Theorem \ref{Theoreme principal}. \qed

\section{Invertibility of an auxiliary integral operator}
\label{Section Invertibility of the operator id moins K}

Let $\op{K}_{\e{tot}}$ be the integral operator, introduced 
in \eqref{definition operateur K sinh}, and associated with 
the function $g$ given in \eqref{definition fonction g globale}. 
It follows from the analysis in Subsection \ref{Sous section 
solvabilite unique du RHP global} that the operator 
$\e{id}-\op{K}_{\e{tot}}$ is invertible on an appropriate 
functional space. 

The goal of this section is to establish, uniformly in 
$w\tend +\infty$, bounds on the inverse of $\e{id}-\op{K}_{\e{tot}}$.  
This will be done by relying on the various results established 
in the previous sections.

\subsection{Decomposition of $\op{K}_{\e{tot}}$}
\label{SousSection decomposition de K tot}

To start with, it is convenient to introduce three intervals
\beq
I_w^{-} \, = \, \intof{-\infty}{-w} \quad , \qquad I_w^{0} \, 
= \, \intoo{-w}{w} \quad , \qquad I_w^{+} \, = \, \intfo{w}{+\infty} \;. 
\enq
Next, introduce three operators on $L^2(\R)$, $\op{L}^{++}_{w}$, 
$\op{L}^{0}_{w}$, $\op{L}^{--}_{w}$ with integral kernels
\beqa
L^{\pm\pm}_{w}(x,y)  & = &  L^{(\e{e})}(x-y) \cdot 
\bs{1}_{I_w^{\pm} \times I_w^{\pm} }(x,y)  
\label{Lpmpmw}\\ 
L^{0}_{w}(x,y)  & = &  L^{(0)}(x-y) \cdot 
\bs{1}_{I_w^{0} \times I_w^{0} }(x,y) \;, 
\label{L0w} 
\eeqa
where 
\beqa
L^{(\e{e})}(x-y)  & = &    \f{1 }{2 \i \tau } \Bigg\{   
\f{  1  }{ \sinh\Big[ \f{\pi}{\tau} (y   - x  - \i\a  ) \Big]  } \, 
- \,  \f{  1  }{ \sinh\Big[ \f{\pi}{\tau} (y   - x  + \i\a  ) \Big]  }   
\Bigg\} \;, \\
L^{(0)}(x-y)  & = &    \f{1 }{2 \i \tau } \Bigg\{   \f{  1  }
{ \sinh\Big[ \f{\pi}{\tau} (y  +\varkappa - x  - \i\a  ) \Big]  } \, 
- \,  \f{  1  }{ \sinh\Big[ \f{\pi}{\tau} (y  - \varkappa  
- x  + \i\a  ) \Big]  }   \Bigg\} \; . 
\label{definition noyau integral op WH tronque interval central}
\eeqa

From now on we agree to denote by $\op{K}_{R}$ and $\op{B}_{R}$, 
resp. $\op{K}_{L}$ and $\op{B}_{L}$, the operators $\op{K}$ as 
introduced in \eqref{definition operateur K sinh} and $\op{B}$ 
as introduced in \eqref{definition operateur B} which are subordinate 
to the function $g_{R}$, resp. $g_{L}$, introduced in 
\eqref{definition gL et gR}. 

Define the operator $\op{B}_{\e{tot}}$ on $L^{2}(\R)$ 
\bem
 B_{\e{tot}}(x,y)\, = \, \bs{1}_{\R^+}(x)\bs{1}_{\intoo{-w}{+\infty}}(y) 
\cdot B_{R}(x-w,y-w) \, + \, \bs{1}_{\R^-}(x)\bs{1}_{\intoo{-\infty}{w}}(y) 
\cdot B_{L}(x+w,y+w)  \\
 \, + \, \bs{1}_{ \R^+ }(x) G_{R}(x,y) \, 
+ \,  \bs{1}_{ \R^- }(x) G_{ L }(x,y) \;. 
\end{multline}
Upon using the function $m_{\a}$ introduced in 
\eqref{definition fonction frak d de base}, the functions $G_{R/L}$ 
are expressed as
\bem
 G_{ L }(x,y) \, = \, \bigg\{ \mf{m}_{\a}\Big( g_R(y-w)+w-x \Big) 
g_R^{\prime}(y-w)  \, + \,  \mf{m}_{0}\Big( g_R(y-w)+2w - g_L(x+w) 
\Big) g_R^{\prime}(y-w) \, - \, \mf{m}_{0}(y-x) \\ 
 \, +\,  K_{L;12}(x+w, y+w) \bigg\}\bs{1}_{ \intoo{w-M_R}{+\infty} }(y)  
\, - \,  K_L(x+w,y+w)\bs{1}_{\intoo{w-M_R}{w}}(y) 
\label{definition GL}
\end{multline}
and 
\bem
 G_{ R }(x,y) \, = \, \bigg\{ \mf{m}_{\a}\Big( g_L(y+w)-w-x 
\Big) g_L^{\prime}(y+w)  \, + \,  \mf{m}_{0}\Big( g_L(y+w)-2w 
- g_R(x-w) \Big) g_L^{\prime}(y+w) \, - \, \mf{m}_{0}(y-x) \\
\, +\,   K_{R;12}(x-w, y-w)  \bigg\} \bs{1}_{ \intoo{-\infty}{-w+M_L} }(y) 
\, - \,  K_R(x-w,y-w)\bs{1}_{\intoo{-w}{-w+M_L}}(y) \;. 
\label{definition GR}
\end{multline}
It is clear that the functions $G_{L/R}$ satisfy  
the bounds 
\beq
 G_{ L }(x,y) \; = \;  \e{O}\Big( \ex{- \f{\pi}{\tau} |x-y| } \Big) 
\cdot \bs{1}_{ \intoo{w-M_R }{+\infty} }(y)  \qquad \e{and} \qquad 
 G_{ R }(x,y) \; = \;  \e{O}\Big( \ex{- \f{\pi}{\tau} |x-y| } \Big) 
\cdot \bs{1}_{ \intoo{-\infty}{M_L-w } }(y)  \;. 
\label{ecriture bornes decroissance fcts G L et R}
\enq
Moreover, the remainders appearing above also hold for the derivatives, 
namely,  for any $(k,\ell)\in \mathbb{N}^2$, one has 
\beq
 \Dp{x}^{\ell}\Dp{y}^{k}G_{ L }(x,y) \; = \;  \e{O}\Big( 
\ex{- \f{\pi}{\tau} |x-y| } \Big) \cdot 
\bs{1}_{ \intoo{w-M_R }{+\infty} }(y)  \qquad \e{and} \qquad 
 \Dp{x}^{\ell}\Dp{y}^{k}G_{ R }(x,y) \; = \;  \e{O}\Big( 
\ex{- \f{\pi}{\tau} |x-y| } \Big) \cdot 
\bs{1}_{ \intoo{-\infty}{M_L-w } }(y)  \;. 
\label{ecriture bornes decroissance fcts G L et R derivatives}
\enq

\begin{lemme}
 \label{Lemme decomposition noyau integral K}
One has
\beq
\op{K}_{\e{tot}}\, = \, \op{L}^{++}_{w} \,  +\,   \op{L}^{0}_{w} \, 
+ \, \op{L}^{--}_{w}\;+ \; \op{B}_{\e{tot}}
\enq

\end{lemme}

\Proof

Recall that $g(x)=g_L(x+w)-w$ whenever $-\infty \,  < \, x < w-M_{R}$.
Hence, for  $x \leq w - M_R$,  
\beq
K_{\e{tot};12}(x,y) \, = \, K_{L;12}(x+w,y+w) \qquad \e{for} \, 
\e{any} \quad y \in \R\,. 
\enq
Also,  
in the same range of $x$'s,  
\beq
K_{\e{tot};21}(x,y) \, = \, \left\{ \ba{ccc}   K_{L;21}(x+w,y+w)    
&  \e{if}  &  y \leq w-M_R \vspace{2mm}\\ 
 \mf{m}_{\a}\Big( g_R(y-w)+w-x \Big) g_R^{\prime}(y-w)    
&  \e{if}  &  y \geq w-M_R   \ea \right. 
\enq
and 
\beq
K_{\e{tot};11}(x,y) \, = \, \left\{ \ba{ccc}   K_{L;11}(x+w,y+w)    
&  \e{if}  &  y \leq w-M_R \vspace{2mm} \\ 
\mf{m}_{0}\Big( g_R(y-w)+2w - g_L(x+w) \Big) g_R^{\prime}(y-w) \, 
- \, \mf{m}_{0}(y-x)   &  \e{if}  &  y \geq w-M_R   \ea \right. \;.   
\enq
Thus, adding up the three pieces, one gets that for $x \leq w - M_R$, 
\beq
K_{\e{tot}}(x,y) \; = \; K_L(x+w,y+w)\bs{1}_{\intoo{-\infty}{w}}(y) \; + \;  G_{L}(x,y) 
\enq
with $G_L$ as defined in \eqref{definition GL}.

Analogously, $g(x)=g_R(x-w)+w$ for  $-w + M_L  \,  \leq  \, x $.
Hence, 
in this range of $x$'s, 
\beq
K_{\e{tot};12}(x,y) \, = \, K_{R;12}(x-w,y-w) \qquad \e{for} \, 
\e{any} \quad y \in \R\,. 
\enq
Also,  
in the same range of $x$'s,  
one has
\beq
K_{\e{tot};21}(x,y) \, = \, \left\{ \ba{ccc}  \mf{m}_{\a}
\Big( g_L(y+w)-w-x \Big) g_L^{\prime}(y+w)    &  \e{if}  &  
y \leq -w+M_L   \vspace{2mm}   
\\
K_{R;21}(x-w,y-w)    &  \e{if}  &  y \geq -w+M_L \ea \right.  
\enq
and 
\beq
K_{\e{tot};11}(x,y) \, = \, \left\{ \ba{ccc}    
\mf{m}_{0}\Big( g_L(y+w)-2w - g_R(x-w) \Big) g_L^{\prime}(y+w) \, 
- \, \mf{m}_{0}(y-x)   &  \e{if}  &  y \leq -w+M_L  \vspace{2mm} \\ 
K_{R;11}(x-w,y-w)    &  \e{if}  &  y \geq -w+M_L    \ea \right. \;.   
\enq
Thus, adding up the three pieces, one gets that for $x \leq w - M_R$, 
\beq
K_{\e{tot}}(x,y) \; = \; K_R(x-w,y-w)\bs{1}_{\intoo{-w}{+\infty}}(y) \; 
+ \;  G_{R}(x,y) 
\enq
with $G_R$ as defined in \eqref{definition GR}.

Observe that the kernels $K_{L/R}$ may be further expressed, 
following \eqref{decomposition operateur K en L pm pls compact}, as
\beqa
 K_L(x+w,y+w) & = & L^{--}_{L;w}(x,y) \, + \, L^{++}_{L;w}(x,y) \, 
+ \, B_{L}(x+w,y+w) \\
  K_R(x-w,y-w) & = & L^{--}_{R;w}(x,y) \, + \, L^{++}_{R;w}(x,y) \, 
+ \, B_{R}(x-w,y-w) \;. 
\eeqa
There, one has 
\beq
L^{--}_{L;w}(x,y) \; = \; L^{(\e{e})}(x-y)  \cdot 
\bs{1}_{I_w^{-} \times I_w^{-} }(x,y)  \qquad \e{and} \qquad 
 L^{++}_{L;w}(x,y) \; = \; L^{(0)}(x-y)  \cdot 
\bs{1}_{\intoo{-w}{+\infty}\times \intoo{-w}{+\infty} }(x,y)  \;. 
\enq
Likewise, 
\beq
L^{--}_{R;w}(x,y) \; = \; L^{(0)}(x-y)  \cdot 
\bs{1}_{\intoo{-\infty}{w}\times \intoo{-\infty}{w} }(x,y)  
\qquad \e{and} \qquad 
 L^{++}_{R;w}(x,y) \; = \; L^{(\e{e})}(x-y)  \cdot 
\bs{1}_{I_w^{+} \times I_w^{+} }(x,y)   \;. 
\enq

The rest follows upon straightforward calculations starting from 
the decomposition, valid almost everywhere, 
\bem
K_{\e{tot}}(x,y) \, = \, \bs{1}_{\R^{+}}(x) \cdot 
\bigg\{ K_R(x-w,y-w)\bs{1}_{\intoo{-w}{+\infty}}(y) \; + \;  
G_{R}(x,y)  \bigg\}  \\
\; + \;  \bs{1}_{\R^{-}}(x) \cdot \bigg\{ K_L(x+w,y+w)
\bs{1}_{\intoo{-\infty}{w}}(y) \; + \;  G_{L}(x,y)  \bigg\} \;. 
\end{multline}
\qed

\vspace{2mm}

From now on, it appears useful to introduce the  
following notation for the projections of $f\in L^{2}(\R)$ subordinate 
to the intervals $I_{w}^{\a}$, $\a \in \{+,-,0\}$:
\beq
f^{-}=f \bs{1}_{ I_{w}^{-} } \; , \qquad f^{0}=f \bs{1}_{ I_{w}^{0} }  \; 
, \qquad f^{+}=f \bs{1}_{ I_{w}^{+} } \; . 
\enq
The decomposition of the operator $\op{K}_{\e{tot}}$ achieved in 
Lemma \ref{Lemme decomposition noyau integral K} allows one to decompose 
naturally $\e{id} \, - \, \op{K}_{\e{tot}}$ into a matrix bloc operator 
relative to the direct sum decomposition  of the space $\mc{E}(\R)$ 
of \eqref{definition espace E de R} 
induced by the above projection operators:
\beq
\mc{E}(\R) \; = \; L^2_{\e{C}}(I_w^{-}) \, \oplus  \,  L^2(I_w^{0}) \, \oplus  \,  L^2(I_w^{+}) \, ,  
\label{ecriture decomposition directe espace entirer mc E de R}
\enq
where  
\beq
L^2_{C}(I_w^{-}) \, = \, \bigg\{ f \in L^2_{\rm loc}
(I_w^{-}) \; : \;  \exists C_f \; \e{and}\, \a > 0\quad  f(x) \; = \; C_f \, + \, \e{O}\Big( \ex{\a x}  \Big)  \bigg\} \;. 
\enq

The main reason for doing so is that the operators $ \e{id} \, 
- \, \op{L}^{--}_{w}$, $ \e{id} \, - \, \op{L}^{0}_{w}$, 
$ \e{id} \, - \, \op{L}^{++}_{w}$ which encapsulate 
the continuous part of the spectrum of $\e{id} \, - \, \op{K}_{\e{tot}}$ 
and arise in the diagonal block subordinate to the splitting 
\eqref{ecriture decomposition directe espace entirer mc E de R}
may be explicitly inverted. This simplifies the analysis of the original 
equation $\Big( \e{id} \, - \, \op{K}_{\e{tot}} \Big)[f]=h$ 
permitting to map it
into a one whose non-trivial piece is governed by a compact operator 
whose large-$w$ behaviour may be controlled.

Furthermore, 
denoting by
\beq
 B_{\e{tot}}^{ \eps \sg }(x,y) \, = \,  B_{\e{tot}}(x,y) 
\bs{1}_{ I_{w}^{\eps} \times   I_{w}^{\sg}  }(x,y) \qquad 
\e{with} \qquad \eps,\sg \in \{+,-,0\} \;, 
\label{definition projection operateur Btot}
\enq
the integral kernels of the appropriate projections of the operator 
$\op{B}_{\e{tot}}$, 
the equation $\,\Big( \e{id} \, - \, \op{K}_{\e{tot}} \Big)[f]=h\,$
may be recast  
into a block-matrix form subordinate to the direct sum decomposition 
\eqref{ecriture decomposition directe espace entirer mc E de R}
\beq
\left( \ba{ccc} \e{id} \, - \, \op{L}^{--}_{w}   
- \op{B}_{\e{tot}}^{--}        &    - \op{B}_{\e{tot}}^{-0}    
&     - \op{B}_{\e{tot}}^{-+ }    \\  
                - \op{B}_{\e{tot}}^{0-}        &     \e{id} \, 
- \, \op{L}^{0}_{w}   - \op{B}_{\e{tot}}^{00}    &     
- \op{B}_{\e{tot}}^{0+ }   \\ 
                - \op{B}_{\e{tot}}^{+-}        &        
- \op{B}_{\e{tot}}^{+0}    &    \e{id} \, - \, \op{L}^{++}_{w}  
- \op{B}_{\e{tot}}^{++ } \ea \right)
                \cdot\left( \ba{c}  f^{-} \\ f^{0} \\ f^{+} 
\ea \right) \; = \; \left( \ba{c}  h^{-} \\ h^{0} \\ h^{+} 
\ea \right) \;. 
\enq
The above matrix operator equations may be
rewritten as 
\beqa
\Big( \e{id} \, - \, \op{L}^{--}_{w} \Big)\big[ f^{-} \big] & = & 
H^{-} \; = \;   h^{-} \, + \, \sul{ \sg \in  \{\pm, 0 \}  }{} 
\op{B}_{\e{tot}}^{-\sg}\big[ f^{\sg} \big] \;,  
\label{definition H moins}  \\
\Big( \e{id} \, - \, \op{L}^{0}_{w} \Big)\big[ f^{0} \big] & = & 
H^{0} \; = \;   h^{0} \, + \, \sul{ \sg \in  \{\pm, 0 \}  }{} 
\op{B}_{\e{tot}}^{0\sg}\big[ f^{\sg} \big] \; , 
\label{definition H zero} \\
\Big( \e{id} \, - \, \op{L}^{++}_{w} \Big)\big[ f^{+} \big] & = & 
H^{+} \; = \;   h^{+} \, + \, \sul{ \sg \in  \{\pm, 0 \}  }{} 
\op{B}_{\e{tot}}^{+\sg}\big[ f^{\sg} \big]    
\label{definition H plus} \;. 
\eeqa
Since $\big( \e{id} \, - \, \op{K}_{\e{tot}}  \big)[f]=0$ for any 
constant function, it is convenient, owing to the setting that 
was analysed in the previous sections, to extend the equation to the space 
\beq
 \mc{E}^{\prime}(\R) \, = \, L^2_{\e{C}}(I_w^{-}) \, \oplus  
\,  L^2(I_w^{0}) \, \oplus  \,  L^2_{C}(I_w^{+}) \;, 
\label{ecriture decomposition directe espace entirer modifie mc E prime 
de R}
\enq
where 
\beq
L^2_{C}(I_w^{+}) \, = \, \bigg\{ f \in L_{\rm loc}^2(I_w^{+}) \; 
: \;  \exists C_f \; \e{and}\, \a > 0\quad  f(x) \; = \; C_f \, 
+ \, \e{O}\Big( \ex{-\a x}  \Big)  \bigg\} \;. 
\enq
Clearly, any solution obtained in $ \mc{E}^{\prime}(\R)$ gives rise 
to the solution in $ \mc{E}(\R)$ by performing a global translation 
by a constant. The main point is that one may apply the results 
of the previous analysis in the case of $ \mc{E}^{\prime}(\R)$, 
as the invertibility of  $\Big( \e{id} \, - \, \op{L}^{\pm\pm}_{w} 
\Big)$ has been formulated on the spaces $  L^2_{C}(I_w^{\pm})$.
Hence, considering \eqref{definition H moins}, \eqref{definition H zero} 
and  \eqref{definition H plus} as a system of equations on the space 
\eqref{ecriture decomposition directe espace entirer modifie mc E prime 
de R} and observing that the functions $H^{\pm}, H^{0}$ do enjoy 
the properties stated in Propositions 
\ref{Proposition inversion Wiener Hopf pour L+} and 
\ref {Proposition inversion Wiener Hopf pour L-} , as may be inferred 
from direct bounds and Proposition 
\ref{Proposition Caractere Hilbert Schmidt operateur B}, one may apply 
the results of Sections 
\ref{Appendix section inversion operateurs id - L ups ups} and 
\ref{Appendix Secion inversion operateur L0} of the appendix  
in order to invert the operators appearing in the \textit{lhs} of 
\eqref{definition H moins}, \eqref{definition H zero}, 
\eqref{definition H plus} so as to get 
\beqa
\mc{F}\big[f^-](k) & = &  - \a_{\da}^{(\e{e})}(k) \ex{-\i k w } 
\Int{\R + \i v }{} \hspace{-2mm} \f{ \dd s }{ 2\i \pi }  \,  
\f{  \big\{ \a_{\ua}^{(\e{e})}(s) \big\}^{-1}  \cdot   
\mc{F}\big[ H^{-} \big](s)  }{ s-k } \cdot  \ex{ \i s w } 
\qquad \e{with} \qquad k \in \R - \i v \, ,   
\label{ecriture forme Fourier de f moins via H moins} \\
\mc{F}[f^0](k) & = & \mc{F}[H^0](k)\; - \; \Int{ \R + \i v }{} 
\hspace{-2mm} \dd \mu \, R(k,\mu) \mc{F}[H^0](\mu)    \hspace{2.5cm} 
\e{with} \qquad k \in \R + \i v \, , 
\label{ecriture forme Fourier de f zero via H zero}  \\
\mc{F}\big[f^+](k) & = & \f{  \ex{ \i k w }  }{ \a_{\ua}^{(\e{e})}(k) } 
\Int{\R - \i v }{}\hspace{-2mm}  \f{ \dd s }{ 2\i \pi }  \, 
\f{  \a_{\da}^{(\e{e})}(s)  \cdot   \mc{F}\big[ H^{+} \big](s)  }{ s-k }  
\cdot \ex{-\i s w }
 \hspace{2cm}\e{with} \qquad k \in \R + \i v \, . 
 \label{ecriture forme Fourier de f plus via H plus} 
\eeqa
Here, $\a_{\da/\ua}^{(\e{e})}$ are given by 
\eqref{forme explicite alpha up kappa ups}-\eqref{forme explicite alpha da kappa ups} 
upon the substitution $\varkappa^{\ups}\mapsto 0$ 
and $R$ is the resolvent kernel of the operator 
$\e{id}+\op{V}$ on $L^2(\R+\i v)$ introduced in 
\eqref{definition noyau integral de V}-\eqref{ecriture lin int eqn avec operateur V}, \textit{c.f.} Theorem \ref{Theorem structure inverse Id + V}.

\subsection{Preliminary estimates for $\op{B}_{\e{tot}}$}
\label{SousSection preliminary estimates on B tot}

In this subsection, we provide estimates for the Fourier transform 
of the $\pm$ and $0$ projections of the operator $\op{B}_{\e{tot}}$
which will then allow to study the large-$w$ behaviour of the solutions 
to the system \eqref{ecriture forme Fourier de f moins via H moins}, 
\eqref{ecriture forme Fourier de f zero via H zero}, 
\eqref{ecriture forme Fourier de f plus via H plus}.

\begin{prop}

Let
\beq
\wh{B}^{\eps \sg}_{\e{tot}}(k,s)  \, = \,  \Int{ \R }{} \dd x  \Int{ \R }{} \f{ \dd y }{ 2\pi } \ex{\i k x - \i s y }  B^{\eps \sg}_{\e{tot}}(x,y) \; .  
\label{definition operateur hat B eps sg}
\enq
with $B^{\eps \sg}_{\e{tot}}(x,y)$ as introduced in 
\eqref{definition projection operateur Btot}. \,One has

\bem
\left(\ba{ccc} \wh{B}^{--}_{\e{tot}}(k,s)  &  \wh{B}^{-0}_{\e{tot}}(k,s)   
&    \wh{B}^{-+}_{\e{tot}}(k,s)     \vspace{2mm} \\ 
		\wh{B}^{\, 0-}_{\e{tot}}(k,s)  &  
\wh{B}^{\, 00}_{\e{tot}}(k,s)   &     \wh{B}^{ \, 0+}_{\e{tot}}(k,s)  
\vspace{2mm} \\    
		\wh{B}^{+-}_{\e{tot}}(k,s)  &  \wh{B}^{+0}_{\e{tot}}(k,s)   
&     \wh{B}^{++}_{\e{tot}}(k,s)  \ea \right) \\
		\; = \; 
\left(\ba{cc }  \ex{\i (s-k) w} \cdot \wh{B}^{--}_{L}(k,s)  &     
\ex{\i (s-k) w} \cdot \wh{B}^{-+}_{L}(k,s) \, + \,    \mf{B}^{-0}(k,s)
        \vspace{2mm}  \\
  \mf{B}^{0-}(k,s) \, + \, \ex{\i (s-k) w} \cdot \wh{B}^{+-}_{L}(k,s)  
&   \hspace{3mm}  \ex{\i (s-k) w} \cdot \wh{B}^{++}_{L}(k,s) \, 
+ \, \ex{\i (k-s) w} \cdot \wh{B}^{--}_{R}(k,s) \, + \,   \mf{B}^{00}(k,s)
       \vspace{2mm}  \\
		 \mf{B}^{+-}(k,s)    			      
& \ex{\i (k-s) w} \cdot \wh{B}^{+-}_{R}(k,s) \, + \,  \mf{B}^{+0}(k,s)    
\ea \right.    \\
 \left. \ba{c }     \mf{B}^{-+}(k,s)  \vspace{2mm}  \\   
 \ex{\i (k-s) w} \cdot \wh{B}^{-+}_{R}(k,s) \, + \,  \mf{B}^{0+}(k,s)   
\vspace{2mm}  \\    
  \ex{\i (k-s) w} \cdot \wh{B}^{++}_{R}(k,s)     \ea \right). 
\label{ecriture decomposition precise de B selon intervalles -0+}
\end{multline}
Above, the kernels $\wh{B}^{\sg \eps}_{L/R}(k,s)$, with $\sg, 
\eps \in \{\pm\}$, are as introduced in 
\eqref{definition Fourier transform de B sur demi axes}, under the 
substitution $g \mapsto g_{L/R}$ with $g_{L/R}$ given by \eqref{definition gL et gR}. 
The kernels $\mf{B}^{\sg \eps}(k,s)$, with $\sg, \eps \in \{\pm,0\}$, 
are all holomorphic in an open, $w$-independent, neighbourhood of $\R^2$ 
and they satisfy the bounds 
\beq
\mf{B}^{\sg \eps}(k,s) \; = \; \e{O} \Bigg( \f{   
\ex{ -2 w \eta  }  }{  \big( 1+ |k| \big) \big( 1+|s|\big)  } \Bigg)
\label{ecriture estimee asymptotique decoissance pour B perturbatif}
\enq
uniformly in $(k,s)\in \Cx^2$ such that $|\Im(k)|<2v$,  $|\Im(s)|<2v$, 
with $v$ small enough, and for some $\eta$ much larger than $v$.

\end{prop}

\Proof 

We only discuss the case of the coefficient 
$\wh{B}^{\, 0-}_{\e{tot}}(k,s) $ which already contains all the features 
of the analysis. Indeed, one has 
\bem
\wh{B}^{\, 0-}_{\e{tot}}(k,s) \, = \, \Int{-w}{w} \dd x  \Int{-\infty}{-w} \f{ \dd y }{ 2 \pi } \ex{\i k x - \i s y} 
\bigg\{  \bs{1}_{\R^-}(x) \cdot B_{L}(x+w,y+w) \, + \, \bs{1}_{ \R^+ }(x) G_{R}(x,y)  \bigg\}  \\
\, = \, \ex{\i w (s-k) } \wh{B}^{+-}_{L}(k,s)  \, + \,   \mf{B}^{0-}(k,s)
\qquad
\label{ecriture Fourier pour hat B 0 moins}
\end{multline}
with 
\beq
\mf{B}^{0-}(k,s) \, = \,  \ex{\i s w} \Int{0}{w} \dd x  
\Int{-\infty}{0} \f{ \dd y }{ 2 \pi } \ex{\i k x - \i s y} G_{R}(x,y-w)
\; - \;  \ex{\i (s-k) w} \Int{w}{+\infty} \dd x  
\Int{-\infty}{0} \f{ \dd y }{ 2 \pi } \ex{\i k x - \i s y} B_{L}(x,y) \;. 
\enq

Note that all integrals do converge either due to the exponential 
decay of $B_{L}$, \textit{c.f.} 
\eqref{ecriture borne noyau integral B decroissance exponentielle}, 
or to the estimates \eqref{ecriture bornes decroissance fcts G L et R} 
for the decay of $G_{R}(x,y)$. We now discuss how to estimate the first 
term appearing in the definition of $\mf{B}^{0-}(k,s)$.   

For $\e{max}\big\{ |\Im(k)|, |\Im(s)| \big\} < 2 v$, one sets $\wt{G}_{R}(x,y-w) \, =\, \ex{ 3 v (x-y+w) } G_{R}(x,y-w)$, so that, by carrying out integrations by parts,  
\bem
 \ex{\i s w} \Int{0}{w} \dd x  \Int{-\infty}{0} \f{ \dd y }{ 2 \pi } \ex{\i k x - \i s y} G_{R}(x,y-w)  \; = \; 
\f{ \ex{\i (s+3\i v) w}     }{  2\pi (s + 3 \i v) (k + 3\i v) } \Bigg\{    \ex{\i (k + 3\i v) w}  \wt{G}_{R}(w,-w) \, - \;  \wt{G}_{R}(0,-w)  \\
\, - \,  \Int{-\infty}{0}   \dd y  \,  \ex{- \i (s + 3 \i v ) y}  \Dp{y} \bigg[   \ex{  \i (k + 3\i v ) w}  \wt{G}_{R}(w,y-w) \, - \, \wt{G}_{R}(0,y-w)   \bigg]
\, - \,  \Int{0}{w} \dd x \, \ex{  \i (k + 3\i v ) x}  \Dp{x} \wt{G}_{R}(x,-w)  \\
\, + \, \Int{0}{w} \dd x  \Int{-\infty}{0}  \dd y  \, \ex{\i (k + 3\i v)  x - \i (s + 3\i v) y} \Dp{x}\Dp{y}\wt{G}_{R}(x,y-w)   \Bigg\} \;. 
\label{ecriture IPP integrale de GR}
\end{multline}
Then, one estimates each term separately by using directly the bounds \eqref{ecriture bornes decroissance fcts G L et R}, where we remind that the remainder is 
differentiable. For instance, one has 
\bem
 \Bigg| \f{ \ex{\i (s+3\i v) w}     }{  (s + 3 \i v) (k + 3 \i v) } \cdot \Int{0}{w} \dd x  \Int{-\infty}{0}  \f{ \dd y }{  2\pi }  \, \ex{\i (k + 3\i v)  x - \i (s + 3\i v) y} \Dp{x}\Dp{y}\wt{G}_{R}(x,y-w)   \Bigg|  \\
 \, \leq \,   \f{ C \ex{-v w }     }{ |s+ 3 \i v| \, |k + 3 \i v| } \cdot \Int{0}{w} \dd x  \Int{-\infty}{0}  \dd y  \, \ex{-v(x-y)} \ex{ -\big(\frac{\pi}{\tau}-3v \big)(x-y+w) }  
\; =\; \e{O}\Bigg(  \f{ \ex{-2\eta w} }{ (1+|k|)(1+|s|) } \Bigg) \, 
\end{multline}
for some $\eta>0$ and for $\nu$ small enough. 
The remaining terms in \eqref{ecriture IPP integrale de GR} are estimated along the same lines. Finally, the estimation of  the last term appearing in the \textit{rhs} of \eqref{ecriture Fourier pour hat B 0 moins} 
follows exactly the same philosophy. \qed

\subsection{Finer direct sum decomposition of the Hilbert space}
\label{SousSection finer direct sum partition of hilbert space}

While effective for the operator inversion, the direct sum decomposition 
\eqref{ecriture decomposition directe espace entirer modifie mc E prime de R}
is however not fine enough to effectively grasp the large-$w$ asymptotics 
of the integral operators appearing  in 
\eqref{ecriture forme Fourier de f moins via H moins}-\eqref{ecriture forme Fourier de f plus via H plus}. 
For this, as will become apparent in the following, one should further 
partition the central interval $I_w^{0}$ as
\beq
I_w^{0} \, = \, I_w^{L}\cup I_w^{R} \qquad \e{with} \qquad 
I_w^{L}=\intoo{-w}{0} \quad \e{and} \quad I_w^{R}=\intfo{0}{w} \;. 
\enq
Accordingly, we introduce a notation for the projections of 
$f\in L^{2}(\R)$ subordinate to the new intervals $I_{w}^{L/R}$:
\beq
f^{L}=f \bs{1}_{ I_{w}^{L} }  \qquad \e{and}  \qquad f^{R}=f 
\bs{1}_{ I_{w}^{R} }  \; . 
\enq
Since the equations 
\eqref{ecriture forme Fourier de f moins via H moins}-\eqref{ecriture forme Fourier de f plus via H plus} are already in Fourier space, 
it is convenient to introduce the projection operators from 
$\mc{F}\big[ L^2(I_w^0)\big] \subset L^2(\R+\i v)$ onto $\mc{F}
\big[ L^2(I_w^{R/L})\big]$ which, for the moment, 
we continue to think of as a subspace of $L^2(\R+\i v)$
\beq
\op{P}^{R}[f] \; = \; \op{C}^{(+)}_{+}[f] \qquad \e{and} \qquad 
\op{P}^{L}[f] \; = \; -\op{C}^{(+)}_{-}[f] 
\enq
in which $\op{C}^{(+)}$ is the Cauchy transform on $L^2(\R+\i v )$
\beq
\op{C}^{(+)}[f](k) \, = \, \Int{ \R - \i v}{} \f{\dd s }{2\i\pi} 
\f{ f(s) }{ s-k } \; . 
\enq

Finally, it will appear convenient to introduce the shorthand notation 
\beq
u^{\a}(k)= \mc{F}[f^{\a}](k) \qquad \e{for} \qquad \a \in 
\{\pm, 0, L, R\}\; .
\enq
for the Fourier transforms appearing in 
\eqref{ecriture forme Fourier de f moins via H moins}-\eqref{ecriture forme Fourier de f plus via H plus}. 
Obviously,  
$u^0=u^L+u^R$. Furthermore, $u^{L/R}$ are entire 
and, in particular, analytic in a tubular neighbourhood of $\R$. 
This makes it possible to identify $\mc{F}\big[ L^2(I_w^{L})\big]$ as 
a subspace of $L^2(\R+\i v)$ and $\mc{F}\big[ L^2(I_w^{R})\big]$
as a subspace of $L^2(\R-\i v)$, even though the splitting $u^0=u^L+u^R$ 
would suggest an identification of $\mc{F}\big[ L^2(I_w^{R})\big]$ as 
being  a subspace of $L^2(\R+\i v)$. The former, however, appears to be 
more  
useful for the purposes of the analysis to come. On the practical 
side, this identification with a subspace of $L^2(\R-\i v)$ simply means 
a shift of the integration domain in the terms involving $u^R$ from 
$\R+\i v$ to $\R - \i v$ what is possible owing to the analyticity
of the integrand.

\subsection{Decomposition in the $-$ sector}
\label{SousSection Decomposition in the - sector}

In this subsection, we recast the equation in the $-$ sector, 
\textit{viz}. \eqref{ecriture forme Fourier de f moins via H moins}, 
in a form convenient for the further analysis,   
In particular, we explicitly implement the changes issuing from the use 
of the decomposition $u^0=u^L+u^R$.

It is readily seen that 
\beq
\mc{F}[H^-](k) \;= \; \mc{F}[h^-](k) \, + \, \sul{ \a 
\in \{\pm, 0\} }{} \wh{\op{B}}_{\e{tot}}^{\, -\sg}[u^{\sg}](k)
\label{ecriture Fourier de H moins en temes des ops B tot}
\enq
with 
\begin{itemize}
 
 \item $H^{-}$ as defined in \eqref{definition H moins};
 \item  $\wh{\op{B}}_{\e{tot}}^{\, -\sg}: L^2(\R+\i v) \tend 
L^2(\R-\i v)$, $\sg\in \{0,+\}$,  acting with the integral kernel 
$\wh{B}_{\e{tot}}^{\, -\sg}(k,s)$ as defined in 
\eqref{definition operateur hat B eps sg};
\item    $\wh{\op{B}}_{\e{tot}}^{\,--}: L^2(\R-\i v) \tend 
L^2(\R-\i v)$ acting with the integral kernel 
$\wh{B}_{\e{tot}}^{\, --}(k,s)$. 
 
\end{itemize}

Observe that, upon using $\op{P}^R[u^R]=u^R$ one has 
\bem
  \Psi^{-R}[u^R](k) \, = \, - \a_{\da}^{(\e{e})}(k) \ex{-\i k w } 
\Int{\R + \i v }{} \hspace{-2mm} \f{ \dd t }{ 2\i \pi }  \,  
\f{  \big\{ \a_{\ua}^{(\e{e})}(t) \big\}^{-1} \,   \ex{ \i t w }  
\, \wh{B}^{-0}_{\e{tot}}[u^R](t) }{ t-k } \\
 \, = \, 
\lim_{\eps\tend 0^+} \Bigg\{ - \a_{\da}^{(\e{e})}(k) \ex{-\i k w }  
\Int{\R + \i v }{} \hspace{-2mm} \f{ \dd t }{ 2\i \pi }  \,  
\f{  \big\{ \a_{\ua}^{(\e{e})}(t) \big\}^{-1}   }{ t-k } 
\Int{\R +\i v }{} \dd s      \Bigg[ \ex{ \i t w }  \mf{B}^{-0}(t,s)  \, 
+ \, \Int{\R +\i v}{} \f{\dd x}{2\i\pi} \f{ \wh{B}^{-+}_L(t,x) 
\ex{\i x w} }{ s-x-\i\eps } \Bigg]  u^R(s) \Bigg\}
\end{multline}
Using that $u_R$ is entire, and that $ \mf{B}^{-0}(t,s)$ is analytic 
in a tubular neighbourhood of $\R^2$, one may deform the $s$ integrals 
to $\R-\i v$. Furthermore, the analytic structure of the integrand 
allows one to deform the $x$-integrations to $\R+ \i \varrho$ for some 
fixed $\varrho>0$  
that is $v$-independent. This entails that 
\beq
  \Psi^{-R}[u^R](k) \, = \, \Int{ \R -\i v }{} \Psi^{-R}(k,s) u^R(s) 
\quad, \qquad k \in \R-\i v\;, 
\enq
with 
\beq
\Psi^{-R}(k,s) \; = \; - \a_{\da}^{(\e{e})}(k) \ex{-\i k w }  
\Int{\R +   \i v }{} \f{ \dd t }{ 2\i \pi }   
\f{  \big\{ \a_{\ua}^{(\e{e})}(t) \big\}^{-1}   }{ t-k } 
\Bigg[ \ex{ \i t w }  \mf{B}^{-0}(t,s)  \, + \, \Int{\R +\i \varrho }{} 
\f{\dd x}{2\i\pi} \f{ \wh{B}^{-+}_L(t,x) \ex{\i x w} }{ s-x  } \Bigg].
\enq
The decay estimates for $\wh{B}^{\sg\eps}_L$ \eqref{ecriture estimee 
decroissance asymptotiques hat B} and $\mf{B}^{\sg \eps}$ 
\eqref{ecriture estimee asymptotique decoissance pour B perturbatif}
along with Lemma \ref{Lemme bornes sur croissance via transfo Cauchy} 
readily entail that, for $0<v$ small enough, 
\beq
\Psi^{-R}(k,s) \; = \; \e{O}\Bigg( \ex{ - \eta w }   \f{ \ln  (1+|s|)  
\cdot  \ln (1+|k|) }{  (1+|s|)\cdot (1+|k|)  }  \Bigg) \qquad \e{with} 
\qquad (k,s)\in \big\{ \R-\i v\big\}^2  \;,  
\label{ecriture borne sur Psi -R}
\enq
where $\eta>0$ is fixed and $v$ independent. 
To take into account the other quantities arising in 
\eqref{ecriture Fourier de H moins en temes des ops B tot}, we 
introduce the integral kernels 
\beq
\left( \ba{c} \Psi^{- L}(k,s) \\  \Psi^{-+}(k,s)      \ea \right) 
\; = \;  - \a_{\da}^{(\e{e})}(k) \ex{-\i k w }  \Int{\R +   \i v }{} 
\f{ \dd t }{ 2\i \pi }   \f{  \big\{ \a_{\ua}^{(\e{e})}(t) \big\}^{-1}  
\ex{ \i t w }   }{ t-k } 
\left( \ba{c} \mf{B}^{- 0}(t,s) \\  \mf{B}^{-+}(t,s)      \ea \right)
\enq
which, owing to Lemma \ref{Lemme bornes sur croissance via transfo Cauchy} 
and \eqref{ecriture estimee asymptotique decoissance pour B perturbatif}, 
enjoy the bounds, $\sg \in \{L,+\}$,
\beq
\Psi^{-\sg}(k,s) \; = \; \e{O}\Bigg( \f{ \ex{ -\eta w }\cdot 
\ln  (1+|k|)  }{  (1+|s|)\cdot (1+|k|)  }  \Bigg) \qquad \e{with} 
\qquad (k,s)\in \big\{ \R-\i v\big\} \times \big\{ \R + \i v\big\} \;. 
\label{ecriture borne sur Psi -L ou -+}
\enq
The functions $\Psi^{-\sg}(k,s)$, $\sg \in \{+, L, R\}$, then allow 
one to introduce the integral operators $\bs{\Psi}^{-\sg}: L^{2}(\R
+\i \veps_{\sg} v) \tend L^2(\R+\i v)$
where $\veps_{\sg}$ is defined as 
\beq
\veps_{-} \, =\, \veps_{R} \, =\, - \qquad \e{and} \qquad 
\veps_{+}\, =\, \veps_{L} \, =\,  +\, .
\label{definition des signes veps sigma}
\enq
Also, one introduces 
\beq
M^{-\sg}_{L}(k,s) \; = \;  - \a_{\da}^{(\e{e})}(k)   
\Int{\R +  \i \varrho }{} \f{ \dd t }{ 2\i \pi }   
\f{  \big\{ \a_{\ua}^{(\e{e})}(t) \big\}^{-1} \wh{B}_L^{-\sg}(t,s)   
}{ t-k }   \;\;,   \qquad \sg = \pm
\label{noyau integral ML -sigma}
\enq
and
\beq
\mf{d}^{-}_{w}[h^-](k)  \, = \, - \a_{\da}^{(\e{e})}(k) 
\ex{-\i k w } \Int{\R + \i v }{} \f{ \dd s }{ 2\i \pi }   
\f{  \big\{ \a_{\ua}^{(\e{e})}(s) \big\}^{-1}  \cdot  
\ex{ \i s w }    }{  s-k }  \cdot  \mc{F}\big[ h^- \big] (s)\,.
\label{definition operateur d- w}
\enq
Furthermore,  we introduced the integral operator 
$\op{M}^{-\sg}_{L}: L^{2}(\R+\i\veps_{\sg} v) \tend L^2(\R-\i v)$, 
$\sg \in \{\pm\}$,  characterised by the integral kernel 
$M^{-\sg}_{L}(k,s)$.

Finally, denote by $\op{e}$ the operator of multiplication by 
the function $e$, \textit{viz}.
\beq
\op{e}[f](\la)\, = \, e(\la) f(\la)  \qquad \e{with} \qquad 
e(\la) \, = \, \ex{\i \la w} \, .
\label{definition operateur de multiplication par exponentielle}
\enq
Then, by using the decomposition 
\eqref{ecriture Fourier de H moins en temes des ops B tot} one may 
recast  the representation 
\eqref{ecriture forme Fourier de f moins via H moins} in the operator 
form 
\beq
u^-(k) \; = \; \Big( \op{e}^{-1}  \op{M}^{--}_{L}  
\op{e}\Big) \big[ u^-  \big](k) \, + \, \Big( \op{e}^{-1} 
\op{M}^{-+}_{L} \op{e}\Big) \big[ u^L  \big](k) 
\, + \!\! \sul{\sg \in \{L,R,+\} }{} \bs{\Psi}^{-\sg}[u^{\sg}](k)\, 
+ \,  \mf{d}^{-}_{w}[h^-](k) \;.  
\label{equation de decomposition pour u moins}
\enq
Note that equation \eqref{equation de decomposition pour u moins} 
may already be interpreted as holding for the first component 
$u^-$ of the vector 
\beq
\bs{u}\,=\, \left( \ba{c} u^- \\ u^L \\ u^R \\ u^+ \ea \right) \, \in \, L^2(\R-\i v) \oplus L^2(\R + \i v) \oplus L^2(\R-\i v) \oplus L^2(\R + \i v) \;.
\label{ecriture decomposition plus fine de lespace de Hilbert}
\enq

Finally, by introducing the operators 
\beq
\Big( \ba{ccc}   \bs{\Om}^{-L}  &    \bs{\Om}^{-R}  & \bs{\Om}^{-+}  
\ea  \Big) \; = \; 
\Big( \ba{ccc} \op{e} \,   \bs{\Psi}^{-L}  \op{e}^{-1} &    
\op{e}  \,  \bs{\Psi}^{-R}  \op{e}    &    \op{e} \,   
\bs{\Psi}^{-+}  \op{e}  \ea  \Big) 
\label{definition Omega -L -R et -+}
\enq
 one may recast \eqref{equation de decomposition pour u moins} in the form 
of a line vector of operators times a column vector of functions, 
which will be best suited for the later handling:
\beq
\Big( \ba{cccc} \e{id}-\op{e}^{-1}  \op{M}^{--}_{L}  \op{e}   
\; ; &  - \op{e}^{-1} \big[  \op{M}^{-+}_{L}   +  \bs{\Om}^{-L} 
\big] \, \op{e}   \; ; 
												    &   -\op{e}^{-1}  \bs{\Om}^{-R} \op{e}^{-1}   
\; ;  & - \op{e}^{-1} \bs{\Om}^{-+} \op{e}^{-1}  \ea \Big) 
\big[ \bs{u} \big]  \; = \;  \mf{d}^{-}_{w}[h^-](k) \;. 
\label{equation matricielle pour u moins}
\enq

\subsection{Decomposition in the $+$ sector}
\label{SousSection Decomposition in the + sector}
In this subsection, we provide the appropriate operator rewriting 
of the equation in the $+$ sector, \textit{viz}. 
\eqref{ecriture forme Fourier de f plus via H plus},
after implementing the decomposition $u^0=u^L+u^R$.

Analogously to the $-$ sector, one has that 
\beq
\mc{F}[H^+](k) \;= \; \mc{F}[h^+](k) \, + \, \sul{ \a \in \{\pm, 0\} }{} 
\wh{\op{B}}_{\e{tot}}^{\, +\sg}[u^{\sg}](k)
\label{ecriture Fourier de H plus en temes des ops B tot}
\enq
with 
\begin{itemize}
 
 \item $H^{+}$ as defined in \eqref{definition H plus};
 \item  $\wh{\op{B}}_{\e{tot}}^{\, +\sg}: L^2(\R+\i v) \tend 
L^2(\R+\i v)$, $\sg\in \{0,+\}$,  acting with the integral kernel 
$\wh{B}_{\e{tot}}^{\, +\sg}(k,s)$, \textit{c.f.} 
\eqref{definition operateur hat B eps sg};
\item    $\wh{\op{B}}_{\e{tot}}^{\, +-}: L^2(\R-\i v) 
\tend L^2(\R+\i v)$ acting with the integral kernel 
$\wh{B}_{\e{tot}}^{\, + -}(k,s)$. 
 
\end{itemize}

The obvious identity $\op{P}^L[u^L]=u^L$ leads to 
\bem
\Psi^{+L}[u^L](k) \, = \, \f{  \ex{\i k w }  }{ \a_{\ua}^{(\e{e})}(k) } 
\Int{\R - \i v }{} \hspace{-2mm} \f{ \dd t }{ 2\i \pi }  
\f{  \a_{\da}^{(\e{e})}(t)  \,  \ex{- \i t w }    }{ t-k }    
\wh{B}^{+0}_{\e{tot}}[u^L](t)  \\
 \, = \, 
\lim_{\eps\tend 0^+} \Bigg\{  \f{  \ex{\i k w }  }{ \a_{\ua}^{(\e{e})}(k) } 
\Int{\R - \i v }{} \hspace{-1mm} \f{ \dd t }{ 2\i \pi }  
\f{  \a_{\da}^{(\e{e})}(t)  \,      }{ t-k } 
\Int{\R +\i v }{} \hspace{-1mm} \dd s \,      \Bigg[ \ex{- \i t w }  
\mf{B}^{+0}(t,s)  \, - \, \Int{\R +\i v}{} \hspace{-2mm}  
\f{\dd x}{2\i\pi} \f{ \wh{B}^{+-}_R(t,x) \ex{-\i x w} }{ s-x+\i\eps } 
\Bigg]  u^L(s) \Bigg\} \;. 
\end{multline}

One then deforms the $x$-integrations to $\R - \i \varrho$ for some 
fixed $\varrho>0$  
that is $v$-independent leading to 
\beq
  \Psi^{+L}[u^L](k) \, = \, \Int{ \R + \i v }{} \hspace{-2mm} 
\Psi^{+L}(k,s) \,  u^L(s) 
\enq
with 
\beq
\Psi^{+L}(k,s) \; = \; \f{  \ex{\i k w }  }{ \a_{\ua}^{(\e{e})}(k) } 
\Int{\R - \i v }{} \hspace{-1mm} \f{ \dd t }{ 2\i \pi }  
\f{  \a_{\da}^{(\e{e})}(t)  \,      }{ t-k } 
   \Bigg[ \ex{- \i t w }  \mf{B}^{+0}(t,s)  \, - \, \Int{\R  
- \i \varrho }{} \hspace{-2mm}  \f{\dd x}{2\i\pi} \f{ \wh{B}^{+-}_R(t,x) 
\ex{-\i x w} }{ s-x  } \Bigg]  \;. 
\enq
The decay estimates for $\wh{B}^{\sg\eps}_L$ 
\eqref{ecriture estimee decroissance asymptotiques hat B} and 
$\mf{B}^{\sg \eps}$ 
\eqref{ecriture estimee asymptotique decoissance pour B perturbatif}
along with Lemma \ref{Lemme bornes sur croissance via transfo Cauchy} 
readily entail that, for $0<v$ small enough, 
\beq
\Psi^{+L}(k,s) \; = \; \e{O}\Bigg( \ex{ -\eta w } \f{ \ln(1+|s|)  
\cdot  \ln(1+|k|)   }{  (1+|s|)\cdot(1+|k|)  }  \Bigg) \qquad 
\e{with} \qquad (k,s)\in \big\{ \R + \i v\big\}^2  \;. 
\label{ecriture borne sur Psi +L}
\enq
In order to take into account the other terms present in 
\eqref{ecriture Fourier de H plus en temes des ops B tot}, one is 
lead to introduce the integral kernels 
\beq
\left( \ba{c} \Psi^{+-}(k,s) \\  \Psi^{ + R }(k,s)      \ea \right) \; 
= \;  \f{  \ex{\i k w }  }{ \a_{\ua}^{(\e{e})}(k) } \Int{\R - \i v }{} 
\f{ \dd t }{ 2\i \pi }  \f{  \a_{\da}^{(\e{e})}(t)  \,  
\ex{- \i t w }    }{ t-k }   
\left( \ba{c} \mf{B}^{+ - }(t,s) \\  \mf{B}^{ + 0}(t,s)      \ea \right)
\enq
which enjoy the bounds,
\beq
\Psi^{+\sg}(k,s) \; = \; \e{O}\Bigg( \f{ \ex{ - \eta w } \cdot 
\ln(1+|k|) }{  (1+|s|)\cdot(1+|k|)  }  \Bigg) \qquad \e{with} 
\qquad (k,s)\in \big\{ \R + \i v \big\} \times \big\{ \R - \i v\big\}   
\quad \e{and} \quad \sg \in \{- , R\}\;. 
\label{ecriture borne sur Psi +- et +R}
\enq
The functions $\Psi^{+\sg}(k,s)$, $\sg \in \{-, L, R\}$, then allow 
one to introduce the integral operators $\bs{\Psi}^{+\sg}: L^{2}
(\R+\i\veps_{\sg} v) \tend L^2(\R+\i v)$ in which $\veps_{\sg}$ has 
been defined in \eqref{definition des signes veps sigma}. 

Finally, one introduces
\beq
M^{+\sg}_{R}(k,s) \; = \;    \f{ 1  }{ \a_{\ua}^{(\e{e})}(k) } 
\Int{\R - \i v }{} \f{ \dd t }{ 2\i \pi }  \f{  \a_{\da}^{(\e{e})}(t)  
\,  \wh{B}_R^{+\sg}(t,s)      }{ t-k }      \;\;,   \qquad \sg = \pm \;, 
\label{noyau integral MR +sigma}
\enq
and
\beq
\mf{d}^{+}_w[h^+](k)  =   \f{ \ex{\i k w } }{ \a_{\ua}^{(\e{e})}(k) } 
\cdot \Int{\R - \i v }{} \f{ \dd s }{2\i\pi}  \f{  \a_{\da}^{(\e{e})}(s)  
\, \ex{-\i s w} \,    \mc{F}\big[ h^+ \big] (s) }{   s-k  }\,.   
\label{definition operateur d+ w}
\enq
At this stage we introduced the integral operator $\op{M}^{+\sg}_{R}: 
L^{2}(\R+\i\veps_{\sg} v) \tend L^2(\R+\i v)$, $\sg \in \{\pm\}$,  
characterised by the integral kernel $M^{+\sg}_{R}(k,s)$.

Altogether, this recasts 
\eqref{ecriture forme Fourier de f moins via H moins} in the 
following operator form 
\beq
u^+(k) \; = \; \Big( \op{e}\, \op{M}^{++}_{R}  \op{e}^{-1}\Big) 
\big[ u^+  \big](k) \, + \, \Big( \op{e}\, \op{M}^{+-}_{R} 
\op{e}^{-1}\Big) \big[ u^R  \big](k) 
\, + \!\! \sul{\sg \in \{L,R,-\} }{} \bs{\Psi}^{+\sg}[u^{\sg}](k)\, 
+ \,  \mf{d}^{+}_w[h^+](k) \;, 
\label{equation de decomposition pour u plus}
\enq
in which $\Psi^{+\sg}$ are integral operators acting with the integral 
kernels introduced earlier on and the operator $\op{e}$ has been 
introduced in 
\eqref{definition operateur de multiplication par exponentielle}.  
This equation now concerns the last component $u^+$ of the vector 
appearing in \eqref{ecriture decomposition plus fine de lespace de Hilbert}.

 Finally, by introducing the operators 
\beq
\Big( \ba{ccc}   \bs{\Om}^{+-}  &    \bs{\Om}^{+L}  & \bs{\Om}^{+R}  
\ea  \Big) \; = \; 
\Big( \ba{ccc} \op{e}^{-1} \,   \bs{\Psi}^{+-}  \op{e}^{-1} &    
\op{e}^{-1} \, \bs{\Psi}^{+L}  \op{e}^{-1}    &    \op{e}^{-1} \, \,   
\bs{\Psi}^{+R}  \op{e}  \ea  \Big), 
\label{definition Omega +- +L et +R}
\enq
 one may recast \eqref{equation de decomposition pour u plus} in the 
following form that will be best suited for the later handling:
\beq
\Big( \ba{cccc}   - \op{e} \,  \bs{\Om}^{+-} \op{e} \; ;   &   -\op{e} 
\,  \bs{\Om}^{+L} \op{e}    \; ;
												&  - \op{e}\,  \big[  \op{M}^{+-}_{R}   
+  \bs{\Om}^{+R} \big] \, \op{e}^{-1}   \; ;     & \e{id} - \op{e} \,   
\op{M}^{++}_{R}  \op{e}^{-1}    \ea \Big) 
\big[ \bs{u} \big]  \; = \;  \mf{d}^{+}_{w}[h^+](k) \;. 
\label{equation matricielle pour u plus}
\enq
Above, $\,\bs{u}$ is as given by 
\eqref{ecriture decomposition plus fine de lespace de Hilbert}.

\subsection{Decomposition in the $0$ sector} 
 \label{SousSection Decomposition in the 0 sector}
 
 In this subsection, we provide an appropriate rewriting of the equation 
in the $0$ sector, \textit{viz}. 
\eqref{ecriture forme Fourier de f zero via H zero},
after incorporating the decomposition $u^0=u^L+u^R$. In the case 
of the sector subordinate to the interval $I_{w}^{0}$, the Fourier 
transform of the function $H^{0}$ defined in \eqref{definition H zero} 
takes the form 
\beq
\mc{F}[H^0](k) \;= \; \mc{F}[h^0](k) \, 
+ \, \sul{ \a \in \{\pm, 0\} }{} \wh{\op{B}}_{\e{tot}}^{\, 0 \sg}
[u^{\sg}](k)
\enq
with 
\begin{itemize}
 \item  $\wh{\op{B}}_{\e{tot}}^{\, 0 \sg}: L^2(\R+\i v) 
\tend L^2(\R+\i v)$, $\sg\in \{0,+\}$,  acting with the integral kernel 
$\wh{B}_{\e{tot}}^{\, 0\sg}(k,s)$, \textit{c.f.} 
\eqref{definition operateur hat B eps sg};
\item    $\wh{\op{B}}_{\e{tot}}^{\, 0 -}: L^2(\R-\i v) \tend L^2(\R+\i v)$ 
\,acting with the integral kernel $\wh{B}_{\e{tot}}^{\,0 -}(k,s)$. 
 
\end{itemize}

\noindent It then follows from 
\eqref{ecriture decomposition precise de B selon intervalles -0+} 
that the latter may be further expressed as 
\bem
\mc{F}[H^0](k) \;= \; \mc{F}[ h^0](k) \, 
+ \, \Big( \op{e}^{-1} \wh{\op{B}}_{L}^{\, +-} \op{e} \Big)[u^{-}] \, 
+ \, \Big( \op{e}^{-1} \wh{\op{B}}_{L}^{\, ++} \op{e} 
+ \op{e}\wh{\op{B}}_{R}^{\, --} \op{e}^{-1}   \Big)[u^{0}] \\
\, + \, \Big( \op{e} \wh{\op{B}}_{R}^{\, -+} \op{e}^{-1} \Big)[u^{+}] \, 
+ \, \sul{ \a \in \{\pm, 0\} }{} \mf{B}^{  0 \sg}[u^{\sg}](k) \;. 
\end{multline}
Thus equation \eqref{ecriture forme Fourier de f zero via H zero} leads 
to the following expression for $u^0(k)$
\beq
u^0(k) \, = \, \mf{d}^{0}_{w}[ h^0](k) \, + \,  \sul{ \a 
\in \{\pm, 0\} }{} \be^{  0 \sg} [u^{\sg}](k)  \, 
+ \,  \sul{ \a \in \{\pm, 0\} }{} \Big( \wt{\mf{B}}^{  0 \sg} \, 
+ \, \de\be^{  0 \sg}   \Big)[u^{\sg}](k) \;. 
\enq
There, we have introduced 
\beq
\mf{d}^{0}_{w}[ h^0](k) \; = \; \Big( \e{id} \, - \, \op{R}\Big) 
\Big[ \mc{F}[ h^0] \Big](k)
\enq
while the operators $ \wt{\mf{B}}^{  0 \sg}$, $\sg \in \{\pm, 0\}$, act 
with the integral kernels 
\beq
\wt{\mf{B}}^{  0 \sg}(k,s) \, = \,  \Big(\e{id}-\op{R}\Big)\big[ \mf{B}^{  0 
\sg}(*,s)  \big](k) \; = \;   \mf{B}^{ 0 \sg}(k,s) \, - \, \Int{\R 
+  \i v }{}\hspace{-2mm} \dd \mu \,  R(k,\mu)  \mf{B}^{  0 \sg}(\mu,s)\;. 
\label{definition fct tile frak B}
\enq
The expressions for the integral kernels $\be^{  0 \sg}(k,s)$ and 
$\de\be^{  0 \sg}(k,s)$ involve the leading $R_{\infty}(\la,\mu)$ and 
perturbative $\de R (\la, \mu)$ resolvent kernel, as introduced  
in \eqref{definition noyau dominant} and \eqref{definition noyau perturbe}. 
Indeed, 
\beq
\be^{0-}(k,s) \, = \,   \Big(\e{id}-\op{R}_{\infty}\Big)\Big[ e^{-1}(*) 
\wh{B}_{L}^{ +-}(*,s) e(s)  \Big](k) 
\; ,  \qquad 
\be^{0+}(k,s) \, = \,   \Big(\e{id}-\op{R}_{\infty}\Big)\Big[ e(*) 
\wh{B}_{R}^{ -+}(*,s) e^{-1}(s)  \Big](k) \;, 
\enq
as well as 
\beq
\be^{00}(k,s) \, = \,  \Big(\e{id}-\op{R}_{\infty}\Big)\Big[ e^{-1}(*) 
\wh{B}_{L}^{ ++}(*,s) e(s) \, +\,  e(*) \wh{B}_{R}^{ --}(*,s) e^{-1}(s)  
\Big](k) \;. 
\enq
Above, $*$ refers to the running variable on which the operator acts. 
Finally, one has 
\beq
\de\be^{0-}(k,s) \, = \,  -\de \op{R} \Big[ e^{-1}(*) \wh{B}_{L}^{ +-}(*,s) 
e(s)  \Big](k) 
\; ,  \qquad 
\de \be^{0+}(k,s) \, = \,  -\de \op{R}\Big[ e(*) \wh{B}_{R}^{ -+}(*,s) 
e^{-1}(s)  \Big](k) \;, 
\label{definition noyaux de  delta beta 0 plus et moins}
\enq
as well as 
\beq
\de \be^{00}(k,s) \, = \,   -\de \op{R}\Big[ e^{-1}(*) 
\wh{B}_{L}^{ ++}(*,s) e(s) \, +\,  e(*) \wh{B}_{R}^{ --}(*,s) 
e^{-1}(s)  \Big](k) \;. 
\label{definition noyaux de  delta beta 00}
\enq

The rewriting of the operators $\de\be^{0\sg}$ and $\wt{\mf{B}}^{0\sg}$ 
in a form appropriate for the analysis to come is rather direct and 
we shall carry it out first. Then, we focus on the operators $\be^{0\sg}$ 
whose large-$w$ asymptotics demand a deeper investigation.

\subsubsection{Perturbing operators $\Psi_{\mf{B}}^{\sg \tau}$}

It follows from the estimates \eqref{ecriture estimee asymptotique 
decoissance pour B perturbatif} and 
\eqref{ecriture borne sur noyau resolvent}, direct bounds and the 
possibility to deform slightly the $\mu$-integration contour in  
\eqref{definition fct tile frak B} that one has 
\beq
\wt{\mf{B}}^{ 0 \sg}(k,s) \, = \,  \e{O}\bigg(  \f{  \ex{-\eta w} }{ 
(1 + |k|) (1 + |s| )  }  \bigg)   \qquad \e{for} \; \e{any} \;  
\qquad |\Im(k)|\leq 2v ,\;  |\Im(s)|\leq 2v \,, 
\enq
this provided that $0<v$ is sufficiently small. 
Furthermore, one also gets that 
\bem
\wt{\mf{B}}^{  L  \sg}(k,s) \, = \,  \op{P}^{L}\big[ 
\wt{\mf{B}}^{  0 \sg}(*,s)  \big](k) \, =\, - \lim_{\eps\tend 0^+} 
\Int{ \R + \i  v}{} \f{ \dd t }{ 2\i\pi } \f{ 
\wt{\mf{B}}^{ 0 \sg}(t,s) }{ t-k + \i\eps }  \\
\, =\, - \Int{ \R + 2 \i v}{} \f{\dd t }{2\i\pi} \f{ 
\wt{\mf{B}}^{  0 \sg}(t,s) }{ t-k }  \; = \; \e{O}
\Bigg(  \f{  \ex{-\eta w} \ln(1 + |k|)  }{ (1 + |k|) (1 + |s| ) }  \Bigg) \;, 
\end{multline}
by virtue of Lemma \ref{Lemme bornes sur croissance via transfo Cauchy}. 
Likewise, 
\bem
\wt{\mf{B}}^{  R  \sg}(k,s) \, = \,  \op{P}^{R}\big[ 
\wt{\mf{B}}^{  0 \sg}(*,s)  \big](k) \, =\, \lim_{\eps\tend 0^+} 
\Int{ \R + \i  v}{} \f{ \dd t }{ 2\i\pi } \f{  
\wt{\mf{B}}^{ 0 \sg}(t,s) }{ t-k - \i\eps }  \\
\, =\,  \Int{ \R - 2 \i v}{} \f{\dd t }{2\i\pi} \f{  
\wt{\mf{B}}^{  0 \sg}(t,s) }{ t-k }  \; = \; \e{O}
\Bigg(  \f{  \ex{-\eta w} \ln(1 + |k|)  }{ (1 + |k|) (1 + |s| ) }  \Bigg) \;.  
\end{multline}
Since the integral kernel $\mf{B}^{00}(k,s)$ is holomorphic in a tubular 
neighbourhood of $\R^2$, the fact that the resolvent kernel $R$ is also 
analytic in such a neighbourhood and the bounds 
\eqref{ecriture borne sur noyau resolvent} entail that 
$\wt{\mf{B}}^{00}(k,s)$ is also analytic in such a tubular neighbourhood. 
Then, the fact that $u^{L/R}$ are entire allows one to deform the 
integration contours in the action below so as to  get 
\beq
\wt{\mf{B}}^{  0 0}[u^{0}](k) \, = \, \Int{ \R + \i v }{} \dd s 
\wt{\mf{B}}^{00}(k,s)u^{L}(s) \, + \,  \Int{ \R - \i v }{} \dd s 
\wt{\mf{B}}^{00}(k,s)u^{R}(s)  \;. 
\enq

 Therefore, upon defining the integral kernels 
\beq
\Big( \Psi_{\mf{B}}^{\sg -} (k,s) \;\; \Psi_{\mf{B}}^{\sg L} (k,s) \;\; 
\Psi_{\mf{B}}^{\sg R} (k,s) \;\; \Psi_{\mf{B}}^{\sg +} (k,s) \Big) \; = \; 
\Big( \wt{\mf{B}}^{ \sg - }(k,s) \;\;  \wt{\mf{B}}^{ \sg 0 }(k,s) \;\;  
\wt{\mf{B}}^{ \sg 0 }(k,s) \;\;  \wt{\mf{B}}^{ \sg + }(k,s) \Big) \;, 
\enq
with $\sg \in \{L,R\}$, one may introduce the associated operators 
$\Psi_{\mf{B}}^{\sg \tau}: L^{2}(\R+\i  \veps_{\sg} v) 
\tend L^{2}(\R+\i  \veps_{\tau} v) $, with $\sg\in \{L,R\}$ and $\tau 
\in \{\pm, L, R\}$ and where $\veps_{\sg}$ is as given in 
\eqref{definition des signes veps sigma}. 

By virtue of the previous estimates, one has 
\beq
\Psi_{\mf{B}}^{\sg \tau} (k,s)\; = \; \e{O}\bigg(  \f{  \ex{-\eta w} 
\ln(1 + |k|)  }{ (1 + |k|) (1 + |s| ) }  \bigg) \qquad \e{with} \qquad 
(k,s) \in \big\{ \R+\i  \veps_{\sg} v \big\} \times \big\{ \R+\i  
\veps_{\tau} v \big\} \;. 
\label{ecriture borne sur Psi mathfrakB sigma tau}
\enq

\subsubsection{Perturbing operators $\Psi_{\de\be}^{\sg \tau}$}

It follows from the estimates \eqref{ecriture estimee decroissance 
asymptotiques hat B} and \eqref{ecriture borne sur noyau resolvent perturbe}, 
direct bounds and the possibility to deform slightly the integration 
contour in  the action of $\de R$ in 
\eqref{definition noyaux de  delta beta 0 plus et moins}-\eqref{definition noyaux de  delta beta 00}
that one has the bounds
\beq
\de \be^{ 0 \sg}(k,s) \, = \,  \e{O}\bigg(  \f{  
\ex{-\eta w} }{ (1 + |k|) (1 + |s| )  }  \bigg)   \qquad \e{for} \; 
\e{any} \;  \qquad |\Im(k)|\leq 2v , \;  |\Im(s)|\leq 2v \,, 
\enq
this provided that $v$ is taken sufficiently small. Moreover, analogously 
to the previous reasonings 
\beqa
\de \be^{  L  \sg}(k,s) & = &   \op{P}^{L}\big[ \de \be^{  0 \sg}(*,s)  
\big](k) 
\, =\, - \Int{ \R + 2 \i v}{} \hspace{-2mm} \f{\dd t }{2\i\pi}\,  
\f{ \de \be^{  0 \sg}(t,s) }{ t-k }  \; = \; \e{O}\bigg(  
\f{  \ex{-\eta w} \ln(1 + |k|)  }{ (1 + |k|) (1 + |s| ) }  \bigg) \;, \\ 
\de \be^{  R  \sg}(k,s)  &  =    & \op{P}^{R}\big[ \de \be^{  0 \sg}(*,s)  
\big](k) 
\, =\,  \Int{ \R - 2 \i v}{}\hspace{-2mm} \f{\dd t }{2\i\pi}\,  
\f{\de \be^{  0 \sg}(t,s) }{ t-k }  \; = \; \e{O}\bigg(  \f{  \ex{-\eta w} 
\ln(1 + |k|)  }{ (1 + |k|) (1 + |s| ) }  \bigg) \;.  
\eeqa
Finally, one may present the action of $\de \be^{  0 0}$ on $u^0$ as
\beq
\de \be^{  0 0}[u^{0}](k) \, = \, \Int{ \R + \i v }{}\hspace{-2mm} \dd s 
\, \de \be^{00}(k,s)u^{L}(s) \, + \,  \Int{ \R - \i v }{}\hspace{-2mm}  
\dd s  \, \de \be^{00}(k,s)u^{R}(s)  \;. 
\enq
 Therefore, upon defining the integral kernels 
\beq
\Big( \Psi_{ \de \be }^{\sg -} (k,s) \;\; \Psi_{ \de \be }^{\sg L} (k,s) 
\;\; \Psi_{ \de \be }^{\sg R} (k,s) \;\; \Psi_{ \de \be }^{\sg +} (k,s) 
\Big) \; = \; 
\Big( \de \be^{ \sg - }(k,s) \;\;  \de \be^{ \sg 0 }(k,s) \;\;  
\de \be^{ \sg 0 }(k,s) \;\;  \de \be^{ \sg + }(k,s) \Big) \;, 
\enq
with $\sg \in \{L,R\}$, one gets the associated operators 
$\Psi_{\de \be}^{\sg \tau}: L^{2}(\R+\i  \veps_{\sg} v) 
\tend L^{2}(\R+\i  \veps_{\tau} v) $, with 
$\veps_{\sg}$ as defined in 
\eqref{definition des signes veps sigma} and where $\tau \in \{\pm, L, R\}$.

By virtue of the previous estimates, one has 
\beq
\Psi_{ \de \be }^{\sg \tau} (k,s)\; = \; \e{O}\bigg(  \f{  \ex{-\eta w} 
\ln(1 + |k|)  }{ (1 + |k|) (1 + |s| ) }  \bigg) \qquad \e{with} \qquad 
(k,s) \in \big\{ \R+\i  \veps_{\sg} v \big\} \times \big\{ \R+\i  
\veps_{\tau} v \big\} \;. 
\label{ecriture borne sur Psi deltaBeta sigma tau}
\enq

\subsubsection{Operator $\be^{0+}$}
\label{SubSubsection operateur beta0+}

 In order to decompose the integral kernel $\be^{0+}(k,s)$ into 
its dominant and sub-dominant in $w$ parts,  by using the explicit 
expression for the leading resolvent 
\eqref{expression explicite noyau resolvent dominant}, one first computes 
\bem
\op{P}^{L}\Big[ \big(\e{id}- \op{R}_{\infty}\big)\big[e(*)
\wh{B}^{-+}_{R}(*,s)e^{-1}(s)\big](\cdot) \Big](k) \, = \, 
\lim_{\eps\tend 0^+} \Int{\R+\i v }{} \hspace{-2mm} \f{\dd \la}{2\i\pi} 
\f{ e(\la)\wh{B}^{-+}_{R}(\la,s)e^{-1}(s)   }{ k - \la -\i  \eps } 
\; - \; \Int{\R+\i v }{} \hspace{-2mm} \f{\dd \la}{2\i\pi}  
\f{ - \mc{F}[L^{(0)}](\la) }{ k - \la -\i \eps} \\
\times \Int{\R+\i v }{} \hspace{-2mm} \f{\dd \mu }{ 2\i \pi } 
\Big( - \{ \a_{\ua}^{(0)}(\la) e(\la) \}^{-1}  \; , \;  
\a_{\da}^{(0)}(\la) e(\la) \Big) 
\cdot  \bigg(I_{2}\, + \, \f{\la-\mu}{\la\mu \mf{b}^{\prime}(0) } 
\op{D} \bigg) 
\left(\ba{c}  \a_{\da}^{(0)}(\mu) e^2 (\mu)  \\ 
\{ \a_{\ua}^{(0)}(\mu) \}^{-1}    \ea \right) \f{ \wh{B}^{-+}_{R}(\mu,s) 
e^{-1}(s) }{ \la - \mu }  
\end{multline}
 where $\op{D}$ is as defined in \eqref{definition matrice D et prefacteur 
theta}, $\mf{b}$ is as defined in \eqref{definition P infty et mathfrak d}, 
while $\a_{\ua/\da}^{(0)}$ are as described in Subsection 
\ref{Subsection RHP for matrix Xi}. 

To proceed further, one splits the integral as follows 
\beq
\op{P}^{L}\Big[ \big(\e{id}- \op{R}_{\infty}\big)
\big[e(*)\wh{B}^{-+}_{R}(*,s)e^{-1}(s)\big](\cdot) \Big](k)  \, 
= \, e^{-1}(k) \Phi^{L+}(k,s) e^{-1}(s) \, + \, \Psi^{L+}_{R_\infty}(k,s)
\enq
in which, for $\sg \in \big\{ +, R \big\}$ and $\veps_{\sg}$ as in 
\eqref{definition des signes veps sigma}, 
\bem
 \Psi^{L \sg }_{R_\infty}(k,s) \; = \;  \Int{\R+ 2 \i \eta }{} 
\hspace{-2mm} \f{\dd \la}{2\i\pi} \f{ e(\la)\wh{B}^{\, 
- \veps_{\sg} }_{R}(\la,s)e^{-1}(s)   }{ k - \la } \\ 
 + \Int{\R +  3 \i \f{\eta}{2} }{}  \hspace{-2mm}  \f{\dd \la}{2\i\pi}  
\f{  \mc{F}[L^{(0)}](\la) }{ k - \la  } 
\Int{\R + 2 \i \eta }{} \hspace{-2mm} \f{\dd \mu }{ 2\i \pi }  
\Big(0 \; , \;  \a_{\da}^{(0)}(\la) e(\la) \Big) 
\cdot \bigg(I_{2}\, + \, \f{\la-\mu}{\la\mu \mf{b}^{\prime}(0) } 
\op{D} \bigg) 
\left(\ba{c}  \a_{\da}^{(0)}(\mu) e^2 (\mu)  \\ 
\{ \a_{\ua}^{(0)}(\mu) \}^{-1}    \ea \right) 
\f{ \wh{B}^{\, - \veps_{\sg}}_{R}(\mu,s) e^{-1}(s) }{ \la - \mu } \\
\; + \; \Int{\R + 2 \i v }{} \hspace{-2mm} \f{\dd \la}{2\i\pi}  
\f{   \mc{F}[L^{(0)}](\la) }{ k - \la  } 
\Int{\R + 2 \i \eta }{} \hspace{-2mm} \f{\dd \mu }{ 2\i \pi } 
\Big( - \{ \a_{\ua}^{(0)}(\la) e(\la) \}^{-1}  \; , \;  0 \Big) 
\cdot  \bigg(I_{2}\, + \, \f{\la-\mu}{\la\mu \mf{b}^{\prime}(0) } 
\op{D} \bigg) 
\left(\ba{c}  \a_{\da}^{(0)}(\mu) e^2 (\mu)  \\ 0  \ea \right) 
\f{ \wh{B}^{\, - \veps_{\sg}}_{R}(\mu,s) e^{-1}(s) }{ \la-\mu  } \;. 
\end{multline}

Furthermore, upon using 
\beq
\big( -a \, , \, b \big) \, \op{D} \, \bigg( \ba{c} 0 \\ 1   \ea \bigg)  
\, = \,  a+b \;, 
\enq
one entails that 
\beqa
\Phi^{L\sg}(k,s) & = &  e(k)  \Int{\R+ 2 \i v }{}\hspace{-2mm}  
\f{\dd \la}{2\i\pi}  \f{   \mc{F}[L^{(0)}](\la)  
e^{-1}(\la)  }{ (k - \la) \la \,   \a_{\ua}^{(0)}(\la)   } 
\times \Int{\R+\i v }{}  \hspace{-2mm} \f{\dd \mu }{ 2\i \pi }  
\f{ \wh{B}^{\, - \veps_{\sg} }_{R}(\mu,s)  }{  \mu  \,  
\a_{\ua}^{(0)}(\mu) \mf{b}^{\prime}(0)    }   \\
 & = & \Bigg\{  e(k)  \Int{\R  +  \i \frac{ v }{2} }{}\hspace{-2mm}  
\f{\dd \la}{2\i\pi}  \f{   \mc{F}[L^{(0)}](\la)  e^{-1}(\la)  }{ (k - \la) 
\la \,   \a_{\ua}^{(0)}(\la)   } 
 \; + \;   \f{   \mc{F}[L^{(0)}](k)   }{ k \,   \a_{\ua}^{(0)}(k)   }    
\Bigg\}
\times \Int{\R+\i v }{}  \hspace{-2mm} \f{\dd \mu }{ 2\i \pi }  
\f{ \wh{B}^{ \, - \veps_{\sg} }_{R}(\mu,s)  }{  \mu  \,  
\a_{\ua}^{(0)}(\mu) \mf{b}^{\prime}(0)    }  \;. 
\eeqa
It is direct to check that $ \Phi^{L \sg }(k,s)$, $\sg \in 
\big\{ +, R \big\}$   enjoys, for some $c>0$, the bound
\beq
 \Phi^{L \sg}(k,s) \; = \;\e{O}\Bigg( \f{ \ex{- 
\f{v w}{2} } }{ (1+ |k|)(1+|s|) } \; + \;   \f{ \ex{- c |k| } }{ w (1+|s|) }  
\Bigg) \qquad \e{for} \qquad (k,s) \in \big\{ \R + \i v \big\} \times  
\big\{ \R + \i \veps_{\sg} v \big\} \;. 
\label{ecriture borne sur Phi L+ et LR}
\enq

Quite similarly, one has 
\bem
\op{P}^{R}\Big[ \big(\e{id}- \op{R}_{\infty}\big)
\big[e(*)\wh{B}^{-+}_{R}(*,s)e^{-1}(s)\big](\cdot) \Big](k) \, = \, 
- \lim_{\eps\tend 0^+} \Int{\R+\i v }{} \hspace{-2mm} 
\f{\dd \la}{2\i\pi} \f{ e(\la)\wh{B}^{-+}_{R}(\la,s)
e^{-1}(s)   }{ k - \la +\i \eps } 
\; - \; \Int{\R+\i v }{} \hspace{-2mm} \f{\dd \la}{2\i\pi}  
\f{   \mc{F}[L^{(0)}](\la) }{ k - \la + \i \eps } \\
\times \Int{\R+\i v }{} \hspace{-2mm} \f{\dd \mu }{ 2\i \pi } 
\Big( - \{ \a_{\ua}^{(0)}(\la) e(\la) \}^{-1}  \; , \;  
\a_{\da}^{(0)}(\la) e(\la) \Big) 
\cdot  \bigg(I_{2}\, + \, \f{\la-\mu}{\la\mu \mf{b}^{\prime}(0) } 
\op{D} \bigg) 
\left(\ba{c}  \a_{\da}^{(0)}(\mu) e^2 (\mu)  \\ 
\{ \a_{\ua}^{(0)}(\mu) \}^{-1}    \ea \right) 
\f{ \wh{B}^{-+}_{R}(\mu,s) e^{-1}(s) }{ \la - \mu }  \\
  \, = \, U^{R+}(k,s) \, + \, \Psi^{R+}_{R_\infty}(k,s)  \;. 
\end{multline}
There, with $\sg \in \big\{ +, R \big\}$ and $\veps_{\sg}$ as 
in \eqref{definition des signes veps sigma}, we set
\beq
U^{R \sg }(k,s) \, = \, e(k)\wh{B}^{\, - \veps_{\sg}}_{R}(k,s)e^{-1}(s)  
\, -  \hspace{-2mm}\Int{\R - 2 \i v }{} \hspace{-2mm} \f{\dd \la}{2\i\pi} 
\Int{\R + 2 \i v }{} \hspace{-2mm} \f{\dd \mu }{ 2\i \pi }  
\f{   \mc{F}[ L^{(0)} ]( \la ) \,  \a_{\da}^{(0)}(\la) e(\la)  }{ (k - \la ) 
  \, \a_{\ua}^{(0)}(\mu)  } 
  \Big( 1 \, + \, \f{\la-\mu}{ \la \mu  \mf{b}^{\prime}(0) }   \Big)
 \f{ \wh{B}^{\, - \veps_{\sg}}_{R}(\mu,s) e^{-1}(s) }{ \la - \mu    }  \;. 
\enq
Also, 
\bem
 \Psi^{R \sg }_{R_\infty}(k,s) \; = \;  - \Int{\R + 2 \i \eta }{} 
\hspace{-2mm} \f{\dd \la}{2\i\pi} \f{ e(\la)\wh{B}^{\, 
- \veps_{\sg}}_{R}(\la,s)e^{-1}(s)   }{ k - \la   } \\  
\; - \; \Int{\R - 2 \i \eta }{} \hspace{-2mm} \f{\dd \la}{2\i\pi}  
\f{   \mc{F}[L^{(0)}](\la) }{ k - \la  } 
\cdot \Int{\R  + 2\i \eta }{} \hspace{-2mm} \f{\dd \mu }{ 2\i \pi } 
\Big( - \{ \a_{\ua}^{(0)}(\la) e(\la)  \}^{-1}  \; , \; 0  \Big) 
\cdot  \bigg(I_{2}\, + \, \f{\la-\mu}{\la\mu \mf{b}^{\prime}(0) } 
\op{D} \bigg) 
\left(\ba{c}  \a_{\da}^{(0)}(\mu) e^2 (\mu)  \\ 
\{ \a_{\ua}^{(0)}(\mu) \}^{-1}    \ea \right) \f{ \wh{B}^{\, 
- \veps_{\sg}}_{R}(\mu,s) e^{-1}(s) }{ \la - \mu }  \\
\; - \; \Int{\R - 2 \i v }{} \hspace{-2mm} \f{\dd \la}{2\i\pi}  
\f{   \mc{F}[L^{(0)}](\la) }{ k - \la   } 
\Int{\R + \i \eta }{} \hspace{-2mm} \f{\dd \mu }{ 2\i \pi } 
\Big( 0 \; , \;  \a_{\da}^{(0)}(\la) e(\la) \Big) 
\cdot  \bigg(I_{2}\, + \, \f{\la-\mu}{\la\mu \mf{b}^{\prime}(0) } 
\op{D} \bigg) 
\left(\ba{c}  \a_{\da}^{(0)}(\mu) e^2 (\mu)  \\ 0   \ea \right) 
\f{ \wh{B}^{\, - \veps_{\sg}}_{R}(\mu,s) e^{-1}(s) }{ \la - \mu }  \;. 
\end{multline}

Finally, for $\tau\in \{L,R\}$ and $\sg \in \big\{ +, R \big\}$, direct 
bounds based on Lemma \ref{Lemme bornes sur croissance via transfo Cauchy} 
lead to
\beq
 \Psi^{\tau \sg}_{R_\infty}(k,s) \; = \; \e{O}\bigg(  
\f{  \ex{-\eta w} \ln(1 + |k|)  }{ (1 + |k|) (1 + |s| ) }  \bigg) 
\qquad \e{with} \qquad 
(k,s) \in \big\{ \R + \i \veps_{\tau}  v \big\} \times 
\big\{ \R + \i \veps_{\sg} v \big\}    
\label{ecriture borne sur Psi R Infty tau=L,R et sigma=+,R}
\enq
where $\veps_{\sg}$ are as defined in 
\eqref{definition des signes veps sigma}.  For further convenience, 
we introduce the integral operators 
\beq
 \bs{\Psi}^{\tau \sg }_{R_\infty} \, : \, L^2(\R-\i \veps_{\sg} v) 
\tend L^2(\R + \i \veps_{\tau} v)\, ,  \quad \tau\in 
\{L,R\} \quad \e{and} \quad  \sg \in \big\{ +, R \big\}\; , 
\enq
acting with the integral kernels $ \Psi^{\tau \sg }_{R_\infty}(k,s)$. 
We remind that $\veps_{\sg}$ is as in 
\eqref{definition des signes veps sigma}.

 One may push the chain of transformations further for
$U^{R\sg}(k,s)$. Indeed, for $k \in \R-\i v$ and $\sg \in \{+, R\}$, 
contour deformations yield
\beq
e^{-1}(k) U^{R\sg}(k,s) e(s) \, = \,\Phi^{R \sg}(k,s) \, 
+ \, \wh{B}^{- \veps_{\sg} }_{R}(k,s)  \, 
+ \,   \mc{F}[ L^{(0)} ]( k )  \a_{\da}^{(0)}(k)
\Int{\R +  \i \eta }{} \hspace{-2mm} \f{\dd \mu }{ 2\i \pi }  
\f{ \wh{B}^{- \veps_{\sg} }_{R}(\mu,s)   }{ ( k - \mu  )  
\a_{\ua}^{(0)}(\mu)  } 
     \;. 
\enq
There, we have introduced 
\bem
  \Phi^{R \sg}(k,s) \, = \,   - e^{-1}(k)  
\Int{\R -  \i \f{v}{2} }{} \hspace{-2mm} \f{\dd \la}{2\i\pi} 
\Int{\R + 2 \i v }{} \hspace{-2mm} \f{\dd \mu }{ 2\i \pi }  
\f{   \mc{F}[ L^{(0)} ]( \la )  \a_{\da}^{(0)}(\la) e(\la)  }{ (k - \la )  
\a_{\ua}^{(0)}(\mu)  } 
  \Big( 1 \, + \, \f{\la-\mu}{ \la \mu  \mf{b}^{\prime}(0) }   \Big)
 \f{ \wh{B}^{- \veps_{\sg} }_{R}(\mu,s)   }{ \la - \mu    }   \\
\, + \,   \mc{F}[ L^{(0)} ]( k )  \f{ \a_{\da}^{(0)}(k) }{ k \cdot  
\mf{b}^{\prime}(0) }
\Int{\R +  \i \eta }{} \hspace{-2mm} \f{\dd \mu }{ 2\i \pi }  
\f{ \wh{B}^{-\veps_{\sg}}_{R}(\mu,s)   }{  \mu \a_{\ua}^{(0)}(\mu)  } \;. 
\end{multline}
Then, by observing that $ \mc{F}[ L^{(0)} ]( k )  \a_{\da}^{(0)}(k) \, 
= \,     \a_{\da}^{(0)}(k) \, - \,    \a_{\ua}^{(0)}(k) $ and setting 
\beq
M_R^{ - \pm }(k,s) \, = \,  -   
\Int{\R +  \i \eta }{} \hspace{-2mm} \f{\dd \mu }{ 2\i \pi }  
\f{ \a_{\da}^{(0)}(k) \cdot \wh{B}^{- \pm }_{R}(\mu,s)   }{ ( \mu- k  ) 
\cdot  \a_{\ua}^{(0)}(\mu)  }    
\label{noyau integral MR -sigma}
\enq
as well as 
\beqa
V_R^{-\pm }(k,s) & = &  \wh{B}^{-\pm}_{R}(k,s)  \, + \,    
\Int{\R +  \i \eta }{} \hspace{-2mm} \f{\dd \mu }{ 2\i \pi }  
\f{ \a_{\ua}^{(0)}(k) \cdot \wh{B}^{- \pm }_{R}(\mu,s)   }{ ( \mu- k  ) 
\cdot  \a_{\ua}^{(0)}(\mu)  }    \nonumber   \\ 
 & = &      \a_{\ua}^{(0)}(k) \Bigg\{
\Int{\R - \i \eta }{} \hspace{-2mm} \f{\dd \mu }{ 2\i \pi }  
\f{   \wh{B}^{- \pm }_{R}(\mu,s)   }{ ( \mu- k  ) \cdot  
\a_{\ua}^{(0)}(\mu)  }     \, + \, 
  \f{   \wh{B}^{- \pm }_{R}(0,s)   }{ k \cdot  \a_{0}^{(0)}  } \Bigg\} \;, 
\label{noyau integral VR - pm}
\eeqa
with $\a_{0}^{(0)}$ as defined in 
\eqref{definition constantes alpha zero et alpha zero tilde},  one gets 
that, for   $\sg \in \big\{ +, R \big\}$ and $\veps_{\sg}$ as in 
\eqref{definition des signes veps sigma},
\beq
e^{-1}(k) U^{R\sg}(k,s) e(s) \, = \,M_R^{- \veps_{\sg} }(k,s)   \, 
+ \,  V_R^{- \veps_{\sg} }(k,s)   \, + \,    \Phi^{R \sg} (k,s) \;. 
\enq
It is direct to check that $ \Phi^{R \sg }(k,s)$, $\sg \in 
\big\{ +, R \big\}$,   enjoy for some $c>0$ the bound
\beq
 \Phi^{R \sg}(k,s) \; = \;\e{O}\Bigg( 
\f{ \ex{- \f{v w}{2} } }{ (1+ |k|)(1+|s|) } \; + \;   
\f{ \ex{- c |k| } }{ w (1+|s|) }  \Bigg) \qquad \e{for} 
\qquad (k,s) \in \big\{ \R-\i v \big\} \times  
\big\{ \R + \i \veps_{\sg} v \big\} \;. 
\label{ecriture borne sur Phi R+ et RR}
\enq

Similarly as before, we introduce the integral operators 
\beq
\bs{\Phi}^{\tau \sg }_{R_\infty} \, : \, L^2(\R-\i \veps_{\sg} v) 
\tend L^2(\R + \i \veps_{\tau} v) \;, \quad \tau\in \{L,R\} \quad 
\e{and} \quad  \sg \in \big\{ +, R \big\}\;, 
\enq
for $\veps_{\sg}$ as in \eqref{definition des signes veps sigma}, 
whose integral kernels are $ \Phi^{\tau \sg }_{R_\infty}(k,s)$. 
We also introduce the integral operators 
\beq
\op{M}_{R}^{ -  \pm } \, : \, L^2(\R \pm \i _{\sg} v) 
\tend L^2(\R - \i  v) \qquad \e{and} \qquad  
\op{V}_{R}^{ -  \pm } \, : \, L^2(\R \pm \i _{\sg} v) \tend L^2(\R - \i  v)
\enq
having integral kernels $M_R^{-\pm }(k,s)$ and $V_R^{ - \pm }(k,s)$, 
respectively. 
 
All in all, we have established that
\beqa
\be^{L+}(k,s) &  = &  e(k)  \Phi^{L +} (k,s)e^{-1}(s) \, 
+ \, \Psi^{L + }_{R_\infty}(k,s) \;,   \\
\be^{R +}(k,s)  & = & e(k) \Big( M_R^{- + }(k,s)   \, 
+ \,  V_R^{- + }(k,s)   \, + \,    \Phi^{R +} (k,s) \Big) e^{-1}(s) \, 
+ \, \Psi^{R + }_{R_\infty}(k,s) \;. 
\eeqa

\subsubsection{Operator $\be^{0-}$}
 \label{SubSubsection operateur beta0-}
 
 In order to decompose the integral kernel $\be^{0-}(k,s)$ into its dominant and sub-dominant in $w$ parts, 
 by using the explicit expression for the leading resolvent \eqref{expression explicite noyau resolvent dominant}, one first computes 
\bem
\op{P}^{R}\Big[ \big(\e{id}- \op{R}_{\infty}\big)\big[e^{-1}(*)\wh{B}^{+-}_{L}(*,s)e(s)\big](\cdot) \Big](k) \, = \, 
- \lim_{\eps \tend 0^+} \Int{\R+\i v }{} \hspace{-2mm} \f{\dd \la}{2\i\pi} \f{  e^{-1}(\la)\wh{B}^{+-}_{L}(\la,s)e(s)    }{ k - \la +\i \eps } 
\; + \; \Int{\R+\i v }{} \hspace{-2mm} \f{\dd \la}{2\i\pi}  \f{ - \mc{F}[L^{(0)}](\la) }{ k - \la +\i \eps } \\
\times \Int{\R+\i v }{} \hspace{-2mm} \f{\dd \mu }{ 2\i \pi } \Big( - \{ \a_{\ua}^{(0)}(\la) e(\la) \}^{-1}  \; , \;  \a_{\da}^{(0)}(\la) e(\la) \Big) 
\cdot  \bigg(I_{2}\, + \, \f{\la-\mu}{\la\mu \mf{b}^{\prime}(0) } \op{D} \bigg) 
\left(\ba{c}  \a_{\da}^{(0)}(\mu)  \\ \{ \a_{\ua}^{(0)}(\mu) e^2 (\mu)  \}^{-1}    \ea \right) \f{ \wh{B}^{+-}_{L}(\mu,s) e(s) }{ \la - \mu } \\ 
  \, = \, e(k)\Phi^{R-}(k,s)e(s) \, + \, \Psi^{R-}_{R_\infty}(k,s) \;. 
\end{multline}

In this splitting, for $\sg \in \{-,L\}$, we set 
\bem
 \Psi^{R\sg}_{R_\infty}(k,s) \; = \;   - \Int{\R- 2 \i \eta }{} \hspace{-2mm} \f{\dd \la}{2\i\pi} \f{  e^{-1}(\la)\wh{B}^{+ \veps_{\sg} }_{L}(\la,s)e(s)    }{ k - \la  }  \\
\; - \; \Int{\R - 3 \i \frac{ \eta }{2} }{} \hspace{-2mm} \f{\dd \la}{2\i\pi}  \f{  \mc{F}[L^{(0)}](\la) }{ k - \la  } 
\;  \Int{\R- 2 \i \eta }{} \hspace{-2mm} \f{\dd \mu }{ 2\i \pi } \Big( - \{ \a_{\ua}^{(0)}(\la)e(\la) \}^{-1}  \; , \;0  \Big) 
\cdot  \bigg(I_{2}\, + \, \f{\la-\mu}{\la\mu \mf{b}^{\prime}(0) } \op{D} \bigg) 
\left(\ba{c}  \a_{\da}^{(0)}(\mu)  \\ \{ \a_{\ua}^{(0)}(\mu) e^2 (\mu)  \}^{-1}    \ea \right) \f{ \wh{B}^{ + \veps_{\sg} }_{L}(\mu,s) e(s) }{ \la - \mu }  \\ 
\; - \; \Int{\R - 2 \i v }{} \hspace{-2mm} \f{\dd \la}{2\i\pi}  \f{  \mc{F}[L^{(0)}](\la) }{ k - \la  } 
\Int{\R - 2 \i \eta }{} \hspace{-2mm} \f{\dd \mu }{ 2\i \pi } \Big( 0\; , \;  \a_{\da}^{(0)}(\la) e(\la) \Big) 
\cdot  \bigg(I_{2}\, + \, \f{\la-\mu}{\la\mu \mf{b}^{\prime}(0) } \op{D} \bigg) 
\left(\ba{c}  0 \\ \{ \a_{\ua}^{(0)}(\mu) e^2 (\mu)  \}^{-1}    \ea \right) \f{ \wh{B}^{ + \veps_{\sg} }_{L}(\mu,s) e(s) }{ \la - \mu }  \;. 
\end{multline}
Finally, upon using  
\beq
\big( -a \, , \, b \big) \, \op{D} \, \bigg( \ba{c} 1 \\ 0   \ea \bigg)  \, = \,  a + b \;, 
\enq
$\Phi^{R \sg }(k,s)$, $\sg \in \{-,L\}$,  may be recast as 
\bem
\Phi^{R\sg}(k,s) =  - e^{-1}(k) \Int{\R -  2 \i v }{}\hspace{-2mm}  
\f{\dd \la}{2\i\pi}  \f{   \mc{F}[L^{(0)}](\la) \, \a_{\da}^{(0)}(\la)  
e(\la)  }{ (k - \la) \la \,  \mf{b}^{\prime}(0)     } 
\times \Int{\R- \i v }{}  \hspace{-2mm} \f{\dd \mu }{ 2\i \pi \mu  }  
\a_{\da}^{(0)}(\mu)    \wh{B}^{ + \veps_{\sg} }_{L}(\mu,s)      \\
= - \Bigg\{ e^{-1}(k) \Int{\R -    \i \frac{ v }{2} }{}\hspace{-2mm}  
\f{\dd \la}{2\i\pi}  \f{   \mc{F}[L^{(0)}](\la) \, \a_{\da}^{(0)}(\la)  
e(\la)  }{ (k - \la) \la    }  
\, - \, \f{   \mc{F}[L^{(0)}](k) \, \a_{\da}^{(0)}(k)  }{ k }  \Bigg\}
 \Int{\R- \i v }{}  \hspace{-2mm} \f{\dd \mu }{ 2\i \pi   }  
\f{ \a_{\da}^{(0)}(\mu)    \wh{B}^{ + \veps_{\sg} }_{L}(\mu,s) }{ \mu  
\,  \mf{b}^{\prime}(0)  } \;.  
\end{multline}
It is direct to check that $ \Phi^{R \sg }(k,s)$, 
$\sg \in \big\{ -, L \big\}$,   enjoy for some $c>0$  the bound 
\beq
 \Phi^{R \sg}(k,s) \; = \;\e{O}\Bigg( \f{ \ex{- 
\f{v w}{2} } }{ (1+ |k|)(1+|s|) } \; + \;   
\f{ \ex{- c |k| } }{ w (1+|s|) }  \Bigg) \qquad \e{for} 
\qquad (k,s) \in \big\{ \R - \i v \big\} \times  
\big\{ \R + \i \veps_{\sg} v \big\} \;. 
\label{ecriture borne sur Phi R- et RL}
\enq
Furthermore, one also has 
\bem
\op{P}^{L}\Big[ \big(\e{id}- \op{R}_{\infty}\big)
\big[e^{-1}(*)\wh{B}^{+-}_{L}(*,s)e(s)\big](\cdot) \Big](k) \, = \, 
 \Int{\R+\i v }{} \hspace{-2mm} \f{\dd \la}{2\i\pi} 
\f{  e^{-1}(\la)\wh{B}^{+-}_{L}(\la,s)e(s)    }{ k - \la  - \i 0^+ } 
\; -   \Int{\R+\i v }{} \hspace{-2mm} \f{\dd \la}{2\i\pi}  
\f{ - \mc{F}[L^{(0)}](\la) }{ k - \la - \i 0^+ } \\
\times \Int{\R+\i v }{} \hspace{-2mm} \f{\dd \mu }{ 2\i \pi } 
\Big( - \{ \a_{\ua}^{(0)}(\la) e(\la) \}^{-1}  \; , 
\;  \a_{\da}^{(0)}(\la) e(\la) \Big) 
\cdot  \bigg(I_{2}\, + \, \f{\la-\mu}{\la\mu \mf{b}^{\prime}(0) } 
\op{D} \bigg) 
\left(\ba{c}  \a_{\da}^{(0)}(\mu)  \\ \{ \a_{\ua}^{(0)}(\mu) 
e^2 (\mu)  \}^{-1}    \ea \right) \f{ \wh{B}^{+-}_{L}(\mu,s) e(s) }{ 
\la - \mu } \\ 
\, = \, U^{L-}(k,s) \, + \, \Psi^{L-}_{R_\infty}(k,s) \;. 
\end{multline}
 There, $U^{L\sg}(k,s)$, for $\sg \in \{-,L\}$ and $\veps_{\sg}$ as 
in \eqref{definition des signes veps sigma}, is the integral kernel of 
the operator $\op{U}^{L\sg} : L^2( \R + 
\veps_{\sg} \i v) \tend L^2(\R + \i v)$:
\beq
U^{L\sg}(k,s) \, = \, e^{-1}(k)\wh{B}^{+ \veps_{\sg} }_{L}(k,s)e(s) \; 
-   \Int{\R + 2 \i v }{} \hspace{-2mm} \f{\dd \la}{2\i\pi}  
\Int{\R -  \i \eta }{} \hspace{-2mm} \f{\dd \mu }{ 2\i \pi }  
  \bigg( 1 \, - \, \f{\la-\mu}{\la\mu \mf{b}^{\prime}(0) }  \bigg)  
\f{ \mc{F}[L^{(0)}](\la) \a_{\da}^{(0)}(\mu) 
\wh{B}^{+\veps_{\sg}}_{L}(\mu,s) e(s) }{ ( k - \la)(\la - \mu)  
\, \a_{\ua}^{(0)}(\la) \,  e(\la) } \;. 
\enq
Moreover, we denote 
\bem
 \Psi^{L\sg}_{R_\infty}(k,s) \; = \;    \Int{\R - 2\i \eta }{} 
\hspace{-2mm} \f{\dd \la}{2\i\pi} 
\f{  e^{-1}(\la)\wh{B}^{+\veps_{\sg}}_{L}(\la,s)e(s)    }{ k - \la   }   \\
+  \Int{\R + 2 \i \eta }{} \hspace{-2mm} \f{\dd \la}{2\i\pi}  
\f{   \mc{F}[L^{(0)}](\la) }{ k - \la  } 
 \Int{\R - 2 \i \eta  }{} \hspace{-2mm} \f{\dd \mu }{ 2\i \pi } 
\Big( 0  \; , \;  \a_{\da}^{(0)}(\la) e(\la) \Big) 
\cdot  \bigg(I_{2}\, + \, \f{\la-\mu}{\la\mu \mf{b}^{\prime}(0) } 
\op{D} \bigg) 
\left(\ba{c}  \a_{\da}^{(0)}(\mu)  \\ \{ \a_{\ua}^{(0)}(\mu) 
e^2 (\mu)  \}^{-1}    \ea \right) 
\f{ \wh{B}^{+\veps_{\sg} }_{L}(\mu,s) e(s) }{ \la - \mu } \\ 
 +   \Int{\R + 2 \i v }{} \hspace{-2mm} \f{\dd \la}{2\i\pi}  
\f{   \mc{F}[L^{(0)}](\la) }{ k - \la   } 
 \Int{\R - 2\i \eta }{} \hspace{-2mm} \f{\dd \mu }{ 2\i \pi } 
\Big( - \{ \a_{\ua}^{(0)}(\la) e(\la) \}^{-1}  \; , \;  0 \Big) 
\cdot  \bigg(I_{2}\, + \, \f{\la-\mu}{\la\mu \mf{b}^{\prime}(0) } 
\op{D} \bigg) 
\left(\ba{c}  0  \\ \{ \a_{\ua}^{(0)}(\mu) e^2 (\mu)  \}^{-1}    
\ea \right) \f{ \wh{B}^{+ \veps_{\sg} }_{L}(\mu,s) e(s) }{ \la - \mu }  \;. 
\end{multline}

These representations entail, upon invoking 
Lemma \ref{Lemme bornes sur croissance via transfo Cauchy} that, 
for $\tau\in \{L,R\}$ and $\sg \in \{-,L\}$  with $\veps_{\sg}$  
as defined in \eqref{definition des signes veps sigma},  
one has
\beq
 \Psi^{\tau \sg  }_{R_\infty}(k,s) \; = \; \e{O}\bigg(  
\f{  \ex{-\eta w} \ln(1 + |k|)  }{ (1 + |k|) (1 + |s| ) }  \bigg) 
\qquad \e{with} \qquad 
(k,s) \in \big\{ \R + \i \veps_{\tau}   v \big\} \times 
\big\{ \R + \i \veps_{\sg} v \big\}  \;. 
\label{ecriture borne sur Psi R Infty tau=L,R et sigma=-,L}
\enq

 Transforming the integral kernel  $U^{L\sg}(k,s)$ further, one gets  
for $k \in \R+\i v$,
\beq
e(k)U^{L\sg}(k,s)e^{-1}(s) \, = \, \wh{B}^{+ 
\veps_{\sg} }_{L}(k,s) \; -   \f{   \mc{F}[L^{(0)}](k)   }{   
\a_{\ua}^{(0)}(k)   }\Int{\R -  \i \eta }{} \hspace{-2mm} 
\f{\dd \mu }{ 2\i \pi }  
 \f{  \a_{\da}^{(0)}(\mu) \wh{B}^{ + \veps_{\sg} }_{L}(\mu,s)   }{  
k - \mu   }  \,  + \, \Phi^{ L \sg }(k,s)  \;, 
\enq
 where 
\bem
 \Phi^{L  \sg }(k,s) \; = \;    -  e(k)  \Int{\R +   
\i  \f{v}{2} }{} \hspace{-2mm} \f{\dd \la}{2\i\pi}  \f{   }{  } 
\Int{\R -  \i \eta }{} \hspace{-2mm} \f{\dd \mu }{ 2\i \pi }  
  \bigg( 1 \, - \, \f{\la-\mu}{\la\mu \mf{b}^{\prime}(0) }  \bigg)  
\f{ \mc{F}[L^{(0)}](\la) \a_{\da}^{(0)}(\mu) \wh{B}^{+ 
\veps_{\sg} }_{L}(\mu,s)   }{ ( k - \la)(\la - \mu)  
\, \a_{\ua}^{(0)}(\la) \,  e(\la) }  \\ 
\, + \,    \f{   \mc{F}[L^{(0)}](k)   }{  k \,  
\a_{\ua}^{(0)}(k) \mf{b}^{\prime}(0)   }\Int{\R -  
\i \eta }{} \hspace{-2mm} \f{\dd \mu }{ 2\i \pi  \mu  }     
\a_{\da}^{(0)}(\mu) \wh{B}^{+\veps_{\sg} }_{L}(\mu,s)      \;. 
\end{multline}
It is direct to check that $ \Phi^{L \sg }(k,s)$, $\sg \in 
\big\{ -, L \big\}$,   enjoy for some $c>0$   the bound
\beq
 \Phi^{L  \sg}(k,s) \; = \;\e{O}\Bigg( \f{ \ex{- \f{v w}{2} } }{ (1
+ |k|)(1+|s|) } \; + \;   \f{ \ex{- c |k| } }{ w (1+|s|) }  
\Bigg) \qquad \e{for} \qquad (k,s) \in \big\{ \R + \i v \big\} 
\times  \big\{ \R + \i \veps_{\sg} v \big\} \;. 
\label{ecriture borne sur Phi L- et LL}
\enq
We now introduce the integral operators 
\beq
\bs{\Psi}^{\tau\sg}_{R_\infty}\; ,\;  \bs{\Phi}^{\tau\sg} : 
L^2( \R + \i \veps_{\sg} v) \tend L^2(\R + \i \veps_{\tau} v) \quad 
\e{with} \quad \sg \in \{-,L\}\;, \; \tau\in \{L,R\}
\enq
and $\veps_{\sg}$ as in \eqref{definition des signes veps sigma} 
with the integral kernels given by 
$\Psi^{\tau\sg}_{R_\infty}(k,s)$ and $\Phi^{\tau\sg}(k,s)$, respectively.

Finally, since $ \mc{F}[L^{(0)}](k)    \big\{   \a_{\ua}^{(0)}(k)  \big\}^{-1} \, = \,  \big\{   \a_{\ua}^{(0)}(k)  \big\}^{-1} \, - \,  \big\{   \a_{\da}^{(0)}(k)  \big\}^{-1} $, one infers that 
\beq
e(k)U^{L\sg}(k,s)e^{-1}(s) \, = \, M^{+\veps_{\sg}}_{L}(k,s) \, 
+ \, V^{+\veps_{\sg}}_{L}(k,s) \,  + \, \Phi^{L \sg }(k,s)  \;, 
\enq
where
\beqa
M^{+\pm}_{L}(k,s) & = &     \f{  1   }{   \a_{\ua}^{(0)}(k)   }
\Int{\R -  \i \eta }{} \hspace{-2mm} \f{\dd \mu }{ 2\i \pi }  
   \f{  \a_{\da}^{(0)}(\mu) \wh{B}^{+\pm}_{L}(\mu,s)   }{  \mu - k   } 
\label{noyau integral ML +sigma}
\eeqa
 while 
\beqa
V^{+\pm}_{L}(k,s)  & = &    \wh{B}^{+\pm}_{L}(k,s)\, - \,     
\Int{\R -  \i \eta }{} \hspace{-2mm} \f{\dd \mu }{ 2\i \pi }  
\f{  \a_{\da}^{(0)}(\mu) \, \wh{B}^{+\pm}_{L}(\mu,s)   }{  
\a_{\da}^{(0)}(k)  \cdot ( \mu - k )   }  \\
 & = &  \Bigg\{  \f{  \wt{\a}^{\,(0)}_0 \wh{B}^{+\pm}_{L}(0,s) }{ k   
\a_{\da}^{(0)}(k) }
 \, - \,     \Int{\R +   \i \eta }{} \hspace{-2mm} 
\f{\dd \mu }{ 2\i \pi }  \f{  \a_{\da}^{(0)}(\mu) \wh{B}^{+\pm}_{L}(\mu,s)   
}{  \a_{\da}^{(0)}(k)   ( \mu - k )   }  \Bigg\} \;. 
\label{noyau integral VL + pm}
\eeqa
 $\wt{\a}_{0}^{(0)}$ appearing above is defined 
in \eqref{definition constantes alpha zero et alpha zero tilde}.

We introduce the integral operators 
\beq
\op{M}_{L}^{ +  \pm } \, : \, L^2(\R \pm \i   v) \tend L^2(\R + \i  v) 
\qquad \e{and} \qquad  
\op{V}_{L}^{ +  \pm } \, : \, L^2(\R \pm \i  v) \tend L^2(\R + \i  v)
\enq
having integral kernels $M_L^{+\pm }(k,s)$ and 
$V_L^{ + \pm }(k,s)$, respectively. 
 
All in all, we have established that
\beqa
\be^{L-}(k,s) &  = &  e^{-1}(k) \Big( M_L^{+ - }(k,s)   \, 
+ \,  V_L^{+ -  }(k,s)   \, + \,    \Phi^{L -} (k,s) \Big) e(s) \, 
+ \, \Psi^{L - }_{R_\infty}(k,s)\;,   \\
\be^{R-}(k,s)  & = & e(k)  \Phi^{R -} (k,s)e^{-1}(s) \, 
+ \, \Psi^{R - }_{R_\infty}(k,s)   \;. 
\eeqa

\subsubsection{Operator $\be^{00}$} 
\label{SousSection operateur beta 00}

One starts by decomposing $\be^{00}\big[ u^0\big] = \big(\be^{00}
\op{P}^{R}\big)\big[ u^R \big]\, + \, \big(\be^{00}\op{P}^{L}\big)
\big[ u^L \big] $. Furthermore, 
one has
\beqa
\big( \be^{00}\op{P}^{R}\big) \big[ u^{R} \big](k) &= & 
- \lim_{\eps\tend 0^+} \Int{\R + \i v  }{} \hspace{-2mm}  
\f{\dd s \dd \mu}{2\i\pi} \f{ \be^{00}(k,s) u^{R}(\mu) }{ s-\mu 
+ \i \eps} \; = \;  \Int{ \R - \i  v }{} \hspace{-2mm} \dd s \, 
\be^{0R}(k,s)u^{R}(s) \\
\big( \be^{00}\op{P}^{L}\big) \big[ u^{L} \big](k) &= &   
\lim_{\eps\tend 0^+} \Int{\R + \i v  }{}  \hspace{-2mm}  
\f{\dd s \dd \mu}{2\i\pi} \f{ \be^{00}(k,s) u^{L}(\mu) }{ s-\mu 
- \i \eps } \; = \;  \Int{ \R + \i  v }{} \hspace{-2mm} \dd s \, 
\be^{0L}(k,s)u^{L}(s) \;, 
\eeqa
in which, for $k \in \R+\i v $ and $s \in \R+\i \veps_{\sg} v$, 
$\sg \in \{L,R\}$ and $\veps_{\sg}$ as in 
\eqref{definition des signes veps sigma},
\beq
\be^{0R}(k,s) \; = \; - \Int{ \R + 2 \i v  }{} \f{\dd \mu}{2\i\pi} 
\f{ \be^{00}(k,\mu)  }{ \mu - s  } \qquad \e{and} \qquad 
\be^{0L}(k,s) \; = \;  \Int{ \R - 2 \i v  }{} \f{\dd \mu}{2\i\pi} 
\f{ \be^{00}(k,\mu)  }{ \mu - s  }  \;. 
\enq
This representation allows one to identify the dominant contribution to 
$\be^{0\sg}(k,s)$ in that a contour deformation entails that 
\beqa
\be^{0L}(k,s) & = & \Big(\e{id}-\op{R}_{\infty}\Big)
\Big[e^{-1}(*)\wh{B}^{++}_L(*,s) e(s) \Big](k) \, 
+ \, \Psi^{0L}_1(k,s) 
\label{ecriture projection a droite type L de beta 00} \\
\be^{0R}(k,s) & = & \Big(\e{id}-\op{R}_{\infty}\Big)
\Big[e(*)\wh{B}^{--}_R(*,s) e^{-1}(s) \Big](k) \, 
+ \, \Psi^{0R}_1(k,s) 
\label{ecriture projection a droite type R de beta 00}
\eeqa
in which 
\beq
\Psi^{0L}_1(k,s) \, = \,    \Int{ \R + 2\i \eta  }{} \hspace{-2mm} 
\f{ \dd \mu }{ 2\i\pi } \f{ \Big(\e{id}-\op{R}_{\infty}\Big)
\Big[e^{-1}(*)\wh{B}^{++}_L(*,\mu) e(\mu) \Big](k)   }{ \mu - s  }
 \, + \,  \Int{ \R - 2\i \eta  }{} \hspace{-2mm} \f{ \dd \mu }{ 2\i\pi } 
\f{ \Big(\e{id}-\op{R}_{\infty}\Big)\Big[e(*)\wh{B}^{--}_R(*,\mu) 
e^{-1}(\mu) \Big](k)  }{ \mu - s  }
\enq
while 
\beq
\Psi^{0R}_1(k,s) \, = \,   - \Int{ \R + 2\i \eta  }{} \hspace{-2mm} 
\f{ \dd \mu }{ 2\i\pi } \f{ \Big(\e{id}-\op{R}_{\infty}\Big)
\Big[e^{-1}(*)\wh{B}^{++}_L(*,\mu) e(\mu) \Big](k)   }{ \mu - s  }
 \, - \,  \Int{ \R - 2\i \eta  }{} \hspace{-2mm} \f{ \dd \mu }{ 2\i\pi } 
\f{ \Big(\e{id}-\op{R}_{\infty}\Big)\Big[e(*)\wh{B}^{--}_R(*,\mu) 
e^{-1}(\mu) \Big](k)  }{ \mu - s  }\,.
\enq
Due to \eqref{ecriture borne sur noyau resolvent dominant}, for $\mu \in  
\R \pm  2\i \eta$ and $|\Im(k)|<2v$ with $k$ uniformly away from $0$, 
one has  
\beq
 \Big(\e{id}-\op{R}_{\infty}\Big)\Big[e^{\mp 1}(*)
\wh{B}^{\pm \pm }_{L/R}(*,\mu)  \Big](k)    \; = \; \e{O}  
\Bigg(  \f{ \ex{5 v w} }{ ( 1+|k|)(1+|\mu|) }    \Bigg)  \;. 
\enq
By virtue of Lemma 
\ref{Lemme bornes sur croissance via transfo Cauchy} one infers  that, 
uniformly in $(k,s)$ such that $|\Im(k)|<2v$, $|\Im(s)|<2v$ and $k$ 
uniformly away from $0$
\beq
\Psi^{0\sg}_1(k,s) \, = \, \e{O}  \Bigg(  \f{ \ex{- \eta w} \, 
\ln (1+|s|)  }{ ( 1+|k|)(1+|s|) }    \Bigg)  \,, \quad \sg \in \{L,R\} \; 
\enq
 and thus, uniformly in $(k,s)$ such that $|\Im(k)|\leq v$ and $|\Im(s)| 
\leq v$, 
\beq
\Psi^{\tau \sg}_1(k,s) \, = \, \op{P}^{\tau}\Big[ \Psi^{0\sg}_1(*,s)
\Big](k) \, = \, \e{O}  \Bigg(  \f{  \ex{- \eta w} \,  \ln (1+|s|)   
\,  \ln (1+|k|)  }{ ( 1+|k|)(1+|s|) }    \Bigg)  \,, \quad \tau, 
\sg \in \{L,R\}  \; .
\label{ecriture borne sur Psi 1 tau=L,R et sigma=L,R}
\enq
As earlier on, we introduce the integral operators characterised by 
the above integral kernels  
\beq
\bs{\Psi}^{\tau\sg}_{1}\; ,\;  \bs{\Phi}^{\tau\sg} : 
L^2( \R + \i \veps_{\sg} v) \tend L^2(\R + \i \veps_{\tau} v) \quad 
\e{with} \quad \sg\; ,  \tau \in \{L,R\}
\enq
and $\veps_{\sg}$ as in \eqref{definition des signes veps sigma}.

The $R$ and L projections of the first terms appearing in 
\eqref{ecriture projection a droite type L de beta 00}-\eqref{ecriture projection a droite type R de beta 00} 
can be computed exactly as in  Subsections 
\ref{SubSubsection operateur beta0+}-\ref{SubSubsection operateur beta0-}. 
All in all, one gets that
\bem
\left( \ba{cc} \be^{LL}(k,s)  &  \be^{LR}(k,s)  \vspace{1mm}  \\  
\be^{RL}(k,s)  &  \be^{RR}(k,s)   \ea \right) \; = \; 
\left( \ba{cc} \op{P}^{L}\big[\be^{0L}(*,s)\big](k)  &  
\op{P}^{L}\big[\be^{0R}(*,s)\big](k)  \vspace{1mm}   \\  
\op{P}^{R}\big[\be^{0L}(*,s)\big](k)  &  
\op{P}^{R}\big[\be^{0R}(*,s)\big](k)    \ea \right)  \\
\; = \; 
\left( \ba{c}        e^{-1}(k)\Big[M_{L}^{++} + V_{L}^{++} 
+ \Phi^{LL} \Big](k,s) e(s) \, + \,  \Psi^{LL}_{1}(k,s)   \, 
+ \,  \Psi^{LL}_{R_{\infty}}(k,s)  \vspace{1mm} \\
		         e(k) \Phi^{RL}(k,s) e(s) \, 
+ \,  \Psi^{RL}_{1}(k,s)   \, + \,  \Psi^{RL}_{R_{\infty}}(k,s)   
\ea \right.  \\
\left. \ba{c}       e^{-1}(k) \Phi^{LR}(k,s) e^{-1}(s) \, 
+ \,  \Psi^{LR}_{1}(k,s)   \, + \,  \Psi^{LR}_{R_{\infty}}(k,s)    
\vspace{1mm} \\
		          e(k)\Big[M_{R}^{--} + V_{R}^{--} 
+ \Phi^{RR} \Big](k,s) e^{-1}(s) \, + \,  \Psi^{RR}_{1}(k,s)   \, 
+ \,  \Psi^{RR}_{R_{\infty}}(k,s)  \ea \right)  \;. 
\end{multline}

\subsubsection{Matrix equation arising in the $0$ sector}
\label{SousSection matrix equation in 0 sector}

In order to write down the final form of the equation associated with
the $R$ and $L$ projections of the $0$-sector, by putting together 
the previous results, one first 
obtains the relation
\bem
\hspace{-1cm}\left( \ba{ccc} - \op{e}^{-1} \big[ \op{M}_L^{+-}
  +\op{V}_L^{+-}+\bs{\Phi}^{L-} \big] \, \op{e}
  - \bs{\Psi}_{\mf{B}}^{L-} -  \bs{\Psi}_{\de \be }^{L-}
  -  \bs{\Psi}_{ R_{\infty} }^{L-}  & ; & 
  \e{id} - \op{e}^{-1} \big[ \op{M}_L^{++}+\op{V}_L^{++}
  +\bs{\Phi}^{LL} \big] \, \op{e}  - \bs{\Psi}_{\mf{B}}^{LL}
  -  \bs{\Psi}_{\de \be }^{LL} -  \bs{\Psi}_{ R_{\infty} }^{LL}
  -  \bs{\Psi}_{ 1 }^{LL}   \vspace{1mm}   \\
  - \op{e} \,  \bs{\Phi}^{R-} \op{e}  - \bs{\Psi}_{\mf{B}}^{R-}
  -  \bs{\Psi}_{\de \be }^{R-} -  \bs{\Psi}_{ R_{\infty} }^{R-}   & ; & 
  - \op{e}\,  \bs{\Phi}^{RL}\, \op{e}  - \bs{\Psi}_{\mf{B}}^{RL}
  -  \bs{\Psi}_{\de \be }^{RL} -  \bs{\Psi}_{ R_{\infty} }^{RL}
  -  \bs{\Psi}_{ 1 }^{RL}     \ea \right. \vspace{4mm} \\
\hspace{-1cm} \left. \ba{ccc}   - \op{e}^{-1} \,  \bs{\Phi}^{LR} \op{e}^{-1}
  - \bs{\Psi}_{\mf{B}}^{LR} -  \bs{\Psi}_{\de \be }^{LR}
  -  \bs{\Psi}_{ R_{\infty} }^{LR} -  \bs{\Psi}_{ 1 }^{LR}     & ; & 
  - \op{e}^{-1} \,  \bs{\Phi}^{L+} \op{e}^{-1}    - \bs{\Psi}_{\mf{B}}^{L+}
  -  \bs{\Psi}_{\de \be }^{L+} -  \bs{\Psi}_{ R_{\infty} }^{L+}
  \vspace{1mm}\\
  \e{id} - \op{e}\, \big[ \op{M}_R^{--}+\op{V}_R^{--}+\bs{\Phi}^{RR} \big]
  \, \op{e}^{-1}   - \bs{\Psi}_{\mf{B}}^{RR} -  \bs{\Psi}_{\de \be }^{RR}
  -  \bs{\Psi}_{ R_{\infty} }^{RR}  -  \bs{\Psi}_{ 1 }^{RR}     & ; &
  - \op{e}\, \big[ \op{M}_R^{-+}+\op{V}_R^{-+}+\bs{\Phi}^{R+} \big] \,
  \op{e}^{-1}   - \bs{\Psi}_{\mf{B}}^{R+} -  \bs{\Psi}_{\de \be }^{R+}
  -  \bs{\Psi}_{ R_{\infty} }^{R+}  \ea \right) 
				  \big[\bs{u}\big]   \\
				  \; = \; \left( \ba{cc} \mf{d}^{L}_{w}[h^{0}]
                                    \\ \mf{d}^{R}_{w}[h^{0}]  \ea \right)
                                  \hspace{1cm} \e{where} \hspace{1cm}
                                  \mf{d}^{\a}_{w}[h^{0}]  \, = \, \op{P}^{\a}\Big[ \big(\e{id}-\op{R}
                                  \big)[h^{0}] \Big] \;. 
\label{equation de decomposition pour u zero}
\end{multline}
We remind that $\bs{u}$ appearing above 
was introduced in 
\eqref{ecriture decomposition plus fine de lespace de Hilbert}. Then, one 
defines the operators 
\bem
\left( \ba{cccc}  \bs{\Om}^{L-}  &   \bs{\Om}^{LL}  &    \bs{\Om}^{LR}  
& \bs{\Om}^{L+}  \\ 
               \bs{\Om}^{R-}  &   \bs{\Om}^{RL}  &    \bs{\Om}^{RR}  
& \bs{\Om}^{R+}     \ea  \right)  \\
\; = \;   
\left( \ba{ccc}  \bs{\Phi}^{L-}  +  \op{e}\,  \big[ \bs{\Psi}_{\mf{B}}^{L-}  
+  \bs{\Psi}_{\de \be }^{L-}  +   \bs{\Psi}_{ R_{\infty} }^{L-}  \big] \, 
\op{e}^{-1}  & ; & 
 \bs{\Phi}^{LL}  +  \op{e}\,  \big[ \bs{\Psi}_{\mf{B}}^{LL}  
+  \bs{\Psi}_{\de \be }^{LL}  +   \bs{\Psi}_{ R_{\infty} }^{LL} 
+  \bs{\Psi}_{ 1 }^{LL}    \big] \, \op{e}^{-1}  \vspace{1mm} \\
 \bs{\Phi}^{R-}  +  \op{e}^{-1}\,  \big[ \bs{\Psi}_{\mf{B}}^{R-}  
+  \bs{\Psi}_{\de \be }^{R-}  +   \bs{\Psi}_{ R_{\infty} }^{R-}  \big] 
\, \op{e}^{-1}  & ; & 
 \bs{\Phi}^{RL}  +  \op{e}^{-1}\,  \big[ \bs{\Psi}_{\mf{B}}^{RL}  
+  \bs{\Psi}_{\de \be }^{RL}  +   \bs{\Psi}_{ R_{\infty} }^{RL} 
+  \bs{\Psi}_{ 1 }^{RL}    \big] \, \op{e}^{-1}  \ea \right. \vspace{2mm} \\
 \left. \ba{ccc}  \bs{\Phi}^{LR}  +  \op{e}\,  \big[ \bs{\Psi}_{\mf{B}}^{LR}  
+  \bs{\Psi}_{\de \be }^{LR}  +   \bs{\Psi}_{ R_{\infty} }^{LR} 
+  \bs{\Psi}_{ 1 }^{LR}    \big] \, \op{e}  & ; & 
\bs{\Phi}^{L+}  +  \op{e}\,  \big[ \bs{\Psi}_{\mf{B}}^{L+}  
+  \bs{\Psi}_{\de \be }^{L+}  +   \bs{\Psi}_{ R_{\infty} }^{L+}  \big] \, 
\op{e} \vspace{1mm} \\
 \bs{\Phi}^{RR}  +  \op{e}^{-1}\,  \big[ \bs{\Psi}_{\mf{B}}^{RR}  
+  \bs{\Psi}_{\de \be }^{RR}  +   \bs{\Psi}_{ R_{\infty} }^{RR} 
+  \bs{\Psi}_{ 1 }^{RR}    \big] \, \op{e}   & ; & 
  \bs{\Phi}^{R+}  +  \op{e}^{-1}\,  \big[ \bs{\Psi}_{\mf{B}}^{R+}  
+  \bs{\Psi}_{\de \be }^{R+}  +   \bs{\Psi}_{ R_{\infty} }^{R+}  \big] \, 
\op{e}
\ea \right) \, .
\label{definition de la partie centrale L et R de operateur Omega}
\end{multline}
 This allows one to rewrite 
\eqref{equation de decomposition pour u zero} 
in the form 
\bem
\left( \ba{ccc}  - \op{e}^{-1} \big[ \op{M}_L^{+-}+\op{V}_L^{+-}
+\bs{\Om}^{L-} \big] \, \op{e} & ;   &  \e{id} - \op{e}^{-1} 
\big[ \op{M}_L^{++}+\op{V}_L^{++}+\bs{\Om}^{LL} \big] \, \op{e}   \\ 
 \op{e} \,  \bs{\Om}^{R-} \op{e} & ; &  \op{e} \,  \bs{\Om}^{RL} \op{e}  
\ea \right.  \vspace{2mm}\\
\left. \ba{ccc}
\op{e}^{-1} \,  \bs{\Om}^{LR} \op{e}^{-1}   & ; & \op{e}^{-1} \,  
\bs{\Om}^{L+} \op{e}^{-1}    \vspace{1mm}  \\ 
  \e{id} - \op{e}\, \big[ \op{M}_R^{--}+\op{V}_R^{--}+\bs{\Om}^{RR} \big] 
\, \op{e}^{-1}      & ; &    - \op{e}\, \big[ \op{M}_R^{-+}
+\op{V}_R^{-+}+\bs{\Om}^{R+} \big] \, \op{e}^{-1}      \ea \right) 
\big[ \bs{u} \big]    \; = \; \left( \ba{cc} \mf{d}^{L}_{w}[h^{0}]   
\\ \mf{d}^{R}_{w}[h^{0}]  \ea \right)   \,, 
\label{equation matricielle pour u zero}
\end{multline}
that 
is best suited for the later handling.

\subsection{Final form of the integral equation}
\label{SousSection final form of integral equation}

By 
gathering the results of Subsections   
\ref{SubSubsection operateur beta0+}, 
\ref{SubSubsection operateur beta0-}, 
\ref{SousSection operateur beta 00} and 
\ref{SousSection matrix equation in 0 sector}, in particular the equations 
\eqref{equation matricielle pour u plus}, 
\eqref{equation matricielle pour u moins} and 
\eqref{equation matricielle pour u zero},  one may recast
the original system 
\eqref{ecriture forme Fourier de f moins via H moins}-\eqref{ecriture forme Fourier de f plus via H plus}  
into the following form 
\beq
\bs{\op{E}} \,  \Big( \e{id}\, - \, \op{O} \, - \, \bs{\Om} \Big) \, 
\bs{\op{E}}^{-1} \big[ \bs{u} \big] \; = \;  \bs{\mf{d}}_{w}[h] \; , 
\label{equation integrale sous forme compact bien separee finale}
\enq
in which $\bs{\op{E}}=\e{Diag}\big( \op{e}^{-1}\, , \,   \op{e}^{-1}\, , 
\,  \op{e}\, , \,   \op{e} \big)$, 
$\bs{\mf{d}}_{w}[h] \, = \, \Big(   \mf{d}^{-}_{w}[h^-](k)  \, , \, 
\mf{d}^{L}_{w}[h^0](k) \, , \, \mf{d}^{R}_{w}[h^0](k)  \, , \, 
\mf{d}^{+}_w[h^+](k) \Big)^{\op{t}}$, while 
\beq
\bs{\Om} \, = \, \left( \ba{cccc}  0 &   \bs{\Om}^{- L } &    
\bs{\Om}^{- R}  &   \bs{\Om}^{- + } \\
 \bs{\Om}^{ L -  } &   \bs{\Om}^{L L } &    \bs{\Om}^{L R}  &   
\bs{\Om}^{L + } \\
\bs{\Om}^{ R -  } &   \bs{\Om}^{R L } &    \bs{\Om}^{R R}  &   
\bs{\Om}^{R + } \\
\bs{\Om}^{+-} &  \bs{\Om}^{+L} & \bs{\Om}^{+R} & 0  \ea \right) \,. 
\enq
Finally, one has
\beq
\op{O} \; = \; \left( \ba{cc}   \op{O}_{L} &  \bs{0} \\  \bs{0} & 
\op{O}_R \ea \right)
\enq
in which $\op{O}_{L/R}$ are integral operators on $L^2(\R 
- \i v) \oplus L^2(\R + \i v) $ having the block-matrix form 
\beq
 \op{O}_{L} \; = \; \left( \ba{cc}   \op{M}_{L}^{--} &  \op{M}_{L}^{-+} 
\\   \op{M}_{L}^{+-} + \op{V}_{L}^{+-} &  \op{M}_{L}^{++} 
+  \op{V}_{L}^{++} \ea \right)
\quad \e{and} \quad 
 \op{O}_{R} \; = \; \left( \ba{cc}   \op{M}_{R}^{--} + \op{V}_{R}^{--} 
&  \op{M}_{R}^{-+} +  \op{V}_{R}^{-+} \\   \op{M}_{R}^{+-} &  
\op{M}_{R}^{++} \ea \right) \;. 
\enq

\begin{prop}
\label{Proposition invertibilite operateur Id - O - Omega}  
  
  The operators $ \op{O}$ and $\bs{\Om}$ appearing in 
\eqref{equation integrale sous forme compact bien separee finale} are 
compact Hilbert-Schmidt operators on 
\beq
L^2(\R-\i v) \oplus L^2(\R+\i v) \oplus L^2(\R-\i v) \oplus L^2(\R+\i v) \, .
\enq
Moreover, the operator  $\e{id} \, - \, \op{O} \, - \, \bs{\Om}$  is 
invertible uniformly in $w$ large enough.  Its inverse 
is equal to $\e{id}+\bs{\De}$, 
where the operator $\bs{\De}\,$  
has the block-matrix form 
\beq
\bs{\De} \, = \, \left( \ba{cccc}  \bs{\De}^{--} &   \bs{\De}^{- L } &    
\bs{\De}^{- R}  &   \bs{\De}^{- + } \\
 \bs{\De}^{ L -  } &   \bs{\De}^{L L } &    \bs{\De}^{L R}  &   
\bs{\De}^{L + } \\
\bs{\De}^{ R -  } &   \bs{\De}^{R L } &    \bs{\De}^{R R}  &   
\bs{\De}^{R + } \\
\bs{\De}^{+-} &  \bs{\De}^{+L} & \bs{\De}^{+R} & \bs{\De}^{++}  
\ea \right) \,.  
\enq
The integral kernels associated with this block decomposition enjoy 
the uniform in $w$ bounds:  
\beq
\De^{\sg \tau}(k,s) \, = \, \e{O}\Bigg(  \f{  \ln(1+|k|) 
\cdot \ln(1+|s|)  }{   (1+|k|) \cdot (1+|s|)  } \Bigg) \qquad for
\qquad (k,s)\in \{\R + \i \veps_{\sg} v\} \times \{\R 
+ \i \veps_{\tau} v\}\; ,
\label{ecriture bornes sur entree noyau resolvent final}
\enq
with $\{\sg, \tau\} \in \{\pm, R, L\}$ and $\veps_{\sg}$ is as introduced 
in \eqref{definition des signes veps sigma}.  

\end{prop}

\Proof 

Using Proposition \ref{Proposition Caractere Hilbert Schmidt operateur B} 
and Lemma \ref{Lemme bornes sur croissance via transfo Cauchy} one may bound 
the integral kernels $M^{\sg \eps }_{L/R}$  and $V^{\sg \eps }_{L/R}$ of 
the operators $\op{M}_{L/R}$ and $\op{V}_{L/R}$ building up the operator 
$\op{O}$ as
\beq
M^{\sg \eps }_{L/R}(k,s) \; = \; \e{O}\Bigg( \f{ \ln (1+|k|)  }{ (1+|k|) 
\cdot (1+|s|)  } \Bigg)  \qquad \e{for} \qquad (k,s) \in \{\R + \i \sg v \}
\times  \{\R + \i \eps v \} \quad \e{with} \quad \sg, \eps\in \{\pm \} \;, 
\label{ecriture borne sup sur M L et R}
\enq
and 
\beqa
V^{+ \eps }_{L}(k,s) &  =  & \e{O}\Bigg( \f{ \ln (1+|k|)  }{ (1+|k|) 
\cdot (1+|s|)  } \Bigg)  \qquad \e{for} \qquad (k,s) \in \{\R + \i  v \}
\times  \{\R + \i \eps v \}  \quad \e{with} \quad \eps\in \{\pm \} \;,  
\vspace{3mm} 
\label{ecriture borne sup sur V L} \\
V^{- \eps }_{R}(k,s) & = &  \e{O}\Bigg( \f{ \ln (1+|k|)  }{ (1+|k|) 
\cdot (1+|s|)  } \Bigg)  \qquad \e{for} \qquad (k,s) \in \{\R - \i  v \}
\times  \{\R + \i \eps v \}  \quad \e{with} \quad \eps\in \{\pm \} \;. \;  
\label{ecriture borne sup sur V R}
\eeqa
Also, upon recalling the definition of  
various block operators building up $\,\bs{\Om}$, 
\eqref{definition Omega -L -R et -+}, \eqref{definition Omega +- +L et +R} 
and \eqref{definition de la partie centrale L et R de operateur Omega}, 
one may infer from the bounds \eqref{ecriture borne sur Psi -R}, 
\eqref{ecriture borne sur Psi -L ou -+}, 
\eqref{ecriture borne sur Psi +L}, \eqref{ecriture borne sur Psi +- et +R}, 
\eqref{ecriture borne sur Psi mathfrakB sigma tau},
\eqref{ecriture borne sur Psi deltaBeta sigma tau}, 
\eqref{ecriture borne sur Phi L+ et LR}, 
\eqref{ecriture borne sur Psi R Infty tau=L,R et sigma=+,R}, 
\eqref{ecriture borne sur Phi R+ et RR},
\eqref{ecriture borne sur Phi R- et RL}, 
\eqref{ecriture borne sur Psi R Infty tau=L,R et sigma=-,L}, 
\eqref{ecriture borne sur Phi L- et LL}, 
\eqref{ecriture borne sur Psi 1 tau=L,R et sigma=L,R}  
the relations 
\beq
\Om^{\sg \tau }(k,s) \; = \; \e{O}\Bigg( \f{ \ln (1+|k|)  
\ln (1+|s|)  }{w \cdot  (1+|k|) \cdot (1+|s|)  } \Bigg) 
\qquad \e{for} \qquad (k,s) \in \{\R + \i \veps_{\sg} v \}\times  
\{\R + \i \veps_{\tau} v \} \quad \e{with} \quad \sg, \tau \in 
\{\pm, R, L \} \;,
\label{ecriture borne sup sur Omega kernel}
\enq
where $\veps_{\sg}$ is as introduced in 
\eqref{definition des signes veps sigma}.

Put together, these pieces of information ensure that the operators 
$\op{O}$ and $\bs{\Om}$ appearing in 
\eqref{equation integrale sous forme compact bien separee finale} 
are compact, Hilbert-Schmidt operators having  
trace and leading to well-defined Fredholm determinants. 

Since the  
Fredholm $2$-determinants are continuous in the Hilbert-Schmidt norm 
\cite{GohbergGoldbargKrupnikTracesAndDeterminants} in the sense that 
for any  $C^{\prime}>0$ there exists $C>0$ such that  
\beq
\Big| \e{det}_2\big[\e{id}-\op{A} \big] \,  - \, \e{det}_2\big[\e{id}
-\op{B}\big] \Big| \, \leq \; C \cdot \norm{ \op{A}-\op{B} }_{\e{HS} } 
\qquad \e{for} \; \e{any} \qquad 
 \norm{  \op{A} }_{\e{HS} } +  \norm{  \op{B} }_{\e{HS} }<C^{\prime} \;, 
\enq
and since $ \e{det}_2\big[\e{id}-\op{O} \big] \, =  \, \e{det}\big[\e{id}
-\op{O} \big]\ex{\e{tr}[ \op{O}] }$, it is enough to show that
 $\e{det}\big[\e{id}-\op{O} \big]$ is away from $0$ so as to have 
the uniform in $w$ invertibility of $ \e{id}\, - \, \op{O} \, - \, \bs{\Om}$. 
For this purpose, we compute the Fredholm determinant of $\e{id}-\op{O}$ 
in terms of other determinants which we know to be non-vanishing. 

Clearly, 
one has 
\beq
\e{det}\big[\e{id}-\op{O} \big] \ = \, \e{det}\big[\e{id}-\op{O}_L \big] \cdot \e{det}\big[\e{id}-\op{O}_R \big] \;.  
\enq
In order to estimate those determinants, one first observes the block 
operator factorisation 
\beqa
\e{id}-\op{O}_L & = & \left( \ba{cc}  \e{id} - \op{M}_{L}^{--} &  
-\op{M}_{L}^{-+} \\   -\op{M}_{L}^{+-} - \op{V}_{L}^{+-} &  \e{id} 
-\op{M}_{L}^{++} -  \op{V}_{L}^{++} \ea \right) \\
 &= & \left( \ba{cc}  \e{id} - \op{M}_{L}^{--} &  -\op{M}_{L}^{-+} \\   
-\op{M}_{L}^{+-}   &  \e{id} -\op{M}_{L}^{++}  \ea \right) \cdot 
 \left( \ba{cc}  \e{id}  &  0 \\    - \op{V}_{L}^{+-} &  \e{id}   
-  \op{V}_{L}^{++} \ea \right)  \,. 
\eeqa
The latter is a consequence of the identities  
\beq
\op{M}_{L}^{\sg +} \cdot \op{V}_{L}^{+ \tau } \, = \, 0 \qquad \e{for} \, 
 \qquad \sg, \tau \in \{\pm\} \;. 
\enq
established in Lemma \ref{Lemme identite operatoire}  
below.

Now, it follows from Proposition 
\ref{Proposition solubilite unique RHP non local} 
that $\det[\e{id}-\op{M}_{L}]\, \not= \, 0$ where 
\beq
\op{M}_{L/R} \; = \; \left( \ba{cc}    \op{M}_{L/R}^{--} &  
\op{M}_{L/R}^{-+} \\   \op{M}_{L/R}^{+-}   &  \op{M}_{L/R}^{++}  
\ea \right) \,. 
\enq
Moreover,  owing to the identity $\op{V}_{L}^{+ +} \cdot \op{V}_{L}^{+ + }=0$ 
and the Plemelj-Smithies expansion for the determinant 
\cite{GohbergGoldbargKrupnikTracesAndDeterminants}: 
\beq
\e{det}\Big[ \e{id} + \op{A} \Big] \; = \; \sul{n \geq 0}{} \f{1}{n!} 
\det_n\left(\ba{cccccc} \e{tr}\big[ \op{A} \big]     &   
n-1 	
		& 		0 		&		 
\dots		   & 0 \\ 
			\e{tr}\big[ \op{A}^2 \big] & \e{tr}\big[ \op{A} \big]
 	&	n-2        &   \dots     & 0                            
     \\  
			\vdots & \vdots &  \vdots & \ddots &	\vdots    
\\ 	
			  \e{tr}\big[ \op{A}^{n-1} \big] & 
\e{tr}\big[ \op{A}^{n-2} \big]  &  \e{tr}\big[ \op{A}^{n-3} \big] & \cdots 
&   1  \\ 
			\e{tr}\big[ \op{A}^{n} \big] & 
\e{tr}\big[ \op{A}^{n-1} \big]  &  \e{tr}\big[ \op{A}^{n-2} \big] 
& \cdots &   \e{tr}\big[ \op{A} \big] 	  \ea \right) \; , 
\enq
one has
\beq
 \e{det}\left( \ba{cc}  \e{id}  &  0 \\    - \op{V}_{L}^{+-} &  \e{id}   
-  \op{V}_{L}^{++} \ea \right)  \; = \; 1 \, - \, 
\e{tr}\big[ \op{V}_{L}^{++}  \big] \;. 
\enq
The latter trace can be
shown to vanish by deforming the integration contours to $- \i \infty$ in 
the expression below 
\beq
 \e{tr}\big[ \op{V}_{L}^{++}  \big]  \; = \; \Int{\R + \i v }{} 
\hspace{-2mm} \dd k \,   \Bigg\{  \f{  \wt{\a}^{\,(0)}_0 
\wh{B}^{+ + }_{L}(0,k) }{ k   \a_{\da}^{(0)}(k) }
 \, - \,     \Int{\R +   \i \eta }{} \hspace{-2mm} \f{\dd \mu }{ 2\i \pi }  
\f{  \a_{\da}^{(0)}(\mu) \wh{B}^{+ + }_{L}(\mu, k )   }{  \a_{\da}^{(0)}(k)
   ( \mu - k )   }  \Bigg\} \,, 
\enq
since the integrand is analytic in $\mathbb{H}_{2v}^{-}=\mathbb{H}^{-}+2\i v$ and goes to zero as $\e{o}(k^{-3/2})$ at $\infty$ in that domain. 

\vspace{2mm}

Thus, all in all, one   
infers that $\det\big[ \e{id} - \op{O}_L \big]  \, = \,
\det[\e{id}-\op{M}_{L}]\, \not= \, 0$.  

\vspace{2mm}

\noindent Very analogous considerations lead to 
\beqa
\e{id}-\op{O}_R & = & \left( \ba{cc}  \e{id} - \op{M}_{R}^{--} 
- \op{V}_{R}^{--} &  -\op{M}_{R}^{-+}  - \op{V}_{R}^{-+} \\   
-\op{M}_{R}^{+-}  &  \e{id} -\op{M}_{R}^{++}  \ea \right) \\
 &= & \left( \ba{cc}  \e{id} - \op{M}_{R}^{--} &  -\op{M}_{R}^{-+} \\   
-\op{M}_{R}^{+-}   &  \e{id} -\op{M}_{R}^{++}  \ea \right) \cdot 
 \left( \ba{cc}  \e{id} - \op{V}_{R}^{--} &  - \op{V}_{R}^{-+} \\ 0 &  
\e{id}    \ea \right) 
\eeqa
as follows from the identities 
\beq
\op{M}_{R}^{\sg -} \cdot \op{V}_{R}^{- \tau } \, = \, 0 \qquad \e{for} \, 
 \qquad \sg, \tau \in \{\pm\} \;. 
\enq
established  in Lemma \ref{Lemme identite operatoire} 
below.
Now, 
Proposition \ref{Proposition solubilite unique RHP non local} 
implies that $\det[\e{id}-\op{M}_{R}]\, \not= \, 0$. 
The identity $\op{V}_{R}^{--} \cdot \op{V}_{R}^{-- }=0$ and the 
Plemelj-Smithies formula lead to 
\beq
 \e{det} \left( \ba{cc}  \e{id} - \op{V}_{R}^{--} &  - \op{V}_{R}^{-+} \\ 0 &  \e{id}    \ea \right)  \; = \; 1 \, -\, \e{tr}\big[ \op{V}_{R}^{--}  \big] \;. 
\enq
The latter trace can be 
shown to vanish by deforming the contour to $+\i \infty$ in the integral 
below 
\beq
 \e{tr}\big[ \op{V}_{R}^{--}  \big]  \; = \; \Int{\R - \i v }{}  \hspace{-2mm} \dd k \,   \a_{\ua}^{(0)}(k) \Bigg\{
\Int{\R - \i \eta }{} \hspace{-2mm} \f{\dd \mu }{ 2\i \pi }  \f{   \wh{B}^{- - }_{R}(\mu,k)   }{ ( \mu- k  ) \cdot  \a_{\ua}^{(0)}(\mu)  }     \, + \, 
  \f{   \wh{B}^{- - }_{R}(0,k)   }{ k \cdot  \a_{0}^{(0)}  } \Bigg\}\,, 
\enq
where the integrand is analytic in $\mathbb{H}^{+}-2\i v$ and goes to zero 
as $\e{o}(k^{-3/2})$ at $\infty$ in that domain.   

\vspace{2mm}
Thus, all in all, $\,\det\big[ \e{id} - \op{O}_R \big]  \, = \,\det[\e{id}-\op{M}_{R}]\, \not= \, 0$.  
\vspace{2mm}

It remains to establish the estimates 
\eqref{ecriture bornes sur entree noyau resolvent final} on the entries 
of the resolvent operator $\bs{\De}$. 
 Set $\op{Q}=\op{O}+\bs{\Om}$ for short. The  
blocks of the  resolvent kernel 
may be expressed as \cite{GohbergGoldbargKrupnikTracesAndDeterminants}
\beq
\De^{\tau \sg}(\la,\mu) \; = \; 
\f{ \De^{\tau \sg}_{\e{n}}(\la,\mu)  }{  \det\big[ \e{id}\, 
- \, \op{Q} \big]   }
\enq
in which the numerator is expressed in terms of the Fredholm series 
\beq
 \De^{\tau \sg}_{\e{n}}(\la,\mu) \, = \, \sul{ n \geq 0}{}  
\f{ (-1)^n }{ n! } \sul{ \substack{ \vsg_1,\dots,\vsg_n  \\ 
\in  \{\pm, R, L \} } }{}  \pl{a=1}{n} \Int{ \R + \i \veps_{\vsg_a} v }{} 
\hspace{-3mm} \dd \la_a \; 
\det_{n+1} \left[ \ba{cc}   Q^{\tau \sg}(\la,\mu)    &   
Q^{\tau \vsg_b}(\la,\la_b)  \\
			       Q^{\vsg_a \sg}(\la_a,\mu)    &   
Q^{\vsg_a \vsg_b}(\la_a,\la_b)  \ea  \right]  \;. 
\enq
Then, introduce the auxiliary kernel 
\beq
\wt{Q}^{\tau \sg}(\la,\mu) \; = \; Q^{\tau \sg}(\la,\mu) \cdot 
\f{  (1+|\la|) \cdot (1+|\mu|)  }{ \ln (1+|\la|)  \ln (1+|\mu|)  } 
\enq
which, by virtue of \eqref{ecriture borne sup sur M L et R}, 
\eqref{ecriture borne sup sur V L}, \eqref{ecriture borne sup sur V R} 
and \eqref{ecriture borne sup sur Omega kernel}
is bounded on $\{ \R + \i \veps_{\tau} v \} \times \{\R+\i \veps_{\sg} v \} $,
uniformly in $w$. This yields the representation 
\bem
 \De^{\tau \sg}_{\e{n}}(\la,\mu) \, = \, \f{ \ln (1+|\la|)  
\ln (1+|\mu|)  } {  (1+|\la|) \cdot (1+|\mu|)  } \\ 
 \times \sul{ n \geq 0}{}  \f{ (-1)^n }{ n! } \sul{ \substack{ 
\vsg_1,\dots,\vsg_n  \\ \in  \{\pm, R, L \} } }{}  \pl{a=1}{n} 
\Int{ \R + \i \veps_{\vsg_a} v }{}  \hspace{-3mm} \dd \la_a \;  
 \pl{a=1}{n} \bigg\{ \f{ \ln (1+|\la_a|)   } {   1+|\la_a|  } 
\bigg\}^2 \cdot \det_{n+1} \left[ \ba{cc}   \wt{Q}^{\tau \sg}(\la,\mu)    
&   \wt{Q}^{\tau \vsg_b}(\la,\la_b)  \\
			       \wt{Q}^{\vsg_a \sg}(\la_a,\mu)    &   
\wt{Q}^{\vsg_a \vsg_b}(\la_a,\la_b)  \ea  \right] \;. 
\end{multline}
A direct application of Hadamard's inequality for determinants allows 
one to infer that the above series converges on 
$\{ \R + \i \veps_{\tau} v \} \times \{\R+\i \veps_{\sg} v \}  $, 
which also yields 
\eqref{ecriture bornes sur entree noyau resolvent final}. \qed 

\begin{lemme}
 \label{Lemme identite operatoire}

Let $\op{M}_{R}^{\sg \tau}$, resp. $\op{M}_{L}^{\sg \tau}$, be the integral 
operators $L^{2}(\R+\i \tau v) \tend L^{2}(\R+\i \sg v) $ defined by 
the integral kernels \eqref{noyau integral MR +sigma}, 
\eqref{noyau integral MR -sigma}, resp. 
 \eqref{noyau integral ML -sigma}, \eqref{noyau integral ML +sigma}. 
Furthermore, let $\op{V}_{L}^{+ \tau }$, resp. $\op{V}_{R}^{- \tau }$, 
be the integral operators $L^{2}(\R+\i  v) \tend L^{2}(\R+\i \tau v) $, 
resp.  $L^{2}(\R-\i  v) \tend L^{2}(\R+\i \tau v) $, defined by the integral 
kernels \eqref{noyau integral VL + pm}, resp. \eqref{noyau integral VR - pm}. 
Then
\beqa
\op{M}_{L}^{\sg +} \cdot \op{V}_{L}^{+ \tau } & = & 0 \qquad for \  any 
\qquad \sg, \tau \in \{\pm\} \;   \\
\op{M}_{R}^{\sg -} \cdot \op{V}_{R}^{- \tau } & = & 0 \qquad for \  
any \qquad \sg, \tau \in \{\pm\} \;. 
\eeqa

\end{lemme}

\Proof 

It follows from equation \eqref{noyau integral VL + pm} that 
$ k \mapsto \op{V}_{L}^{+ \tau }(k,s)$ is analytic on 
$\mathbb{H}^{-}_{2 v}=\mathbb{H}^{-} + 2 \i v$ and that it falls-off 
at infinity in this domains like $\e{O}\Big( \ln(1+|k|)/|k| \Big)$. 
Furthermore, $ s \mapsto \op{M}_{L}^{ \sg + }(k,s)$ is also analytic 
on $\mathbb{H}^{-}_{2 v}$ and falls-off at infinity 
in this domains like $\e{O}\Big( 1/|s| \Big)$. Hence, a direct contour 
deformation up to $\R-\i\infty$ entails that 
\beq
\Int{ \R + \i v  }{} \dd s M^{\sg + }_{L}(k,s) \op{V}_{L}^{+ \tau }(s,t) \; 
= \; 0 \,. 
\enq
Similarly, it follows from equation \eqref{noyau integral VR - pm} that 
$ k \mapsto \op{V}_{R}^{- \tau }(k,s)$ is analytic on 
$\mathbb{H}^{+}_{-2 v}=\mathbb{H}^{+} - 2 \i v$ and that it falls off 
at infinity in this domains like $\e{O}\Big( \ln(1+|k|)/|k| \Big)$. 
Furthermore, $ s \mapsto \op{M}_{R}^{ \sg - }(k,s)$ is also analytic 
on $\mathbb{H}^{+}_{-2 v}$ and falls-off at infinity 
in this domains like $\e{O}\Big( 1/|s| \Big)$. Hence, a direct contour 
deformation up to $\R+\i\infty$ entails that 
\beq
\Int{ \R - \i v  }{} \dd s M^{\sg - }_{R}(k,s) \op{V}_{R}^{- \tau }(s,t) \; 
= \; 0 \,. 
\enq
\qed

\begin{theorem}
\label{Theorem bornage solution eqn integrale}
 
 The solution $\bs{u}$ to the integral equation 
\eqref{equation integrale sous forme compact bien separee finale} 
satisfies to the bounds 
\beq
u^{\sg}(k) \; = \; \e{O}\bigg(  \f{ C_{h} \ex{ 2v w }
\ln( 1 + |k| ) }{ 1 + |k|  } \bigg)  \qquad with \qquad 
k\in \R+\i\veps_{\sg} v \;, 
\enq
provided that 
\beq
\mf{d}_{w}^{\a}[h](k) \; \leq \; \f{ C_{h} \ln( 1 + |k| ) }{  1 + |k|  } \;. 
\enq

\end{theorem}

\Proof 

This is an obvious consequence of Proposition 
\ref{Proposition invertibilite operateur Id - O - Omega}. \qed

\section{Conclusion}

In this paper we have carried out the $w\tend +\infty$ asymptotic analysis 
of the solution to a non-local Riemann-Hilbert problem characterising 
the conformal map from a welded cylinder onto the standard one in the case 
where the 
welding diffeomorphism is composed of two non-trivial bumps 
separated from one another by distance $w$. 
This problem was motivated by the study
of the large-time behaviour of the generating function of full counting 
statistics of energy transfers in 1+1 dimensional 
non-equilibrium
conformal field theories
discussed in \cite{KozGawedzkiFullCountingStatisticsIThermoLimAndConjecture}. 
Our results allowed us to establish 
the large-time asymptotics
of the generating function rigorously. 

On technical ground, we have developed 
methods that allow to establish
the existence and the uniqueness of solutions to non-local Riemann--Hilbert 
problems in the case where the "compact operator" arguments developed in 
the literature cannot be applied directly. 
Our analysis shows that 
it is still possible to study the asymptotic behaviour of solutions 
of such problems even if this is much more involved than in the local case.

\section*{Acknowledgements}

K.K.K. is supported by CNRS.

\appendix

\section{Conformal map of the welded cylinder}
\label{Appendix Conformal Map}

In this section we prove Proposition \ref{proposition ouverture}. 
Introduce the functions 
\beq
\om^{(L)}(z) \; = \; \f{ \wt{\ga}_{+} \cdot z }{ 1 
+ \ex{- \f{2\pi }{ \tau } z } } \qquad \e{with} \qquad 
\om^{(R)}(z) \; = \; \f{ \wt{\ga}_{-} \cdot z }{ 1 
+ \ex{ \f{2\pi }{ \tau } z } } \;, 
\enq
with $\wt{\ga}_{\pm}$ as given in the statement of Proposition 
\ref{proposition ouverture} and where $\tau>2\a$. 
These admit the decomposition
\beq
\om^{(L)}(z) \; = \;\wt{\ga}_{+} \cdot z \, + \, \om^{(L)}_R(z)  
\quad \e{and} \quad \om^{(R)}(z) \; = \;\wt{\ga}_{-} \cdot z \, 
+ \, \om^{(R)}_L(z)  \;. 
\enq
Let us set $\Om(z\mid \varkappa^{+}, \varkappa^{-}) \,  
= \, \Ups(z) \, + \,  \om^{(L)}(z)  \, + \, \om^{(R)}(z)$. 
Then, $\Ups $ solves the non-local Riemann-Hilbert problem: 
find $\Ups \in \mc{O}(\mc{S}_{\a})$ such that
\begin{itemize}
 \item $\Ups$ has smooth $-$, resp. $+$, boundary values on $\R$, 
resp. $\R-\i\a$; 
 \item $\Ups_{+}\big(g(x)-\i\a \big) \, = \, \Ups_{-}\big( x \big)\, 
+ \, G_{\Ups}(x)$, with $x \in \R$; 
\item there exists a constants $C_{\Ups}$ and $\eta>0$  such that 
\beq
\Ups(z) =\left\{ \ba{ccc} -\om^{(L)}_R(z)  \, -\, \om^{(R)}(z)  \, 
+ \,  \e{O}\Big( \ex{ - \eta   \Re(z) } \Big)  \quad &  
\e{when}  \quad & \Re(z) \tend + \infty  \vspace{2mm}\\
-\om^{(L)}(z)  \, -\, \om^{(R)}_L(z)  \, + \,   C_{\Ups} \de_{\pm; -} \, 
+ \,  \e{O}\Big( \ex{ \eta  \Re(z) } \Big)  \quad &  
\e{when}  \quad & \Re(z) \tend - \infty  \ea \right. \; , 
\enq
with an asymptotic expansion that is valid uniformly up to the boundary, 
\end{itemize}
and where 
\beq
G_{\Ups}(x) \; = \; \left\{ \ba{ccc}  \om^{(L)}(x) \, 
- \, \om^{(L)}\big( g(x) - \i\a \big)  \, + \, \om^{(R)}_L(x) \, 
- \, \om^{(R)}_L\big( g(x) - \i\a \big)    & \e{for} & x \leq - M \\
-\i\a \, + \, \om^{(L)}(x) \, - \, \om^{(L)}\big( g(x) - \i\a \big)  \, 
+ \, \om^{(R)}(x) \, - \, \om^{(R)}\big( g(x) - \i\a \big)     
& \e{for} & |x|  \leq  M  \\
\om^{(L)}_R(x) \, - \, \om^{(L)}_R\big( g(x) - \i\a \big)  \, 
+ \, \om^{(R)}(x) \, - \, \om^{(R)}\big( g(x) - \i\a \big)    
& \e{for} & x \geq  M  \ea \right. 
\enq

By virtue of Proposition 
\ref{Proposition solubilite unique RHP non local}, this non-local 
Riemann-Hilbert problem admits a unique solution. Hence, so does the one 
of $\Om$. 

Since  $\Om \in \mc{O}(\mc{S}_{\a})$, $\Om$ is open and thus 
$\Dp{} \Om(\mc{S}_{\a}\mid \varkappa^{+},\varkappa^{-})
=\Om_{+}(\R-\i\a\mid \varkappa^{+},\varkappa^{-})\cup\Om_{-}(\R\mid 
\varkappa^{+},\varkappa^{-})$. 
Thus, clearly, 
\beq
  \Om(\mc{S}_{\a}\mid \varkappa^{+},\varkappa^{-}) \ni  \om \; 
\mapsto \; \# \Big\{ \Om^{-1}(\om\mid \varkappa^{+},\varkappa^{-}) \Big\} 
  \; = \; \Int{ \big\{ \R-\i\a \} \cup -\R }{} \f{ \dd s  }{2\i\pi } 
\f{ \Om^{\prime}(s\mid \varkappa^{+},\varkappa^{-}) }{ \Om(s\mid \varkappa^{+},
\varkappa^{-})- \om }
\enq
is continuous in $\om$ on $\Om(\mc{S}_{\a}\mid \varkappa^{+},\varkappa^{-})$. 
Being integer valued, it is constant. The asymptotic behaviour of $\Om$ 
at infinity entails that $\# \Big\{ \Om^{-1}(\om\mid \varkappa^{+},
\varkappa^{-}) \Big\} \; = \; 1$
for $\Re(\om)$ large enough and such that $\om \in \Om(\mc{S}_{\a}\mid 
\varkappa^{+},\varkappa^{-})$. Hence, $\Om$ is injective and thus a 
biholomorphism on its image. \qed

\section{Inversion of the operators $\e{id}-\op{L}^{\ups\ups}$ on 
$L^{2}(\R^{\ups})$}
\label{Appendix section inversion operateurs id - L ups ups} 

We now discuss the invertibility of $\op{id}-\op{L}^{\ups\ups}$ with the help 
of the Wiener-Hopf technique, see \textit{e.g.} \cite{EstradaKanwalSingularIntegralEqnsAndWienerHopf},  as will be detailed in the two next subsections.  
The method builds on the solution of a multiplicative Riemann-Hilbert problem 
involving the Fourier transform of the kernel $L^{\ups}(x)$.

\subsection{Inversion of the operators $\e{id}-\op{L}^{++}$}
\label{Soussection inversion L ++}

For the purpose of the present section, we introduce the space 
\beq
L^2_{C}(\R^+) \, = \, \bigg\{ f \in L^2_{\rm loc}(\R^+) \; : \;  \exists C_f \; \e{and}\, \a > 0\quad  f(x) \; = \; C_f \, + \, \e{O}\Big( \ex{-\a x}  \Big)  \bigg\} \;. 
\label{definition espace L2 de R plus} 
\enq

\begin{prop}
\label{Proposition inversion Wiener Hopf pour L+}
 
 Let $L^{+}$ be as defined through \eqref{definition noyau integral L++} and consider the integral equation  
\beq
f(x) \, - \, \Int{0}{+\infty} L^{+}(x-y) f(y) \dd y \, = \, h(x)  \;, \qquad x \in \R^+
\label{ecriture Wiener-Hopf sur R+}
\enq
 on $L^2_{C}(\R^+)$ with $h$ such that there exist $\eta>0$ so that 
\beq
h(x) \; = \;  \e{O}\Big( \ex{-\eta x}  \Big) \;, \qquad when  \quad x \tend + \infty \, .
\label{ecriture cmptement asymptotique de h}
\enq
Then equation \eqref{ecriture Wiener-Hopf sur R+} is uniquely solvable on $L^2_{C}(\R^+)$ and the Fourier transform of the solution takes the form 
\beq
\mc{F}\big[f](k) \; = \; \f{1}{ \a_{\ua}^{(+)}(k) } \Int{\R - \i\eta^{-} }{} \f{ \dd s }{ 2\i \pi }  \f{  \a_{\da}^{(+)}(s)  \cdot   \mc{F}\big[ h \bs{1}_{\R^+} \big](s) }{ s-k }
\qquad with \qquad k \in \R + \i v 
\label{ecriture expression TF de solution WH sur R plus}
\enq
for any $0<\eta^{-}<\eta$ and with $v >0$.

\end{prop}

\Proof 

Following the strategy of the Wiener-Hopf method, one starts by extending 
$f$ and $h$ to $\R$ in such a way that equation 
\eqref{ecriture Wiener-Hopf sur R+} now holds on $\R$. 
We shall make the choice
\beq
h(x)=0 \qquad \e{and} \qquad f(x)\, = \, \Int{0}{+\infty} L^{+}(x-y) f(y) 
\dd y  \qquad \e{for} \;\;  x <0 \;. 
\enq
Given the behaviour of $f$ on $\R^+$ and the explicit expression 
\eqref{definition noyau integral L++} for $L^{+}(x)$, it is easy to convince 
oneself that, upon reducing $\eta$ if need be, these extensions satisfy
\beq
f(x) \; = \; \e{O}\Big( \ex{\eta x}  \Big) \qquad \e{and} \qquad 
h(x) \; = \;\e{O}\Big( \ex{\eta x}  \Big) \;, 
\enq
when $x \tend - \infty$. Actually, for the purpose of the analysis to come, 
it is convenient to introduce a specific notation for the restrictions 
of a function on $\R$ to $\R^{\pm}$: $f^{\pm}=f\bs{1}_{\R^{\pm}}$.
In particular, by construction, we have that $h=h^{+}$. 
The properties of the extended functions allow one to compute a well-defined 
Fourier transform provided that $k \in \R + \i v$, $0<v < \eta$. Thus Fourier 
transforming \eqref{ecriture Wiener-Hopf sur R+}
leads to 
\beq
\Big( 1 \, - \, \mc{F}\big[   L^{+}  \big](k)  \Big) \cdot  \mc{F}\big[ f^{+} 
\big](k)  \, + \,  \mc{F}\big[ f^{-} \big](k) \, = \, \mc{F}\big[ h^+ \big](k) 
\qquad \e{with} \qquad  k \in \R + \i v \;. 
\enq
Then, by using the Wiener-Hopf factorisation of $1 \, - \, 
\mc{F}\big[   L^{++}  \big]$ given in 
\eqref{ecriture eqn de WH pour 1-L eps eps}, one may recast the equation as
\beq
 \a_{\ua}^{(+)}(k)  \cdot   \mc{F}\big[ f^{+} \big](k)  \, 
+ \,  \a_{\da}^{(+)}(k)  \cdot   \mc{F}\big[ f^{-} \big](k) \; 
= \;  \a_{\da}^{(+)}(k)  \cdot   \mc{F}\big[ h^+ \big](k)  \;. 
\enq
Given half-planes $\mc{B}_{\ua/ \da}^{(+)}$ as introduced in 
\eqref{definition domaines B up down plus moins}, one may define 
$U\in \mc{O}\Big(  \mc{B}_{\ua}^{(+)}  \cup  \mc{B}_{\da}^{(+)} 
\setminus \{0\} \Big)$
and having a simple pole at $0$ by the piecewise formula
\beq
U(z) \; = \; \left\{  \ba{cc}      \a_{\ua}^{(+)}(z)  \cdot   
\mc{F}\big[ f^{+} \big](z) \, - \, \op{C}^{(+)}\Big[ \a_{\da}^{(+)}  
\cdot   \mc{F}\big[ h^+ \big]  \Big]  (z)     \quad &, \quad  z \in 
\mc{B}_{\ua}^{(+)} \vspace{2mm} \\
   -  \a_{\da}^{(+)}(z)  \cdot   \mc{F}\big[ f^{-} \big](z) \, 
- \, \op{C}^{(+)}\Big[ \a_{\da}^{(+)}  \cdot   \mc{F}\big[ h^+ \big]  
\Big]  (z)     &, \quad   z \in \mc{B}_{\da}^{(+)}    \ea \right. 
\label{definition fonction U cas ++}
\enq
where $\op{C}^{(+)}$ is the Cauchy transform on $L^2(\R+\i v )$:
\beq
 \op{C}^{(+)}\big[u \big]  (z)  \; = \; \Int{ \R + \i  v }{} 
\f{ \dd s }{ 2\i\pi } \f{ u(s) }{  s-z }   \qquad \e{for} \qquad z 
\in \Cx \setminus \big\{ \R+\i v \big\}\;. 
\label{definition transfo Cauchy sur R shifte vers haut}
\enq
Then, by using the relation valid for any $u\in L^p(\R+\i v)$, 
$+\infty>p>1$, 
\beq
 \op{C}_{+}^{(+)}\big[u \big](k) - \op{C}_{-}^{(+)}\big[u \big](k) \, 
= \, u(k) \;, 
\label{ecriture formule Sokhotsky Plejmel}
\enq
one gets that $U_+=U_-$ on $\R+\i v$ and hence $U$ extends into a 
meromorphic function on $\Cx$ whose single pole is located at $0$ 
and is simple. Moreover, it follows from 
\eqref{definition fonction U cas ++} that
\beq
U(z) \; = \; -\f{ \wt{\a}_{0}^{\,(+)} }{ z } \mc{F}[f^{-}](0) \; 
+ \; \e{O}(1) \;. 
\enq
It is easy to see that $ \mc{F}\big[ f^{+/-} \big](k) \tend 0$
when $k\tend \infty$ in $\ov{\mc{B}_{\ua/\da}^{(+)}}$ and this up to 
the boundary. Hence, $U(k) \tend 0$ as $k\tend \infty$. Since the 
constant $\mc{F}[f^{-}](0)$ is part of the unknowns in the problem, 
we conclude that there exists a constant $\mc{K}^{(+)}$ such that 
\beq
U(z) \, = \,  - \f{ \wt{\a}^{\, (+)}_{0}  \mc{K}^{(+)} }{ z } \;. 
\enq

This explicit expression for $U$ entails that, for any $k \in 
\mc{B}_{\ua}^{(+)}$,  
\beq
 \mc{F}\big[ f^{+} \big](k) \; = \;  \f{1}{ \a_{\ua}^{(+)}(k) } \cdot 
\bigg\{  \op{C}^{(+)} \Big[ \a_{\da}^{(+)}  \cdot   \mc{F}\big[ h^+ \big]  
\Big]  (k)  \, - \, \f{ \wt{\a}^{\, (+)}_{0}  \mc{K}^{(+)} }{ k } \bigg\}  \;. 
\label{expression F de f+ dans B plus up}
\enq
Note that, owing to \eqref{ecriture cmptement asymptotique de h}, one may 
meromorphically continue  $\mc{F}\big[ f^{+} \big](k)$ from 
$\mc{B}_{\ua}^{(+)}$ up to $\Big\{ z \in \Cx \, : \, \Im(z)> - \eta \Big\}$ 
by the expression 
\beq
 \mc{F}\big[ f^{+} \big](k) \; = \;  \f{1}{ \a_{\ua}^{(+)}(k) } \cdot 
\bigg\{  \op{C}^{(+)} \Big[ \a_{\da}^{(+)}  \cdot   \mc{F}\big[ h^+ \big]  
\Big]  (k)\, + \,  \a_{\da}^{(+)}(k)  \cdot   \mc{F}\big[ h^+ \big](k)   \, 
- \, 
 \f{ \wt{\a}^{\, (+)}_{0}  \mc{K}^{(+)} }{ k } \bigg\}  \;. 
\enq
Since $\a_{\ua}^{(+)}(k)$ has a simple zero at $k=0$, 
the expression above entails that $\mc{F}\big[ f^{+} \big]$ 
may have 
a double pole at $k=0$, and that it is its sole pole in the domain 
$\Im(k)>-\eta^{\prime}$, for some $\eta^{\prime}>0$ and small enough.

Now assume that one is given a meromorphic function $w$ in the tubular 
neighbourhood $|\Im(z)|<2\eta^{\prime}$ of $\R$ having one pole of order 
$r+1$ at $k=0$:
\beq
w(k) \, = \, \sul{p=0}{r} \f{ w_p }{ k^{p+1} } \; + \; \e{O}(1) \qquad 
k \tend 0\;, 
\enq
and decaying at least as $1/k$ at infinity. Then, it is easy to convince 
oneself that, for $x \not=0$,  
one has 
\beq
\Int{\R + \i \eta^{\prime} }{} \f{ \dd k }{ 2\pi } \ex{-\i k x } w(k) \; 
= \;  -  \i \sul{p=0}{r} \f{ w_p }{ p! } \cdot (-\i x)^{ p } \; + \; 
\Int{\R - \i \eta^{\prime} }{} \f{ \dd k }{ 2\pi } \ex{-\i k x } w(k)   \;. 
\label{ecriture Fourier H+ vs H- plus power law}
\enq
The integral appearing on the \textit{rhs} of the above identity produces 
a $\e{O}\big( \ex{- \eta^{\prime} x } \big)$ behaviour when $x\tend +\infty$.

$f^{+}$ can be reconstructed from \eqref{expression F de f+ dans B plus up} 
by taking the inverse Fourier transform on $\R+\i v$. One infers from 
\eqref{ecriture Fourier H+ vs H- plus power law}  that the 
only way to give rise to a solution $f^+$ to \eqref{ecriture Wiener-Hopf 
sur R+} enjoying the asymptotic behaviour that is compatible with 
$f \in L^2_{C}(\R^+)$, \textit{c.f.} \eqref{definition espace L2 de R plus},  
is that the meromorphic continuation of $ \mc{F}\big[ f^{+} \big](k)$ has 
at most a simple pole at $k=0$. This entails that 
\beq
\wt{\mc{K}}^{(+)} \; = \;  \mc{F}\big[ h^+ \big](0) \;. 
\enq
Thus, if a solution to \eqref{ecriture Wiener-Hopf sur R+} exists in 
the class \eqref{definition espace L2 de R plus}, then it is unique and  
necessarily takes the form 
\beq
 \mc{F}\big[ f^{+} \big](k) \; = \;  \f{1}{ \a_{\ua}^{(+)}(k) } \cdot 
\bigg\{  \op{C}^{(+)} \Big[ \a_{\da}^{(+)}  \cdot   \mc{F}\big[ h^+ \big]  
\Big]  (k)   \, - \, 
 \f{ \wt{\a}^{\, (+)}_{0}   \mc{F}\big[ h^+ \big](0) }{ k } \bigg\}  \;, 
\label{ecriture Fouier forme finale solution WH R plus}
\enq
with $k \in \mc{B}^{(+)}_{\ua}$. By deforming the contour in the Cauchy 
transform from $\R+\i v $ to $\R-\i\eta^-$ with $0<\eta^{-}<\eta$ one 
obtains the representation 
\eqref{ecriture expression TF de solution WH sur R plus}. 

Reciprocally, it is easy to see that the function $f$ defined as 
\beq
f^{\pm}(x) \; = \; \Int{ \R + \i v }{} \f{ \dd k}{2\pi} \ex{-\i k x} 
\ga^{\pm}(k) \qquad \e{with} \qquad
\left\{ \ba{ccc}  \ga^{+}(k) & = &     \f{1}{ \a_{\ua}^{(+)}(k) } 
\cdot \bigg\{  \op{C}^{(+)}_+ \Big[ \a_{\da}^{(+)}  \cdot   
\mc{F}\big[ h^+ \big]  \Big]  (k)    \, - \, 
 \f{ \wt{\a}^{\, (+)}_{0}   \mc{F}\big[ h^+ \big](0) }{ k } \bigg\}    \\ 
  \ga^{-}(k) & = &     \f{1}{ \a_{\da}^{(+)}(k) } \cdot \bigg\{   
\f{ \wt{\a}^{\, (+)}_{0}   \mc{F}\big[ h^+ \big](0) }{ k }   \, 
- \, \op{C}^{(+)}_- \Big[ \a_{\da}^{(+)}  \cdot   
\mc{F}\big[ h^+ \big]  \Big]  (k) \bigg\} 
   \ea \right. 
\enq
solves the linear integral equation \eqref{ecriture Wiener-Hopf sur R+} 
on $\R^{+}$. 

Indeed, since $\ga^{+}$, resp. $\ga^{-}$, admits a holomorphic continuation 
to  $\mc{B}_{\ua}^{(+)}$, resp. $\mc{B}_{\da}^{(+)}$, that decays as 
$\e{O}(1/k)$ at infinity, one readily shows that, indeed, the function
\beq
x \mapsto \Int{ \R + \i v }{} \f{ \dd k}{2\pi} \ex{-\i k x} \ga^{\pm}(k)
\enq
are supported on $\R^{\pm}$ and that they exhibit the  
required asymptotic behaviour. The previous reasonings taken backwards then 
ensure that  
\beq
\Big( 1 \, - \, \mc{F}\big[   L^{+}  \big](k)  \Big) \cdot  \ga^{+}(k)  \, 
+ \,  \ga^{-}(k) \, = \, \mc{F}\big[ h^+ \big](k) \qquad \e{for} \qquad  
k \in \R + \i v \;. 
\enq
Upon taking the inverse Fourier transform, the above  
relation  
leads to equation \eqref{ecriture Wiener-Hopf sur R+}, hence proving the 
existence of solutions in $L^2_{C}(\R^+)$. \qed

\vspace{3mm}




\subsection{Inversion of the operators $\e{id}-\op{L}^{--}$}
\label{Soussection inversion L --}

\noindent Analogously to the previous setting, we introduce the space 
\beq
L^2_{C}(\R^-) \, = \, \bigg\{ f \in L^2(\R^-) \; : \;  
\exists \; C_f \;\;  \e{and}\;\;  \a > 0 \quad  f(x) \; = \; C_f \, 
+ \, \e{O}\big( \ex{\a x}  \big)  \bigg\} \;. 
\enq

\begin{prop}
\label{Proposition inversion Wiener Hopf pour L-}
 Let $L^{-}$ be as defined through 
\eqref{definition noyau integral L++} and consider the integral equation  
\beq
f(x) \, - \, \Int{-\infty}{0} L^{-}(x-y) f(y) \dd y \, 
= \, h(x)  \qquad  for  \qquad x \in \R^- \;, 
\label{ecriture equation WH sur R moins}
\enq
 on $L^2_{C}(\R^-)$ with $h$ such that there exist $\eta>0$ so that 
\beq
h(x) \; = \;  \e{O}\Big( \ex{\eta x}  \Big) \;
\enq
when $x \tend - \infty$. 

Then,  equation \eqref{ecriture Wiener-Hopf sur R+} is uniquely solvable 
on $L^2_{C}(\R^-)$ and the Fourier transform of the solution takes the form 
\beq
\mc{F}\big[f](k) \; = \;  - \a_{\da}^{(-)}(k)  
\Int{\R + \i \eta^{-} }{} \f{ \dd s }{ 2\i \pi }   
\f{  \big\{ \a_{\ua}^{(-)}(s) \big\}^{-1}  \cdot   
\mc{F}\big[ h \bs{1}_{\R^-} \big](s) }{ s-k }
\qquad with \qquad k \in \R - \i v 
\label{ecriture expression TF de solution WH sur R moins}
\enq
for any $0<\eta^{-}<\eta$ and with $v >0$.

\end{prop}

\Proof 

One extends the functions $f$ and $h$ to $\R$ as
\beq
f(x) \, = \, \Int{-\infty}{0} L^{-}(x-y) f(y) \dd y  
\qquad \e{and} \qquad h(x) \, = \,  0  \qquad \e{for} \qquad x>0 \;, 
\enq
so that, reducing $\eta>0$ if need be, these extensions 
possess the $x \tend + \infty$ asymptotic behaviour  
\beq
f(x) \; = \;  \e{O}\Big( \ex{ - \eta x} \Big) \qquad \e{and} \qquad 
h(x) \; = \; \e{O}\Big( \ex{ - \eta x} \Big) \;. 
\enq
One may then take the Fourier transform of 
\eqref{ecriture equation WH sur R moins} extended to $\R$, provided 
that the Fourier variable $k$ satisfies $k \in \R-\i v$, with $0<v\ll 1$. 
This leads to 
\beq
 \mc{F}\big[ f^{+} \big](k)  \, + \, \Big( 1 \, 
- \, \mc{F}\big[   L^{-}  \big](k)  \Big) \mc{F}\big[ f^{-} \big](k) \, 
= \, \mc{F}\big[ h \big](k)\;. 
\enq
Using the Wiener-Hopf factorisation of $1 \, 
- \, \mc{F}\big[   L^{-}  \big] $ relatively to $\R-\i v$, one may recast 
the last equation as
\beq
\Big\{  \a_{\ua}^{(-)}(k)  \Big\}^{-1} \cdot   \mc{F}\big[ f^{+} \big](k)  \, + \,  \Big\{ \a_{\da}^{(-)}(k) \Big\}^{-1} \cdot   \mc{F}\big[ f^{-} \big](k) \; = \;  \Big\{ \a_{\ua}^{(-)}(k) \Big\}^{-1}  \cdot   \mc{F}\big[ h \big](k) \;. 
\enq
Define $U\in \mc{O}\Big(  \mc{B}_{\ua}^{(-)}  \cup  \mc{B}_{\da}^{(-)} \setminus \{0\}  \Big)$ by the piecewise formula
\beq
U(k) \; = \; \left\{  \ba{cc}      \op{C}^{(-)}\Big[  \big\{ \a_{\ua}^{(-)} \big\}^{-1} \cdot   \mc{F}\big[ h \big]  \Big]  (k)    - \Big\{  \a_{\ua}^{(-)}(k)  \Big\}^{-1} \cdot   \mc{F}\big[ f^{+} \big](k) \qquad 
&   , \quad z \in \mc{B}_{\ua}^{(-)} \vspace{2mm} \\
    \Big\{ \a_{\da}^{(-)}(k) \Big\}^{-1} \cdot   \mc{F}\big[ f^{-} \big](k) \, + \, \op{C}^{(-)}\Big[ \Big\{ \a_{\ua}^{(-)} \Big\}^{-1}  \cdot   \mc{F}\big[h \big]  \Big]  (k)     & , \quad z \in \mc{B}_{\da}^{(-)}    \ea \right. 
\enq
where $\op{C}^{(-)}$ is the Cauchy transform on $L^2(\R-\i v)$:
\beq
 \op{C}^{(-)}\big[u \big]  (z)  \; = \; \Int{ \R - \i  v }{} \f{ \dd s }{ 2\i\pi } \f{ u(s) }{  s-z }   \qquad \e{for} \qquad z \in \Cx \setminus \big\{ \R  - \i v \big\}\;. 
\label{definition transfo Cauchy sur R shifte vers bas}
\enq
Since $ \a_{\ua}^{(-)}(k)$ admits a simple zero at $k=0$, one 
gets that $U$ is meromorphic on $\mc{B}_{\ua}\cup \mc{B}_{\da}$. Its sole 
pole is located at $k=0$ and is simple. Moreover $U$  vanishes 
at $\infty$ and satisfies $U_+=U_-$ on $\R-\i v$. All of this allows 
one to infer that 
\beq
U(k)  \; = \; - \f{ \mc{F}[f^{+}](0) }{  k \a_0^{(-)}  } \;. 
\label{expression explicite pour U cas moins}
\enq
However, since $\mc{F}[f^{+}](0)$ is part of the unknowns in the problem, 
it is more convenient to set $\mc{K}^{(-)}=\mc{F}[f^{+}](0)$.

The expression \eqref{expression explicite pour U cas moins} allows one 
to reconstruct the Fourier transform of $f^{-}$ for $k \in \mc{B}_{\da}^{(-)}$ 
as:
\beq
 \mc{F}\big[ f^{-} \big](k) \; = \; -   \a_{\da}^{(-)}(k)  
\bigg\{    \f{ \mc{K}^{(-)}  }{k \cdot \a_0^{(-)} } \, + \,     
\op{C}^{(-)} \Big[ \big\{  \a_{\ua}^{(-)} \big\}^{-1} \cdot   
\mc{F}\big[ h^- \big]  \Big]  (k)  \bigg\} \;. 
\enq
The meromorphic continuation of $ \mc{F}\big[ f^{-} \big](k) $
to $\mc{B}_{\ua}^{(-)}$ takes the form 
\beq
 \mc{F}\big[ f^{-} \big](k) \; = \; -   \a_{\da}^{(-)}(k)  
\bigg\{      \f{ \mc{K}^{(-)}  }{k \cdot \a_0^{(-)} }  \,
- \, \big\{ \a_{\ua}^{(-)} (k) \big\}^{-1} \cdot   \mc{F}\big[ h^- \big](k) 
 \, + \,     \op{C}^{(-)} \Big[ \big\{  \a_{\ua}^{(-)} \big\}^{-1} 
\cdot   \mc{F}\big[ h^- \big]  \Big]  (k)  \bigg\}  \qquad 
\e{for} \qquad k \in \mc{B}_{\ua}^{(-)} \;. 
\enq
The function $\a_{\da}^{(-)}(k) $ admits a simple pole at $k=0$. 
For generic  $\mc{K}^{(-)}$, the term under the bracket also admits 
a simple pole at $k$, so that the meromorphic continuation has a double 
pole at $k=0$. As in the case of the Wiener-Hopf equation on $\R^+$, 
contour displacements in the inverse Fourier transform ensure that if 
$f^{-}$ has at most constant asymptotics at $-\infty$ then the meromorphic 
continuation of $ \mc{F}\big[ f^{-} \big](k)$ must have at most a simple 
pole at $k=0$. This unambiguously fixes the unknown constant as 
$\mc{K}^{(-)}=\mc{F}[h^{-}](0)$, leading to
\beq
 \mc{F}\big[ f^{-} \big](k) \; = \; -   \a_{\da}^{(-)}(k)  
\bigg\{    \f{\mc{F}[h^{-}](0)  }{k \cdot \a_0^{(-)} } \, 
+ \,     \op{C}^{(-)} \Big[ \big\{  \a_{\ua}^{(-)} \big\}^{-1} \cdot   
\mc{F}\big[ h^- \big]  \Big]  (k)  \bigg\}   
\enq
 for any $k \in \mc{B}^{(-)}_{\da}$. Upon deforming the contour in 
the Cauchy transform $\op{C}^{(-)} $ up to $\R+\i\eta^{-}$ with 
$0< \eta^{-} <\eta$, one arrives to 
 \eqref{ecriture expression TF de solution WH sur R moins}.

It is easy to see, 
proceeding similarly as before,
that the above expression does give rise to a solution to  
\eqref{ecriture equation WH sur R moins}.  

\qed

\section{Inversion of $\,\e{id}-\op{L}_w^{(0)}$}
\label{Appendix Secion inversion operateur L0}

\subsection{Characterisation in terms of a Riemann-Hilbert 
problem}

The operator $\e{id}-\op{L}_w^{(0)}$ on $L^{2}(\intoo{-w}{w})$, as defined 
through \eqref{L0w} and 
\eqref{definition noyau integral op WH tronque interval central},  
is a truncated Wiener-Hopf operator and, as such, can be explicitly 
inverted in terms of the solution to an auxiliary Riemann-Hilbert problem. 
Consider the operator $\op{V}$ on $L^2(\R+\i v)$ with the kernel 
\beq
V(k,s) \, = \, -\mc{F}[L^{(0)}](k) \cdot \f{ \ex{\i(k-s)w } \, 
- \,  \ex{-\i(k-s)w }   }{ 2\i\pi (k-s) } \qquad \e{where} \qquad 
\mc{F}[L^{(0)}](k) \; = \; \f{ \cosh\big[k(\tau/2-\a-\i\varkappa) \big] }{  \cosh\big[k\tau/2 \big]  } \;. 
\label{definition noyau integral de V}
\enq
Then, it is easy to see that $\mc{F}^{-1}\Big( \e{id}+\op{V} \Big)\mc{F} 
\, = \, \e{id}-\op{L}_w^{(0)}$ or, more precisely, if $f$ solves 
$\big(\e{id}-\op{L}_w^{(0)}\big)[f]=h$ with $h\in L^{2}(\intoo{-w}{w})$,
then 
\beq
\Big( \e{id}+\op{V} \Big)\Big[ \mc{F}[f]\Big](k) \, = \, \mc{F}[h](k) 
\label{ecriture lin int eqn avec operateur V}
\enq
for an appropriate extension of $f$ outside $\intff{-w}{w}$. 

Observe that 
\beq
V(\la,\mu) \; = \; \f{ \Big( \bs{E}_{L}(\la), \bs{E}_R(\mu) \Big)  }{ \la-\mu } 
\label{ecriture noyau integral type IIO pour V}
\enq
where, upon setting $e(\la)=\ex{\i w \la}$, 
\beq
\bs{E}_{R}(\mu)\;= \; \f{1}{2\i\pi}\left(\ba{c} e(\mu) \\ e^{-1}(\mu) \ea \right)
\qquad \e{and} \qquad \bs{E}_{L}(\la)\;= \; -\mc{F}[L^{(0)}](\la) \left(\ba{c} -e^{-1}(\la) \\ e(\la) \ea \right) \;, 
\enq
so that $\Big( \bs{E}_{L}(\la), \bs{E}_R(\la) \Big) \; = \; 0$. This means that $\op{V}$ is an integrable integral operator. As such, it can be studied by means of an associated Riemann--Hilbert problem
as first observed in \cite{ItsIzerginKorepinSlavnovDifferentialeqnsforCorrelationfunctions}.

Assume that $\e{id}+\op{V}$ is invertible. Then, define the functions $  \bs{F}_{R/L}(\la)$ as the solutions to the linear integral equations
\beq
\big[\bs{F}_{R}\big] \Big( \e{id}+\op{V} \Big)(\la)  \, = \,     \bs{E}_{R}(\la) \qquad \e{and} \qquad
\Big( \e{id}+\op{V} \Big) \big[\bs{F}_{L}\big](\la)  \,  =  \,  \bs{E}_{L}(\la)  \;. 
\enq

The first formula is to be understood as an action of the operator 
to the left and the second one as its action to the right.

We refer the reader to Subsection \ref{SubSection notations} where 
the notations used below are introduced.

\begin{figure}[h]
\begin{center}

\includegraphics[width=.7\textwidth]{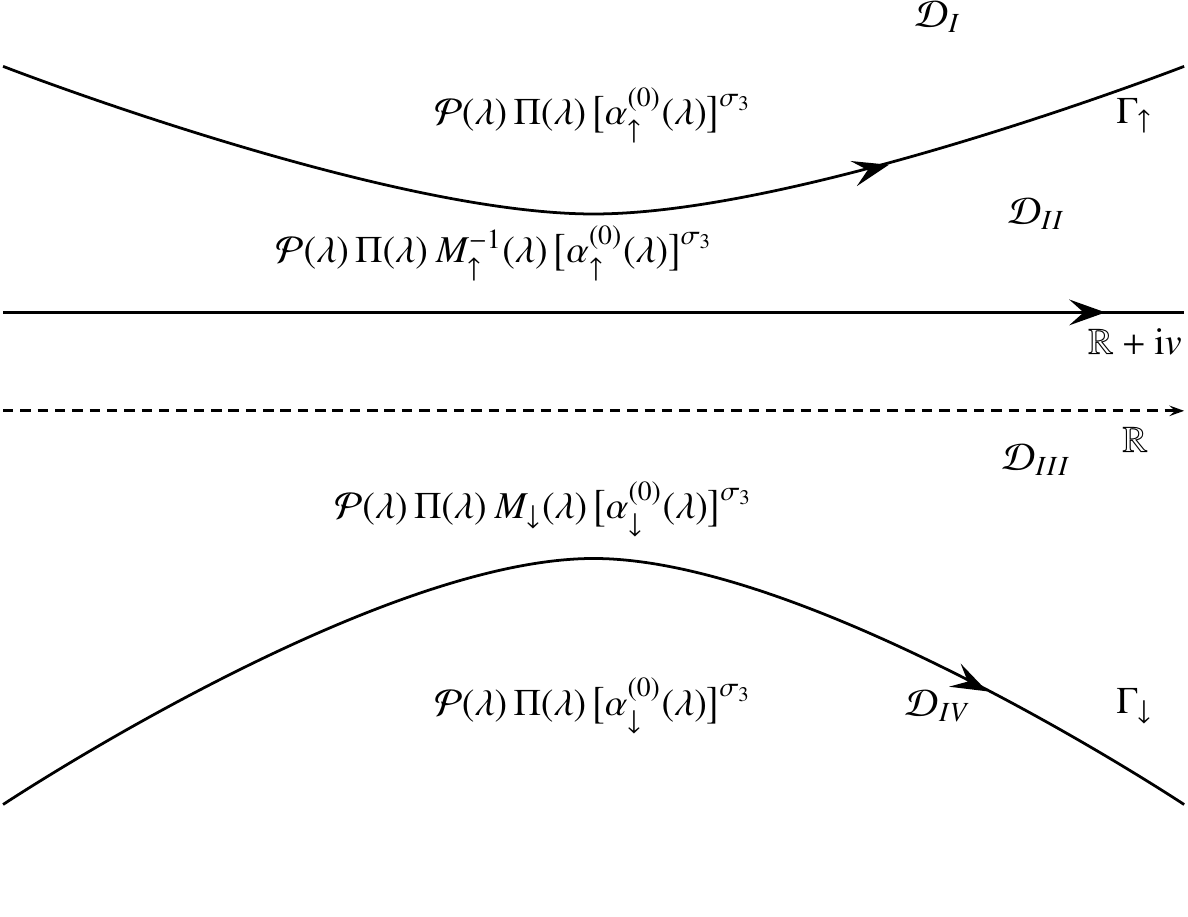}

\caption{Piecewise definition of the matrix $\chi$.
The curves $\Ga_{\ua/\da}$ separate all poles, other that at $0$, 
of $\la \mapsto \mc{F}[L^{(0)}](\la) \cdot \Big\{ 1 - \mc{F}[L^{(0)}](\la) 
\Big\}^{-1}$ from $\R$ and are such that $\e{dist}(\Ga_{\ua/\da}, \R) 
> \varrho$ for some $\varrho >0$. The piecewise holomorphic matrix 
$\Pi$ appearing in the above figure is as defined through 
\eqref{ecriture rep int matrice Pi}. 
\label{Figure definition sectionnelle de la matrice chi}}
\end{center}
\end{figure}

\begin{theorem}
\label{Theorem structure inverse Id + V}

There exists $w_0$ large enough such that the operator $\e{id}+\op{V}$ 
acting on $L^2(\R+\i v)$ with the integral kernel 
\eqref{ecriture noyau integral type IIO pour V} is invertible for any 
$w\geq w_0$ with inverse given by $\e{id}-\op{R}$. \,The integral kernel 
of the  resolvent operator $\op{R}$ is expressed as
\beq
R(\la,\mu) \; = \; \f{ \Big( \bs{F}_{L}(\la), 
\bs{F}_R(\mu) \Big)  }{ \la-\mu } \;. 
\label{ecriture noyau resolvent de Id + V}
\enq
The vectors $\bs{F}_{R/L}(\la)$ are  
given by
\beq
\bs{F}_{R}(\la) \, = \, \chi_{+}(\la)\cdot \bs{E}_{R}(\la) \qquad 
and \qquad \bs{F}_{L}^{\op{t}}(\la) \, = \, \bs{E}_{L}^{\op{t}}(\la) 
\cdot \chi_{+}^{-1}(\la)  \,  .
\label{ecriture solution F R et F L}
\enq
Above, $\op{t}$ is the vector transposition while $\chi$ corresponds 
to the unique solution to the matrix Riemann-Hilbert problem for
\footnote{In Appendix \ref{Appendix Secion inversion operateur L0},
symbols $\chi$, $\Xi$, $\Upsilon$, $\Pi$, $\mc{P}$ and $\mc{G}$ 
stand for matrix-valued functions defined below.} $\chi$: \,find 
$\,\chi \in \mc{M}_2\Big( \mc{O}\big( \Cx \setminus \big\{ \R  + \i v \big\} 
\big) \Big)\,$ such that

\begin{itemize}
 \item $\chi(\la)=I_2+\e{O}\Big(\f{1}{\la}\Big) $ when $\la \tend \infty$;
\item $\chi$ admits continuous $\pm$ boundary values on $\R$ such that 
$ \chi_{\pm}-I_2 \in \mc{M}_2\big( L^{2}(\R+\i v) \big)$. These boundary 
values are related by 
\beq
\chi_+(\la)\,  G_{\chi}(\la) \, = \, \chi_{-}(\la)\, ,
\enq
where the  jump matrix takes the form 
\beq
G_{\chi}(\la) \, = \, I_2 \, + \, 2\i \pi \bs{E}_{R}(\la) \cdot 
\bs{E}_{L}^{\op{t}} (\la) \; = \; \left( \ba{cc}  1+ \mc{F}[L^{(0)}](\la)   
& -\mc{F}[L^{(0)}](\la) \, e^{2}(\la)  \vspace{2mm} \\
														      \mc{F}[L^{(0)}](\la) \, e^{-2}(\la)  
&  1 -  \mc{F}[L^{(0)}](\la)   \ea \right)\;. 
\enq
The 
unique solution $\chi$ takes the explicit form given in 
Fig. \ref{Figure definition sectionnelle de la matrice chi}.
\,It 
admits the integral representations 
\beq
\chi(\la) \, = \; I_2 \, - \, \Int{ \R }{} \f{ \bs{F}_{R}(\mu) \cdot  
\bs{E}_{L}^{\op{t}}(\mu)   }{ \mu - \la } \dd \mu 
\qquad and \qquad 
\chi^{-1}(\la) \, = \; I_2 \, + \, \Int{ \R }{} \f{ \bs{E}_{R}(\mu) 
\cdot  \bs{F}_{L}^{\op{t}}(\mu)   }{ \mu - \la } \dd \mu \; .
\enq

\end{itemize}

\end{theorem}

Most results stated in Theorem 
\ref{Theorem structure inverse Id + V} are classic and  go back to 
the work 
\cite{ItsIzerginKorepinSlavnovDifferentialeqnsforCorrelationfunctions}. 
The representation given in 
Fig.~\ref{Figure definition sectionnelle de la matrice chi} is established 
throughout Subsection 
\eqref{SousSection preuve Factorisation Asymp RHP pour chi} to come
by a rather standard application of the non-linear steepest descent 
method \cite{DeiftZhouSteepestDescentForOscillatoryRHP}. 
It is a standard fact, which follows from $\det G_{\chi}=1$, that 
the Riemann-Hilbert problem for $\chi$ admits a unique solution, 
see \textit{e.g.} \cite{DeiftOrthPlyAndRandomMatrixRHP}.  
Thus, we will not discuss this question further.

\subsection{Asymptotic resolution of the 
Riemann-Hilbert problem}
\label{SousSection preuve Factorisation Asymp RHP pour chi}

\subsubsection{Riemann-Hilbert problem for $\Xi$}
\label{Subsection RHP for matrix Xi} 

First, we consider 
the solution to an auxiliary scalar Riemann-Hilbert problem. 
Let 
\beq
 \mc{B}_{\ua}^{(0)} \, = \, \Big\{ z \in \Cx \; : \;  
\Im z > v  \Big\} \qquad \e{and}\qquad 
 \mc{B}_{\da}^{(0)} \, = \, \Big\{ z \in \Cx \; : \;  \Im z < v  \Big\} \;. 
\enq
One introduces the function
\beq
\a^{(0)}  \in \mc{O}\big( \Cx^* \setminus \{\R + \i v \} \big)  
\qquad \e{with} \qquad 
\a^{(0)}(\la) \, = \, \left\{ \ba{cc}   \a^{(0)}_{\ua}(\la)  & 
\la \in \mc{B}_{\ua}^{(0)} \vspace{2mm}\\  \a^{(0)}_{\da}(\la)  & 
\la \in \mc{B}_{\da}^{(0)}  \ea \right. 
\enq
in which\footnote{Any function holomorphic on a closed set is, in fact understood to be holomorphic on an open neighbourhood thereof}  
$\a^{(0)}_{\ua}\in \mc{O}\Big( \ov{\mc{B}}_{\ua}^{(0)} \Big)$, 
$\,\a^{(0)}_{\da}\in \mc{O}\Big( \ov{\mc{B}}_{\da}^{(0)} 
\setminus \{ 0 \} \Big)$, $\,\a^{(0)}_{\ua/\da}(\la) \tend 1$ 
when $\la \tend \infty$ in $ \ov{\mc{B}}_{\ua/\da}^{(0)}$ and such that 
\beq
 \f{ \a^{(0)}_{\ua}(\la)  }{   \a^{(0)}_{\da}(\la)  } \; = \; 
1 -  \mc{F}[L^{(0)}](\la) \,. 
\enq
$ \a^{(0)}_{\ua/\da}$ admit meromorphic continuations to $\mc{B}_{\da/\ua}^{(0)}$ such that 
\beq
\a_{\ua}^{(0)}(k)  \; \widesim{k\tend 0}  k \, \a_{0}^{(0)} \qquad 
\e{and} \qquad \a_{\da}^{(0)}(k)  \; \widesim{k\tend 0}  
\f{\wt{\a}_{0}^{\, (0)} }{ k  }\;.  
\label{definition constantes alpha zero et alpha zero tilde}
\enq
Note that $k=0$ is the only zero and pole
of $\a_{\ua/\da}^{(0)}$ in a fixed $v$-independent tubular neighbourhood 
of $\R$. 

The functions $\a_{\ua/\da}^{(0)}$ can be read out from equations 
\eqref{forme explicite alpha up kappa ups}-\eqref{forme explicite alpha da kappa ups} upon the substitution $\varkappa^{\ups} \hookrightarrow \varkappa$. 

\vspace{2mm}

Assume that one is given a solution $\chi$ to the Riemann-Hilbert problem for $\chi$, 
and define 
\beq
\Xi(\la) \, =  \, \chi(\la) \cdot \Big( \a^{(0)}(\la) \Big)^{-\sg_3} \; . 
\enq
It is clear that the Riemann--Hilbert problem for $\chi$ is in 
one-to-one correspondence with the Riemann--Hilbert problem for $\Xi$. 
The latter consists in finding $\Xi \in \mc{M}_2\Big( \mc{O}\big( \Cx^{*} 
\setminus \big\{ \R +\i v \big\} \big) \Big)$ such that

\begin{itemize}
 \item $\Xi$ admits a simple pole at $0$; 
 \item $\Xi(\la) = I_2 + \e{O}\Big(\f{1}{\la}\Big) $ when $\la \tend \infty$;
\item $\Xi(\la) \cdot \Big( \a^{(0)}_{\da}(\la) \Big)^{\sg_3}$ is 
regular at $\la=0$; 
\item $\Xi$ admits continuous $\pm$ boundary values on $\R+\i v$ such that 
$ \Xi_{\pm}-I_2 \in \mc{M}_2\big( L^{2}(\R+\i v) \big)$. These boundary 
values are related as 
\beq
\Xi_+(\la)\,  G_{\Xi}(\la) \, = \, \Xi_{-}(\la)\;, \qquad \e{where} 
\qquad G_{\Xi}(\la) \; = \; \left( \ba{cc}  1+ P(\la)Q(\la)  & 
P(\la) e^{2}(\la) \\  
						 Q(\la) e^{-2}(\la)  &  1    
\ea \right)  \,. 
\enq
\end{itemize}

Note that the jump matrix factorises as $\,G_{\Xi}(\la) \; 
= \; M_{\ua}(\la) \cdot M_{\da}(\la)$ in which 
\beq
M_{\ua}(\la) \; = \; 
  \left( \ba{cc}  1  & P(\la) e^{2}(\la) \\  
		    0 &  1    \ea \right)    \qquad \e{and} \qquad 
M_{\da}(\la) \, = \, \left( \ba{cc}  1    &    0   \\  
														      Q(\la) e^{-2}(\la)  &  1    
\ea \right)    \,. 
\enq
The expression for these matrices involve the functions 
\beq
P(\la) \; = \; - \a^{(0)}_{\ua}(\la)\cdot \a^{(0)}_{\da}(\la)
\cdot\mc{F}[L^{(0)}](\la) \qquad \e{and} \qquad 
Q(\la) \; = \;  \f{  \mc{F}[L^{(0)}](\la)  }{  \a^{(0)}_{\ua}(\la)\cdot
\a^{(0)}_{\da}(\la)  } \;. 
\enq
In particular, $Q$ is analytic on a tubular neighbourhood of $\R$ and 
satisfies 
\beq
Q(0)\; = \; \f{1}{\a_0^{(0)}\cdot \wt{\a}_{0}^{(0)} } \, = \, -1 \,, 
\enq
 see \eqref{for b(0)=-1}.

The matrices $M_{\ua/\da}$ are such that their off-diagonal entries are 
exponentially small in $w$ for $\la$ 
belonging to $\mathbb{H}^{\pm}$ and uniformly away from $\R$.

\subsubsection{Riemann-Hilbert problem for $\Ups$}

\begin{figure}[h]
\begin{center}

\includegraphics[width=.7\textwidth]{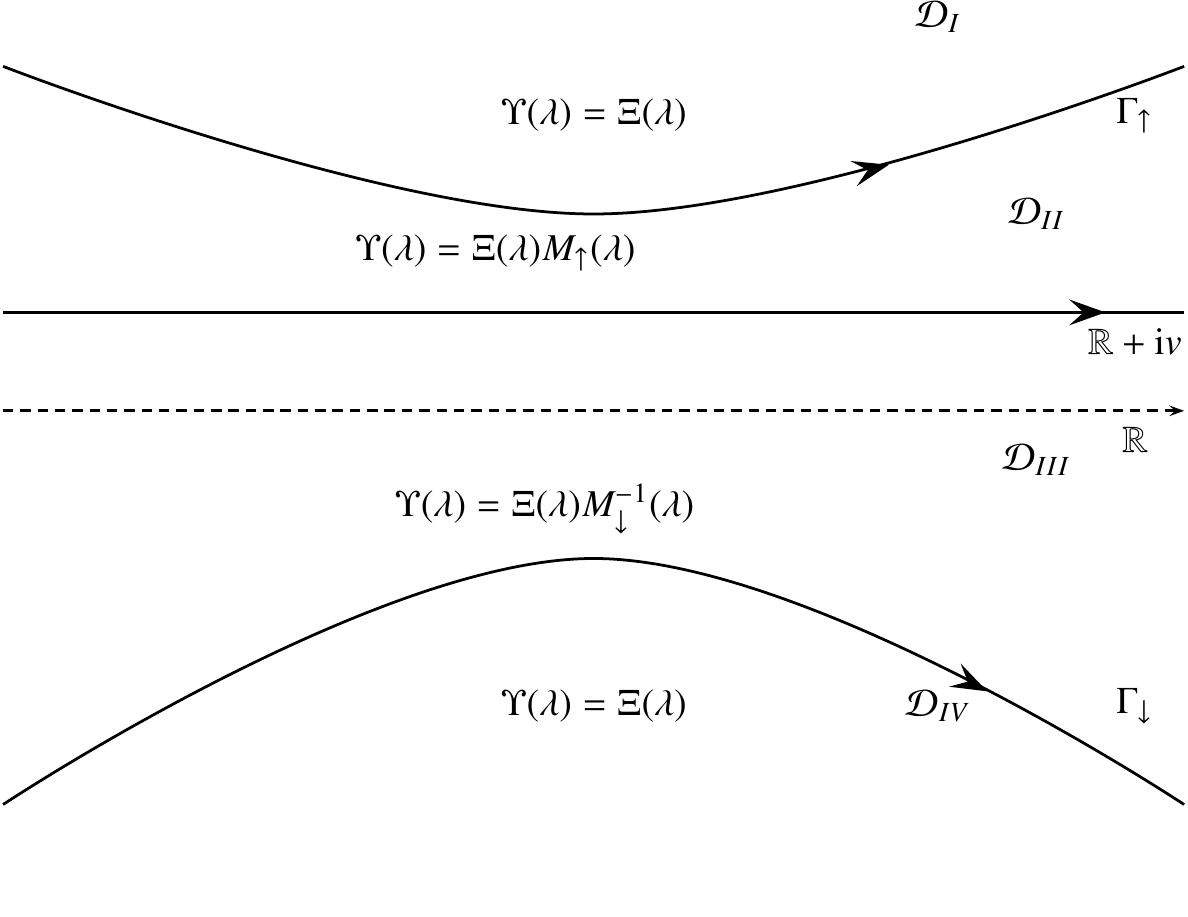}

\caption{Piecewise definition of the matrix $\Ups$ in terms of the 
matrix $\Xi$. The curves $\Ga_{\ua/\da}$ separate all poles, other 
than the one at $0$, of $\la \mapsto \mc{F}[L^{(0)}](\la) \cdot 
\Big\{ 1 - \mc{F}[L^{(0)}](\la) \Big\}^{-1}$ from $\R$ and are such 
that $\e{dist}(\Ga_{\ua/\da}, \R) > \varrho$
for some $\varrho >0$. 
\label{Figure definition sectionnelle de la matrice Ups}}
\end{center}
\end{figure}

Next, one defines $\Ups$ as in Fig. 
\ref{Figure definition sectionnelle de la matrice Ups}. The contours 
$\Ga_{\ua/\da}$ are chosen such that it holds $\Ga_{\ua}=-\Ga_{\da}$. 
It is clear that the Riemann-Hilbert problems for $\Xi$ is in one-to-one 
correspondence with the one for $\Ups$.

Find $\Ups \in \mc{M}_2\Big( \mc{O}\Big( \Cx^{*} \setminus 
\big\{ \Ga_{\ua} \cup   \Ga_{\da} \big\} \Big) \Big)$ 
such that 
\begin{itemize}
 \item  $\Ups$ admits a simple pole at $0$; 
 \item $\Ups(\la)=I_2+\e{O}\Big(\f{1}{\la}\Big) $ when $\la \tend \infty$;
\item $\Ups(\la)\cdot M_{\da}(\la) \cdot \Big( \a^{(0)}_{\da}(\la) 
\Big)^{\sg_3}$ is regular at $\la=0$; 
\item $\Ups$ admits continuous $\pm$ boundary values on $\Ga_{\ua} \cup   \Ga_{\da}$
such that 
$ \Ups_{\pm}-I_2 \in \mc{M}_2\big( L^{2}(\Ga_{\ua} \cup   \Ga_{\da}) \big)$. These boundary values are related by 
\beq
\Ups_+(\la)\,  G_{\Ups}(\la) \, = \, \Ups_{-}(\la)\,  
\qquad \e{with} \qquad G_{\Ups}(\la) \, = \,  
M_{\da}(\la)\cdot \bs{1}_{ \Ga_{\da} }(\la) \, 
+ \, M_{\ua}(\la)\cdot \bs{1}_{ \Ga_{\ua} }(\la) \;. 
\enq

\end{itemize}

\subsubsection{Auxiliary Riemann--Hilbert problem for $\Pi$}

To continue further, one first introduces $\Pi$ as the unique solution 
to the below Riemann-Hilbert problem for $\Pi$. Find $\Pi \in 
\mc{M}_2\Big( \mc{O}\big( \Cx \setminus \big\{ \Ga_{\ua} 
\cup   \Ga_{\da} \big\} \big) \Big)$
such that:
\begin{itemize}
 \item $\Pi(\la)=I_2+\e{O}\Big(\f{1}{\la}\Big) $ when $\la \tend \infty$; 
\item $\Pi$ admits continuous $\pm$ boundary values on  
$\Ga_{\ua} \cup   \Ga_{\da}$ 
such that $ \Pi_{\pm}-I_2 \in \mc{M}_2\big( L^{2}(\Ga_{\ua} \cup   \Ga_{\da}
) \big)$. These boundary values are related by 
\beq
\Pi_+(\la)\,  G_{\Ups}(\la) \, = \, \Pi_{-}(\la) \;. 
\enq

\end{itemize}

Again, there exists at most a one solution to the Riemann-Hilbert problem 
for $\Pi$. Existence may be established by the singular integral equation 
method introduced in 
\cite{BealsCoifmanScatteringInFirstOrderSystemsEquivalenceRHPSingIntEqnMention}. 

Indeed, introduce the singular integral operator on the space 
$\mc{M}_2\big( L^2(  \Ga_{\ua} \cup   \Ga_{\da}  ) \big)$ 
of $2\times 2$ matrix-valued $L^2\big(  \Ga_{\ua} \cup   \Ga_{\da}  
\big)$ functions by 
\beq
\mc{C}^{(+)}_{  \Ga_{\ua} \cup   \Ga_{\da}  }\big[ \Psi \big] (\la) \; 
= \; \lim_{ \substack{ z \tend \la \\ z \in + \e{side}  \, \e{of} \, 
\Ga_{\ua} \cup   \Ga_{\da}  } } 
\int_{  \Ga_{\ua} \cup   \Ga_{\da}  }{} \f{  \Psi(t)\cdot  
(G_{\Ups}-I_2)(t)  }{t-z} \cdot \f{ \dd t}{ 2 \i \pi } \;. 
\enq
Since $ G_{\Ups}-I_2 \in \mc{M}_2\Big( \big(L^{\infty} \cap L^2 \big) 
\big(  \Ga_{\ua} \cup   \Ga_{\da}  \big) \Big)$ and $ \Ga_{\ua} 
\cup   \Ga_{\da} $ is a Lipschitz curve, 
it follows from 
\cite{CalderonContinuityCauchyTransformLipschitzCurves} that 
${\cal C}^{(+)}_{ \Ga_{\ua} \cup   \Ga_{\da}  }$ is   continuous  on 
$\mc{M}_2\Big(L^2\big(  \Ga_{\ua} \cup   \Ga_{\da}  ) \Big)$ 
and fulfils:
\beq
\big| \big| \big|  \mc{C}^{(+)}_{  \Ga_{\ua} \cup   \Ga_{\da}  }   
\big| \big| \big|_{ \mc{M}_2(L^2(  \Ga_{\ua} \cup   \Ga_{\da}  )) }  
\; \leq  \; 
C \ex{- \varrho w} \;. 
\enq
Hence, since 
\beq
G_{\Ups}-I_2 \in \mc{M}_2\Big( L^2\big(  \Ga_{\ua} \cup   \Ga_{\da}  
\big) \Big)  \quad \e{and} \quad 
\mc{C}^{(+)}_{  \Ga_{\ua} \cup   \Ga_{\da}   }[I_2] \in  
\mc{M}_2\Big( L^2\big(  \Ga_{\ua} \cup   \Ga_{\da}  \big) \Big) \;, 
\enq
 provided that $w$ is large enough, it follows that the singular integral 
equation
\beq
\Big(I_2 + \mc{C}^{(+)}_{ \Ga_{\ua} \cup   \Ga_{\da}  }  \Big)
\big[ \Pi_+ \big] \; = \; I_2
\label{ecriture eqn int sing pour matrice Pi moins}
\enq
admits a unique solution $\Pi_{+}$ such that $\Pi_{+} - I_2 \in 
\mc{M}_2\Big(  L^2  \big(  \Ga_{\ua} \cup   \Ga_{\da}  ) \Big)$. 
It is then a standard fact 
\cite{BealsCoifmanScatteringInFirstOrderSystemsEquivalenceRHPSingIntEqnMention} in the theory of Riemann-Hilbert problems that the matrix  
\beq
\Pi(\la) \; = \; I_2 \; - \; \Int{ \Ga_{\ua} \cup   \Ga_{\da} }{} 
\f{  \Pi_+(t) (G_{\Ups}-I_2)(t)  }{ t-\la } \cdot \f{ \dd t }{ 2\i\pi }
\label{ecriture rep int matrice Pi}
\enq
is the unique solution to the Riemann-Hilbert problem for $\Pi$. 
It is a direct consequence of the Neumann expansion of the solution to 
the singular integral equation 
\eqref{ecriture eqn int sing pour matrice Pi moins} for $\Pi_+$ 
and of the local holomorphicity of the jump matrices that,  for some 
$\varrho >0$, 
\beq
\Pi(\la) \; = \; I_2 \, + \, \e{O}\bigg( \f{ \ex{-\varrho w} }{ 1+|\la| } 
\bigg)
\label{ecriture estimee sur Pi uniforme}
\enq
uniformly on $\Cx$ and with a differentiable remainder.

The piecewise holomorphic matrix $\Pi$ thus constructed enjoys a few 
properties that will be useful below. 
Indeed, one readily infers from the identity $M_{\ua}(-\la) \, = 
\, \sg^x \cdot M_{\da}^{-1}(\la) \cdot \sg^x$, adjoined to the contour 
symmetry 
$\Ga_{\ua}\, =\, \big\{ -z \,: \, z \in  \Ga_{\da} \big\}$ and the 
uniqueness of the Riemann--Hilbert problem for $\Pi$ 
that  
the relation
$\Pi(\la) \, = \, \sg^x \cdot \Pi(-\la) \cdot \sg^x$  
holds. In particular,
\beq
\Pi(0) \, = \, \sg^x \cdot \Pi(0) \cdot \sg^x \qquad \e{and} \qquad 
\Pi^{\prime}(0) \, = \, -\sg^x \cdot \Pi^{\prime}(0) \cdot \sg^x \;. 
\enq
These properties lead to the the $\la \tend 0$ expansion 
\beq
\Pi(\la) \; = \; \left(\ba{cc} \Pi_{11}(0)   & \Pi_{21}(0)  \\ 
\Pi_{21}(0)  &  \Pi_{11}(0)  \ea \right) 
\, + \, \la \left(\ba{cc} \Pi_{11}^{\prime}(0)   
&- \Pi_{21}^{\prime}(0)  \\ \Pi_{21}^{\prime}(0)  &  -\Pi_{11}^{\prime}(0)  
\ea \right) \; + \; \e{O}(\la^2) \;. 
\enq
In other words,  by setting
\beq
c_1=\Pi_{11}(0)\Pi_{11}^{\prime}(0) - \Pi_{21}(0)\Pi_{21}^{\prime}(0)  
\qquad \e{and} \qquad  c_2=\Pi_{11}(0)\Pi_{21}^{\prime}(0) 
- \Pi_{21}(0)\Pi_{11}^{\prime}(0)\; ,
\label{definition coefficients c1 et c2}
\enq
since $\,\det\Pi(\la)=1$, \,one infers
that 
\beq
\Pi^{-1}(0) \Pi(\la) \; = \; I_2 \, + \, \la \left(\ba{cc}     c_1   
& -c_2 \\   c_2  &   - c_1 \ea  \right)  \, + \, \e{O}(\la^2)  \;. 
\enq

\subsubsection{Solution of the Riemann--Hilbert problem for $\Ups$}

With $\Pi$ defined, the solution to the Riemann-Hilbert problem for $\Ups$ can 
be constructed as $\Ups(\la) \, = \, \mc{P}(\la) \cdot \Pi(\la)$, 
where $\mc{P}(\la)$ is a meromorphic matrix on $\Cx$ whose only pole 
is located at $\la=0$. Below, we establish that this meromorphic matrix 
takes the form
\beq
\mc{P}(\la) \; = \; \Pi(0) \cdot \bigg(I_2  \, 
+ \, \f{ \th }{ \la } \op{D} \bigg) \cdot \Pi^{-1}(0)
\enq
where
\beq
\op{D} \, = \, \left(\ba{cc}  -1   & - 1 \\ 1   &   1 \ea \right)  
\qquad \e{and} \qquad \th \, = \, \f{  1 }{  (e^{-2}Q)^{\prime}(0)
+2(c_1 +c_2)  } \;,  
\label{definition matrice D et prefacteur theta}
\enq
with $c_1,c_2$ as introduced in \eqref{definition coefficients c1 et c2}. 
The matrix $\mc{P}$ is constructed so that 
\beq
\la \; \mapsto \;  \mc{P}(\la) \Pi(\la)  M_{\da}(\la)  
\Big( \a^{(0)}(\la) \Big)^{\sg_3}
\enq
is regular at $\la=0$.

When looking for $\mc{P}$, it is convenient to parameterise
\beq
\mc{P}(\la) \, = \, \Pi(0) \mc{G}(\la) \Pi^{-1}(0) \qquad \e{with} 
\qquad    \mc{G}(\la)  \, = \,  I_2 \, + \, \f{1}{\la} \left(  \ba{cc} 
g_{11}  & g_{12}  \\  g_{21}  &  g_{22} \ea \right) 
\enq
Then, one has $\mc{P}(\la) \Pi(\la) M_{\da}(\la) 
\Big( \a^{(0)}_{\da}(\la) \Big)^{\sg_3}\, = \, \Pi(0)  H(\la)$, with 
\bem
H_{11}(\la) \; = \; \a^{(0)}_{\da}(\la)   
\Big[  \mc{G}_{11}(\la)+Q(\la)e^{-2}(\la) \mc{G}_{12}(\la) \Big]
\, + \, c_1\la \a^{(0)}_{\da}(\la)   
\Big[  \mc{G}_{11}(\la)-Q(\la)e^{-2}(\la) \mc{G}_{12}(\la) \Big]  \\
\, + \, c_2 \la \a^{(0)}_{\da}(\la)   \Big[  
\mc{G}_{12}(\la)-Q(\la)e^{-2}(\la) \mc{G}_{11}(\la) \Big] + \e{O}(1)   \;,     
\end{multline}
as well as $H_{21}=[H_{11}]_{\mid \mc{G}_{1a} \hookrightarrow \mc{G}_{2a} }$ 
and $ H_{a2}(\la) \, = \, \e{O}(1)$, when $\la \tend 0$.  

In principle, $H_{11}$ admits a second order pole at $\la=0$. By imposing 
that $H_{11}$ is regular at $\la=0$, one obtains the system of equations
on the coefficients $g_{1a}$:
\beq
g_{11}\,=\, -Q(0) g_{12} \qquad \e{and} \qquad 
g_{12}\cdot \bigg[(e^{-2}Q)^{\prime}(0)-2 c_1 Q(0)
+ c_2 \big[1+Q^2(0) \big]  \bigg] = -1 \;. 
\enq
These equations are solvable owing to $|c_1|+|c_2|
=\e{O}(  \ex{-\varrho w } )$, what is in itself a consequence of 
\eqref{ecriture estimee sur Pi uniforme}. 
 
Likewise, by requiring that $H_{21}$ is regular at $\la=0$, one 
obtains the system of equations on the coefficients $g_{2a}$:
\beq
g_{21}\,=\, -Q(0) g_{22} \qquad \e{and} \qquad 
g_{22}\cdot \bigg[(e^{-2}Q)^{\prime}(0)-2 c_1 Q(0)+ 
c_2 \big[1+Q^2(0) \big]  \bigg] = -Q(0) \;. 
\enq
All in all, this yields that 
\beq
\left(  \ba{cc} g_{11}  & g_{12}  \\  g_{21}  &  g_{22} \ea \right)  \; 
= \; \f{ 1 }{ (e^{-2}Q)^{\prime}(0)-2 c_1 Q(0)+ c_2 \big[1+Q^2(0) \big] }  
\cdot 
\left(  \ba{cc} Q(0)  & -1  \\  Q^2(0)  &  -Q(0)  \ea \right)  \;. 
\enq
The form of $\mc{P}(\la)$ then follows upon recalling that $Q(0)=-1$. 

\vspace{2mm}

By tracing backwards the various transformations, one gets that 
the unique solution $\chi$ to the Riemann-Hilbert problem for $\chi$ 
takes the piecewise form as depicted in 
Fig. \ref{Figure definition sectionnelle de la matrice chi}.

\subsection{Resolvent kernel of  $\,\e{id}+\op{V}$}
\label{Appendix SousSection nouyau resolvent}

\noindent It follows from the results of Theorem 
\ref{Theorem structure inverse Id + V} that the solution to 
\eqref{ecriture lin int eqn avec operateur V} takes the form 
\beq
\mc{F}[f](k)\; = \; \mc{F}[h](k)\; - \; \Int{ \R + \i v }{} 
\hspace{-2mm}\dd \mu\,  R(k,\mu) \mc{F}[h](\mu)   \;. 
\enq

Since $\chi_{+}(\la) \bs{E}_{R}(\la)=\chi_{-}(\la) \bs{E}_{R}(\la)$, 
and since the vectors $\bs{E}_{L/R}$ are analytic in a tubular 
neighbourhood of $\R$, it follows that $R(\la,\mu)$ is also analytic 
in some open neighbourhood of $\R^2$.

\subsubsection{Support restrictions}

One may explicitly check that the integral term only involves 
the values of $h$ inside of $\intff{-w}{w}$. Indeed, one has 
\beq
\Int{ \R + \i v }{}\hspace{-2mm}\dd \mu\, \ex{\i \mu x} R(k,\mu)  \, 
= \, \Int{ \R + \i v }{} \hspace{-2mm}\dd \mu \, \ex{\i \mu x} 
\f{ \Big( \bs{E}_{L}(k), \chi_{+}^{-1}(k) \cdot \chi_+(\mu) 
\bs{E}_R(\mu) \Big)  }{ k - \mu } 
\enq
 If $x>w$, then $\ex{\i \mu x}  \bs{E}_{R}(\mu)$ is bounded on 
$\mathbb{H}^{+}_{v}=\mathbb{H}^+ + \i v$, and so, since the integrand 
vanishes at $\infty$ in $\mathbb{H}^+_{v}$, one obtains zero by 
deforming the integration contour to $+\i\infty$. 
One arrives to the same conclusion when $x<-w$ upon using 
$\chi_{+}(\mu) \bs{E}_{R}(\mu)=\chi_{-}(\mu) \bs{E}_{R}(\mu)$. 
Hence, for \textit{any} function $h$ on $\R$ with exponential decay 
at $\pm \infty$, one gets, for $0<v$ small enough, that 
\beq
 \Int{ \R + \i v }{}\hspace{-2mm}\dd \mu\,   R(k,\mu) 
\mc{F}[h](\mu)  \; = \; \Int{ \R + \i v }{} \hspace{-2mm}\dd \mu\,  
R(k,\mu) \mc{F}[h \bs{1}_{\intff{-w}{w}}](\mu)  \;. 
\enq

\subsubsection{Leading asymptotic form of the resolvent}

The resolvent may be approximated, in the large-$w$ limit, by inserting 
the leading behaviour of the matrix $\chi$ into the expression for 
the vectors $\bs{F}_{R/L}$ \eqref{ecriture solution F R et F L}, and 
then inserting the latter into the formula for the resolvent kernel 
\eqref{ecriture noyau resolvent de Id + V}.

For further convenience, given $\op{D}$ as in 
\eqref{definition matrice D et prefacteur theta}, set 
\beq
\mc{P}_{\infty}(\la) \; = \; I_2 \; + \, \f{1}{\la 
\mf{b}^{\prime}(0)}\op{D} \qquad \e{with} \qquad 
\mf{b}(\la) \; = \; \f{ e^{-2}(\la)  }{ \a_{\da}^{(0)}(\la) 
\a_{\ua}^{(0)}(\la) } \;, 
\label{definition P infty et mathfrak d}
\enq
so that, by using that $ (e^{-2}Q)^{\prime}(0) \; = \; 
\mf{b}^{\prime}(0)$, one may decompose
\beq
\mc{P}(\la) \; = \;  \mc{P}_{\infty}(\la) \, + \, 
\de \mc{P}(\la) \qquad \e{with} \qquad  \de \mc{P}(\la) \; 
= \;  \e{O}\bigg(  \f{ \ex{-\varrho w} }{ |\la| } \bigg) \;. 
\enq
It is as well convenient to introduce an analogous parameterisation 
gathering the exponentially small corrections to $\Pi(\la)=I_2 \, 
+ \, \de \Pi(\la)$, where, by virtue of 
\eqref{ecriture estimee sur Pi uniforme}, one has 
\beq
\de \Pi(\la) \; = \; \e{O}\bigg(  \f{ \ex{-\varrho w} }{ 1+|\la| } \bigg)
\enq
uniformly on $\Cx$.

From there, one obtains that, uniformly throughout the region 
$\mc{D}_{II}$, as defined in Fig. 
\ref{Figure definition sectionnelle de la matrice chi}, one has 
\beq
\chi(\la) \; = \; \chi_{\infty}^{(II)}(\la) \, 
+ \, \de \chi^{(II)}(\la) \qquad \e{with} \qquad  
\chi_{\infty}^{(II)}(\la)
\; = \; \mc{P}_{\infty}(\la) M_{\ua}^{-1}(\la) 
\big[ \a_{\ua}^{(0)}(\la) \big]^{\sg_3} \;, 
\enq
and 
\beq
\de \chi^{(II)}(\la)\;=\; \de\mc{P}(\la) \Pi(\la)   M_{\ua}^{-1}(\la) 
\big[ \a_{\ua}^{(0)}(\la) \big]^{\sg_3}
\; + \; \mc{P}_{\infty}(\la)  \de \Pi(\la)  M_{\ua}^{-1}(\la) 
\big[ \a_{\ua}^{(0)}(\la) \big]^{\sg_3}  \;. 
\enq
By direct inspection, one obtains that  uniformly 
in $\la \in \ov{\mc{D}}_{II}$,  
\beq
\de \chi^{(II)}(\la) \; = \; \e{O}\bigg( 
\f{ \ex{-  \varrho w    } }{ 1 + |\la| }  \bigg)\;. 
\enq

Likewise, uniformly throughout the region $\mc{D}_{III}$, one has 
the decomposition
\beq
\chi(\la) \; = \; \chi_{\infty}^{(III)}(\la) \, 
+ \, \de \chi^{(III)}(\la) \qquad \e{with} \qquad  
\chi_{\infty}^{(III)}(\la) \; = \; \mc{P}_{\infty}(\la) M_{\da}(\la) 
\big[ \a_{\da}^{(0)}(\la) \big]^{\sg_3} \;, 
\enq
and 
\beq
\de\chi^{(III)}(\la)\;=\;  
\de\mc{P}(\la) \Pi(\la)  M_{\da}(\la) \big[ \a_{\da}^{(0)}(\la) 
\big]^{\sg_3}  
\; + \; \mc{P}_{\infty}(\la)  \de \Pi(\la)  M_{\da}(\la) 
\big[ \a_{\da}^{(0)}(\la) \big]^{\sg_3} \;. 
\enq
Again, a direct analysis shows that for $\la \in \mc{D}_{III}$ and 
uniformly away from $0$,  
one has  
\beq
\de \chi^{(III)}(\la) \; = \; \e{O}\bigg( 
\f{ \ex{-  \varrho w  + 2 v w  } }{ 1 + |\la| }  \bigg)\;. 
\enq
 Note that the additional term $\ex{ 2 v w  }$ present in the estimates 
on the remainder is due to
the presence of $e^{-2}$ in the off-diagonal entry  of $M_{\da}$ and 
the fact that $\mc{D}_{III}\cap \mathbb{H}^{+} \, 
= \,  \Big\{ \la \in \Cx \, : \, 0 < \Im(\la) < v \Big\}$. 
 
These formulae allow  
to compute the leading behaviour of the vector $\bs{F}_R(\la)$ inside 
each of the domains. One infers
that 
\beq
\bs{F}_R(\la)=\bs{F}_{R;\infty}(\la)\, + \, \de \bs{F}_R^{(A)}(\la)  
\quad \e{with} \quad  \de \bs{F}_R^{(A)}(\la) \, = \, \de\chi^{(A)}(\la) 
\bs{E}_R(\la) 
\quad \e{for} \quad \la \in \mc{D}_{A} \;, \; A \in \{II, III\}\;. 
\enq
We stress that the expression for $\bs{F}_{R;\infty}(\la)$ does not 
depend on whether $\la \in \mc{D}_{II}$ or $\la \in \mc{D}_{III}$. 
A direct calculation shows that 
\bem
\bs{F}_{R;\infty}(\la) = \f{ 1 }{ 2 \i \pi }  \mc{P}_{\infty}(\la) \cdot 
\left(\ba{c}  \a_{\da}^{(0)}(\la) e(\la)   \\  \big\{ \a_{\ua}^{(0)}(\la)  
e(\la) \big\}^{-1}    \ea  \right)  \\
\; =  \; \f{ 1 }{ 2 \i \pi }   
\left(\ba{c}  \a_{\da}^{(0)}(\la) e(\la) \, 
- \,  \f{ 1 }{ \la \mf{b}^{\prime}(0) } 
\Big[ \a_{\da}^{(0)}(\la) e(\la) \,  
+ \, \big\{ \a_{\ua}^{(0)}(\la)   e(\la) \big\}^{-1}  \Big] \\
\big\{ \a_{\ua}^{(0)}(\la)  e(\la) \big\}^{-1}   \, 
+ \,  \f{ 1 }{ \la \mf{b}^{\prime}(0) } 
\Big[ \a_{\da}^{(0)}(\la) e(\la) \,  
+ \, \big\{ \a_{\ua}^{(0)}(\la)   e(\la) \big\}^{-1}  \Big]    
\ea  \right) \\
\; = \;  \f{ 1 }{ 2 \i \pi }   
\left(\ba{c} \a_{\da}^{(0)}(\la) e(\la) \cdot  \Big[ 1 \, 
- \,  \f{ 1 }{ \la \mf{b}^{\prime}(0) } \big( 1+\mf{b}(\la) \big) 
\Big] \\
 \big\{  \a_{\ua}^{(0)}(\la) e(\la) \big\}^{-1} \cdot  
\Big[ 1 \,+ \,  \f{ 1 }{ \la \mf{b}^{\prime}(0) } 
\big( 1 + \big\{ \mf{b}(\la)\big\}^{-1} \big) \Big] \ea \right) 
 \; = \; \f{ 1 }{ 2 \i \pi }  \left(\ba{c}  f_{+;\infty}(\la) \\
   f_{-;\infty}(\la)
\ea  \right) \;. 
\end{multline}
Above, $\mf{b}$ is as introduced in 
\eqref{definition P infty et mathfrak d}. It is easy to see that 
the above expression for $f_{\pm;\infty}(\la)$ is analytic in 
a tubular neighbourhood of $\R$. In particular, there is no pole
at $\la=0$ as follows from $\mf{b}(0)=-1$.

Similarly, using the relation $\det\chi(\lambda)=1$, one infers that
\beq
\bs{F}_L(\la)=\bs{F}_{L;\infty}(\la)\, + \, \de \bs{F}_L^{(A)}(\la)  
\quad \e{with} \quad  \de \bs{F}_L^{(A)}(\la) \, = \,
\e{CoMat}\Big(\de\chi^{(A)}(\la)\Big) 
\bs{E}_L(\la) 
\quad \e{for} \quad \la \in \mc{D}_{A} \;, \; A \in \{II, III\} 
\enq
where
\beq
\bs{F}_{L;\infty}(\la)=- \mc{F}\big[ L^{(0)} \big](\la)
\left(\ba{c} - f_{-;\infty}(\la) \\
   f_{+;\infty}(\la)
\ea  \right)
\enq
and $\e{CoMat}(M)$ stands for the Comatrix of $M$.

From the above one infers that the resolvent admits the following 
expansion 
\beq
R(\la,\mu) \; = \;  R_{\infty}(\la,\mu) \,  + \, \de R(\la,\mu)  \qquad \e{uniformly}\, \e{in} \quad \mc{D}_{II}\cup\mc{D}_{III} \;,  
\enq
where 
\beq
 R_{\infty}(\la,\mu) \; = \; \f{  - \mc{F}\big[ L^{(0)} \big](\la) }{ 2\i \pi (\la-\mu) } \cdot \Big(  -f_{-;\infty}(\la) \; \;  f_{+;\infty}(\la) \Big)\cdot
\left(\ba{c}  f_{+;\infty}(\mu) \\
   f_{-;\infty}(\mu)
\ea  \right)
\label{definition noyau dominant}
\enq
while, for $(\la,\mu) \in \mc{D}_{A}\times \mc{D}_{B}$ 
with
$A,B \in \{II, III\}$,  
\bem
 \de R(\la,\mu)  \; = \; \f{1}{\la-\mu} \Bigg\{   \Big(\bs{E}_L(\la), ^{\op{t}}\e{CoMat}\Big( \de \chi^{(A)}(\la) \Big) \bs{F}_{R;\infty}(\mu) \Big) \; + \;  
 \Big(\bs{F}_{L;\infty}(\la),  \de \chi^{(B)}(\mu)    \bs{E}_{R}(\mu) \Big) \\ 
\; + \;   \Big(\bs{E}_L(\la), ^{\op{t}}\e{CoMat}\Big( \de \chi^{(A)}(\la) \Big) \cdot \de \chi^{(B)}(\mu)    \bs{E}_{R}(\mu)  \Big) \Bigg\} \;.
\label{definition noyau perturbe}
\end{multline}
The leading resolvent may be explicitly cast as
\beq
 R_{\infty}(\la,\mu) \; = \;  \f{ - \mc{F}[L^{(0)}](\la) }{ 2\i\pi (\la 
- \mu) } 
\Big( - \{ \a_{\ua}^{(0)}(\la) e (\la) \}^{-1}  \; , \;  
\a_{\da}^{(0)}(\la) e(\la) \Big) 
\cdot  \bigg(I_{2}\, + \, \f{\la-\mu}{\la\mu \mf{b}^{\prime}(0) } 
\op{D} \bigg) 
\left(\ba{c}  \a_{\da}^{(0)}(\mu) \; e(\mu)  \vspace{2mm} \\ 
\{ \a_{\ua}^{(0)}(\mu)  e (\mu) \}^{-1}    \ea \right)  
\label{expression explicite noyau resolvent dominant}
\enq
as follows from $\op{D}^2=0$.

Obviously, $R_{\infty}(\la,\mu)$ is analytic in a tubular neighbourhood 
of $\R^2$ and satisfies, for some $\a>0$,  
the bounds 
\beq
 \big|  R_{\infty}(\la,\mu) \big| \, = \, \e{O}\Bigg( 
\f{   \ex{-\a |\la|}  }{ |\la-\mu|   } \ex{ w ( |\Im(\la)| 
+ |\Im(\mu)| ) }  \Bigg) \;, 
\label{ecriture borne sur noyau resolvent dominant}
\enq
which is valid throughout  $\Big\{ \mc{D}_{II} \cup  \mc{D}_{III} 
\cup \big\{ \R + \i v \big\} \Big\}^2$,  
\,provided that $\la, \mu$ are both uniformly away from $0$.  

Since $\de R= R-R_{\infty}$, one  
infers 
that $\de R$ is analytic in a tubular neighbourhood of $\R^2$. 
One can bound $\de R$, globally on $\Big\{ \mc{D}_{II} \cup  
\mc{D}_{III} \cup \big\{ \R + \i v \big\} \Big\}^2$ by using its 
patch-wise valid decomposition. This yields that 
\beq
 \big| \de R(\la,\mu) \big| \, = \, \e{O}\Bigg( 
\f{  \ex{-\a |\la|- \varrho w + 4 v w }  }{  | \la - \mu |   } 
\ex{w(|\Im(\la)|+|\Im(\mu)|)}  \Bigg) \;. 
\label{ecriture borne sur noyau resolvent perturbe}
\enq

Upon putting these two bounds together, one  
concludes 
that for $\la, \mu$ throughout $\Big\{ \mc{D}_{II} \cup  
\mc{D}_{III} \cup \big\{ \R + \i v \big\} \Big\}$ but both uniformly 
away from $0$,  
\beq
\big| R(\la,\mu) \big| \, \leq \,  \f{ C \ex{-\a |\la|}  }{ |\la-\mu|   } \ex{w(|\Im(\la)|+|\Im(\mu)|)} \;,  
\label{ecriture borne sur noyau resolvent}
\enq
for some $\a >0$.

\section{Auxiliary Lemma}
\label{Appendix Auxiliary Lemma}

\begin{lemme}
\label{Lemme bornes sur croissance via transfo Cauchy}
 
Given $\sg, v>0$ and $r \in \mathbb{N}$ there exists $C>0$ such that 
one has the upper bound
\beq
 \Int{ \R \pm \i (\sg +v) }{} \hspace{-4mm} \dd t \cdot 
\f{  \big[ \ln (1+|t|) \big]^{r} }{ (1+|t|)\cdot  |k-t|} \; 
\leq  \; C  \cdot \f{ \big[ \ln (1+|k|) \big]^{r + 1}  }{  1+|k|  } \;, 
\label{ecriture borne sur borne sup transfo Cauchy}
\enq
  for any $k\in \Cx$ satisfying $|\Im k| \leq  v$.
 
\end{lemme}

\Proof 

First of all, by changing $\Re(t) \hookrightarrow - \Re(t)$ under 
the integral, one may always assume that $\Re(k)>0$. Furthermore, 
for $|\Re(k)|<M$ for some fixed $M$, the integral is well-defined
and the bound \eqref{ecriture borne sur borne sup transfo Cauchy} 
is obvious. Hence, from now on, one may assume $\Re(k)$ to be large 
enough. 

Given $t=u\pm \i (\sg + v)$, 
one has 
\beq
 \big[ \ln (1+|t|) \big]^{r}  \; \leq  \;  \big[ \ln (1+|u|+\sg +v ) 
\big]^{r} \, \leq \, \sul{\ell=0}{r} C^{r}_{\ell} \,  
\big[ \ln (1+|u|) \big]^{\ell}   \cdot 
\big[ \ln(1+\sg +v ) \big]^{r-\ell} \;, 
\enq
with $C^{r}_{\ell}$ being the binomial coefficients. 

Given the same parameterisation for $t$, since  
\beq
|k-t| \, \geq \, \f{1}{3} \Big\{ \sg + |x-u| \Big\}  
\quad \e{where} \quad  k = x+\i \Im(k)  \quad \e{as}\, 
\e{well}\, \e{as} \quad  1 + |t|\geq 1+|u| \, , 
\enq
one gets the upper bound 
\beq
 \Int{ \R \pm \i (\sg +v) }{} \hspace{-4mm} \dd t \cdot 
\f{  \big[ \ln (1+|t|) \big]^{r} }{ (1+|t|)\cdot  |k-t|}  \, 
\leq  \,  \sul{\ell=0}{r} 3 C^{r}_{\ell} \, 
\big[ \ln(1+\sg +v ) \big]^{r-\ell} \, \mc{I}_{\ell} 
 \quad \e{with} \quad \mc{I}_{\ell}\, = \,  \Int{ \R  }{}  
\f{  \dd u \cdot \big[ \ln (1+|u|) \big]^{\ell}    }{ (1+|u|)\cdot  
(\sg + |x-u|) }  \;. 
\enq
Then, one may decompose $\mc{I}_{\ell}$ as
\beq
\mc{I}_{\ell} \, = \, \underbrace{ \Int{ -\infty }{0}    
\f{  \dd u \cdot \big[ \ln (1-u) \big]^{\ell} }{ (1-u)\cdot  
(\sg + x-u) } }_{ = \mc{I}_{\ell}^{(1)} } \, + \, 
\underbrace{ \Int{ 0}{x}    \f{   \dd u \cdot 
\big[ \ln (1+u) \big]^{\ell} }{ (1+u)\cdot  
(\sg + x-u) } }_{ = \mc{I}_{\ell}^{(2)} } \, + \, 
\underbrace{ \Int{x}{ +\infty }   \f{   \dd u \cdot 
\big[ \ln (1+u) \big]^{\ell} }{ (1+u)\cdot  
(\sg +u- x) } }_{ = \mc{I}_{\ell}^{(3)} } \;. 
\enq
$\mc{I}_{\ell}^{(2)}$ may be estimated by direct bounds
\beq
\mc{I}_{\ell}^{(2)} \, \leq \,  \f{ \big[ \ln (1+x) 
\big]^{\ell} }{  1 + x + \sg }\Int{ 0}{x}  \dd u \cdot 
\bigg( \f{ 1 }{ (1+u) } \, + \, \f{1}{  (\sg + x-u) } \bigg)
\, = \, \f{ \big[ \ln (1+x) \big]^{\ell} }{  1 + x + \sg } 
\cdot \Big( \ln(1+x) \, - \,  \ln \sg \, + \, \ln (\sg + x) \Big) \;. 
\enq
To estimate $\mc{I}_{\ell}^{(3)}$ one first decomposes it as 
\beq
\mc{I}_{\ell}^{(3)} \, = \, \Int{0}{ +\infty }    
\f{  \dd u \cdot \big[ \ln (1+u+x) \big]^{\ell} }{ (1+u+x)
\cdot  (\sg +u) }  \, = \, 
\Int{0}{ x  }   \dd u    \f{  \big[ \ln (1+u+x) 
\big]^{\ell} }{ 1+x-\sg} \bigg( \f{ 1 }{ \sg + u } \, 
- \, \f{1}{ 1 + u + x }  \bigg)  
\, + \, \Int{0}{ +\infty } \f{  \dd u \cdot  \big[ \ln (1 + u + 2 x) 
\big]^{\ell} }{ (1 + u + 2x)\cdot  (\sg +u+x) }\,.\quad
\enq
The logarithmic term in the first integral may be bounded by 
$ \big[ \ln (1+2x) \big]^{\ell}$ while, in the second integral 
one uses the bound valid for $x$ large enough
\beq
\f{ 1 + 2 x + u }{ \sg + x + u } \, \leq  \, \f{ 1+2x}{ \sg + x} \;, 
\enq
so as to integrate only a function of the variable $ 1 + 2 x + u $. 
Then, the identity 
\beq
\Int{0}{+\infty} \dd u \f{ \big[ \ln (A+u)\big]^{\ell} }{ (A+u)^2} \, 
= \, \f{1}{A} \sul{p=0}{\ell}  \f{ \ell ! }{ p! } 
\big[ \ln A\big]^{p} 
\enq
leads to 
\beq
\mc{I}_{\ell}^{(3)} \, \leq  \,  \f{  \big[ \ln (1+u+x) \big]^{\ell} }{ 1+x-\sg} \Big\{ \ln(\sg+x)-\ln(1+2 x) - \ln \sg + \ln(1+x)  \Big\}
\, +\,  \sul{p=0}{\ell} \f{ \ell ! }{ p! }  \f{ \big[ \ln (1+2x) \big]^{ p }  }{ \sg + x }  \;. 
\enq
By using analogous techniques, one may bound $\mc{I}_{\ell}^{(1)}$ 
concluding that
\beq
\mc{I}_{\ell}^{(1)}=\e{O}\Big( \f{ \big[ \ln (1+x) \big]^{\ell+1} }{ 1 + x } \Big)\, .
\enq
All in all, this entails the claim. \qed

\end{document}